\documentclass[fleqn,usenatbib]{mnras}

\usepackage{newtxtext,newtxmath}
\usepackage[T1]{fontenc}
\usepackage{xcolor}
\DeclareRobustCommand{\VAN}[3]{#2}
\let\VANthebibliography\thebibliography
\def\thebibliography{\DeclareRobustCommand{\VAN}[3]{##3}\VANthebibliography}

\usepackage{graphicx}	% Including figure files
\usepackage{amsmath}	% Advanced maths commands
\usepackage[normalem]{ulem} % including possibility of striking through some text with \sout{}

\title[Kilonova modeling with \texttt{SNEC}]{Radiation hydrodynamics modeling of kilonovae with \texttt{SNEC}}

\author[Z.~Wu et al.]{
Zhenyu Wu$^{1}$\thanks{E-mail: zhenyuwu99@gmail.com},
Giacomo Ricigliano$^{2}$,
Rahul Kashyap$^{3,4}$,
Albino Perego$^{2,6}$,
and David Radice$^{3,4,5}$
\\
$^{1}$ School of Astronomy and Space Science, Nanjing University, Nanjing 210023, China\\
$^{2}$ Dipartimento di Fisica, Università di Trento, Via Sommarive 14, 38123 Trento, Italy \\
$^{3}$ Institute for Gravitation and the Cosmos, The Pennsylvania State University, University Park, PA 16802, USA \\
$^{4}$ Department of Physics, The Pennsylvania State University, University Park, PA 16802, USA \\
$^{5}$ Department of Astronomy \& Astrophysics, The Pennsyvlania State University, University Park, PA 16802, USA \\
$^{6}$ INFN-TIFPA, Trento Institute for Fundamental Physics and Applications, ViaSommarive 14, I-38123 Trento, Italy
}

\date{Accepted 2022 February 8. Received 2022 January 10; in original form 2021 November 17}

\pubyear{2021}

\begin{document}
\label{firstpage}
\pagerange{\pageref{firstpage}--\pageref{lastpage}}
\maketitle

\begin{abstract}
    We develop a method to compute synthetic kilonova light curves that combines numerical relativity simulations of neutron star mergers and the \texttt{SNEC} radiation-hydrodynamics code. We describe our implementation of initial and boundary conditions, r-process heating, and opacities for kilonova simulations. We validate our approach by carefully checking that energy conservation is satisfied and by comparing the \texttt{SNEC} results with those of two semi-analytic light curve models. We apply our code to the calculation of color light curves for three binaries having different mass ratios (equal and unequal mass) and different merger outcome (short-lived and long-lived remnants). We study the sensitivity of our results to hydrodynamic effects, nuclear physics uncertainties in the heating rates, and duration of the merger simulations.
    We find that hydrodynamics effects are typically negligible and that homologous expansion is a good approximation in most cases. However, pressure forces can amplify the impact of uncertainties in the radioactive heating rates. We also study the impact of shocks possibly launched into the outflows by a relativistic jet. None of our models match AT2017gfo, the kilonova in GW170817. This points to possible deficiencies in our merger simulations and kilonova models which neglect non-LTE effects and possible additional energy injection from the merger remnant, and to the need to go beyond the assumption of spherical symmetry adopted in this work. 
\end{abstract}


%%%%%%%%%%%%%%%%%%%%%%%%%%%%%%%%%%%%%%%%%%%%%%%%%%%%%%%%%%%%%%%%%%%%%%%%%%%%%

% ===========================================================================
\section{Introduction}
% ===========================================================================
The orbit of compact binary neutron-star neutron-star (NSNS) and neutron-star black-hole (NSBH) systems decays due to the emission of gravitational waves. Eventually, the two components of these binaries collide and merge. This process produces abundant gravitational radiation that can be detected by ground-based observatories such as LIGO, Virgo, and KAGRA \citep{KAGRA:2013rdx}. Tidal torques and shocks during these mergers can eject neutron rich material, the so-called dynamical ejecta \citep{Ruffert:1995fs, Rosswog:1998hy, Rosswog:2001fh, Rosswog:2003rv, Rosswog:2003tn, Oechslin:2006uk, Sekiguchi:2011zd, Rosswog:2012wb, Bauswein:2013yna, Sekiguchi:2015dma, Radice:2016dwd, Lehner:2016lxy, Sekiguchi:2016bjd, Foucart:2016rxm, Bovard:2017mvn, Radice:2018pdn, Vincent:2019kor, Shibata:2019wef, Radice:2020ddv, Perego:2020evn, Nedora:2020hxc, Foucart:2020qjb, Nedora:2020qtd, Kullmann:2021gvo}. Additional outflows are driven from the merger remnant by neutrino heating, magnetic, and other hydrodynamic effects on a timescale of a few seconds, the so called secular ejecta \citep{Dessart:2008zd, Metzger:2008av, Metzger:2008jt, 2009ApJ...699L..93L, Fernandez:2013tya, Siegel:2014ita, Just:2014fka, Metzger:2014ila, Perego:2014fma, Martin:2015hxa, Fujibayashi:2017puw, Siegel:2017jug, Metzger:2018uni,Fernandez:2018kax, Nedora:2019jhl, Miller:2019mfl, Fujibayashi:2020dvr, Ciolfi:2020wfx, Mosta:2020hlh, De:2020jdt, Just:2021cls, Shibata:2021bbj, Metzger:2021grk}. As this material decompresses it undergoes r-process nucleosynthesis producing heavy elements
\citep[see e.g.][for recent reviews]{Cowan:2019pkx, Perego:2021dpw}. The nuclear decays of the unstable isotopes synthesised by the r-process heat the material and produce an electromagnetic transient known as kilonova \citep{Li:1998bw, Kulkarni:2005jw, Metzger:2010sy, Kasen:2013xka, Tanaka:2013ana, Metzger:2019zeh, Hotokezaka:2021ofe}.

This scenario has been confirmed by the multi-messenger observations of GW170817 \citep{LIGOScientific:2017ync, 2017ApJ...848L..33A, Chornock:2017sdf, Cowperthwaite:2017dyu, Coulter:2017wya, Drout:2017ijr, Evans:2017mmy, Hallinan:2017woc, Kasliwal:2017ngb, Murguia-Berthier:2017kkn, Nicholl:2017ahq, Rosswog:2017sdn, Smartt:2017fuw, Soares-Santos:2017lru, Tanvir:2017pws, Tanaka:2017qxj, Troja:2017nqp, Villar:2017wcc, Waxman:2017sqv, Kasliwal:2018fwk, Waxman:2019png, Margutti:2020xbo}. Possible other kilonova detections have been reported in conjunction with some short gamma-ray burst, also thought to be the result of compact binary mergers \citep{Nakar:2007yr, Berger:2013jza}. These include a possible kilonova associated with GRB 130603B, the first claimed detection of a kilonova, and several other sources \citep{Tanvir:2013pia, Berger:2013wna, Hotokezaka:2013kza, Jin:2013jca, Fong:2013lba, Yang:2015pha, Jin:2016pnm, Jin:2019uqr, Troja:2019ccb, Lamb:2019lao, Rossi:2019fnm}. Kilonovae appear to be commonly produced in NS mergers. However, observations also suggest that there might be significant variability between different events, possibly associated with a diversity in the outcome of NSNS and NSBH mergers \citep{Kawaguchi:2019nju} and in the viewing angle \citep{Korobkin:2020spe, Heinzel:2020qlt}. Possibly due to the uncertain sky localization and larger distances, no kilonova counterpart has been reported for the second binary NS merger observed by LIGO and Virgo, GW190425 \citep{LIGOScientific:2020aai}, or for GW200105 and GW200115, the first two NSBH merger events detected by LIGO and Virgo \citep{LIGOScientific:2021qlt}.

Kilonova emission is produced by an expanding cloud of radioactive ejecta. The dynamics is not unlike that of type Ia (thermonuclear) supernovae. Indeed, analogous analytic arguments can be used to predict the basic features of the light curve in both cases \citep{1980ApJ...237..541A,Li:1998bw, Kulkarni:2005jw, Metzger:2010sy, Chatzopoulos:2011vj, Kashyap:2019ypm}. However, there are some important differences between kilonovae and type Ia supernovae. The expansion velocities of the kilonova outflows can be much larger than those of the supernova ejecta \citep{Hotokezaka:2018gmo, Radice:2018pdn, Nedora:2021eoj, Dean:2021gpd}. The radioactive heating of the kilonova material is not dominated by the decay chain of ${}^{56}{\rm Ni}$ as in supernovae, but it is the result of the individual decays of thousands of unstable nuclides, resulting in a characteristic power law decay \citep{Metzger:2010sy, Roberts:2011xz, Korobkin:2012uy, Lippuner:2015gwa, Hotokezaka:2017dbk}. The thermalization efficiency is also very different among different decay channels \citep{Hotokezaka:2015cma, Barnes:2016umi, Kasen:2018drm, Hotokezaka:2019uwo}. Finally, the opacity of r-process elements produced in NS mergers is much higher than that of the iron produced in type Ia supernovae, particularly when lanthanides are produced \citep{Kasen:2013xka, Tanaka:2013ana, Barnes:2013wka, Fontes:2019tlk, Tanaka:2019iqp}.

The broad features of the color light curves of kilonovae can be reproduced with simple, one zone, semi-analytical models \citep{Li:1998bw, Kulkarni:2005jw, Metzger:2010sy, Villar:2017wcc, Waxman:2017sqv}, using parametrized heating rates and effective grey opacities obtained with Monte Carlo calculations. One of them is that of \citet{Perego:2017wtu} who developed a multi-dimensional semi-analytical framework that included multiple outflow components and geometry information from ab-initio simulations. This model was later used by \citet{Breschi:2021tbm} to perform a joint electromagnetic, gravitational wave parameter estimation for GW170817. More advanced models use moment based \citep{Just:2021vzy} or multi-frequency Monte Carlo radiative transfer calculations \citep{Kasen:2013xka, Tanaka:2013ana, Wollaeger:2017ahm, Kawaguchi:2018ptg, Korobkin:2020spe, Bulla:2020jjr}. Surrogate models that can interpolate detailed Monte Carlo calculations have also been proposed \citep{Coughlin:2018miv}. However, most previous works have ignored the hydrodynamics of the ejecta and adopted the assumption of homologous expansion. A notable exception is the work of \citet{Rosswog:2013kqa} and \citet{Grossman:2013lqa} which performed long-term smoothed particle hydrodynamics (SPH) simulations of the expanding tidal tail ejected in a NS merger. However, those simulations were based on the output of Newtonian NS merger simulations and did not include the contribution from the secular ejecta, which is currently thought to be dominant \citep{Siegel:2019mlp}. Later works combined hydrodynamics simulations of the early phase of the outflows and homologous expansion Monte Carlo radiative transfer calculations \citep{Kawaguchi:2020vbf, Klion:2021jzr, Klion:2020efn}. The studies of \citet{Ishizaki:2021qne} performed long term simulations of the ejecta in a NS merger focusing on the impact of the radioactive heating on the fallback, but did not model the radiative transfer and the light curve from such flows.

In this work, we implement appropriate radioactive heating rates and opacities into the publicly available radiation hydrodynamics code \texttt{SNEC} (SuperNova Explosion Code; \citealt{Morozova:2015bla}) to perform self-consistent calculation of kilonova light curves starting from the output of ab-initio numerical relativity NS merger simulations. This approach allows us to study hydrodynamic effects on kilonova signals that have so far been neglected in calculation employing more sophisticated radiative transfer approaches. \texttt{SNEC} also provides a test platform for the development of microphysics routines that we ultimately plan to include in multi-dimensional calculations. Here, we discuss the implementation details of our code, we validate it against semi-analytic light curve models and by carefully monitoring energy conservation. We use \texttt{SNEC} to study kilonova signals from realistic ejecta profiles obtained from merger simulations and we study the importance of hydrodynamic effects and the sensitivity of our results to nuclear physics uncertainties and to the duration of the simulations. Finally, we study the impact of shocks launched by the GRB jet into the ejecta on the light curves.

The rest of this paper is organized as follows. In Sec.~\ref{section_methods} we describe all of the modifications we have made to the \texttt{SNEC} code to simulate kilonovae. In Sec.~\ref{section_code_validation}  we validate the code by checking energy conservation and comparing the results with two alternative semi-analytic models. In Sec.~\ref{section_first_applications_of_SNEC} we introduce the general features of the light curves from three realistic profiles. Then, we study the effects of various factors, including hydrodynamics, uncertainties in heating rates, duration of binary neutron-star (BNS) merger simulations, and the presence of shocks. We summarize and conclude in Sec.~\ref{section_conclusion_and_discussion}.

% ===========================================================================
\section{Methods}
\label{section_methods}
% ===========================================================================

    \subsection{Brief overview of \texttt{SNEC}} 
    \texttt{SNEC}, the SuperNova Explosion Code, is a spherically symmetric (1D) Lagrangian radiation-hydrodynamics code, primarily used to simulate core-collapse supernova explosions and generate synthetic color light curves \citep{Piro:2015kro, Morozova:2015bla, Morozova:2016asf, Morozova:2016efp, Morozova:2017hbk, Morozova:2019hiu}. 
    The \texttt{SNEC} code mainly uses Paczynski equation of state (EOS) \citep{paczynski1983models, weiss2004cox}, which includes the contributions from ions, electrons and radiation. To get the fractions of atoms in different ionization states, \texttt{SNEC} solves the Saha equations. The code uses matter opacities $\kappa$ from existing tables of Rosseland mean opacities \citep{Iglesias:1996bh} as a function of composition, temperature and density. \texttt{SNEC} accounts for the radioactive heating due to $^{56}$Ni and ${}^{56}{\rm Co}$ and implements a simplified treatment of the associated $\gamma$-ray emission and thermalization. More details on the code can be found on \texttt{SNEC}'s website\footnote{\url{https://stellarcollapse.org/index.php/SNEC.html}}.

    Kilonovae are powered by the radioactive decay of r-process elements synthesized in the ejecta. We use some of the \texttt{SNEC} modules, but modify others to model kilonova emission. The main differences between the original \texttt{SNEC} code and our kilonova code are the opacities (\S \ref{subsection_opacity}), heating rates (\S \ref{subsection_heating}) and initial conditions (\S \ref{subsection_initial_boundary}). Other differences are described in \S \ref{subsection_other_differences}. \S \ref{subsection_lightcurve} gives the formulae to calculate light curves in our model.

    \subsection{Opacities} 
    \label{subsection_opacity}
    
    Unlike supernovae, which are powered by iron group elements, r-process can generate heavier elements, including lanthanides and actinides. If present, lanthanides and actinides can increase the ejecta opacity by more than one order of magnitude to $\sim$ 10 cm$^2$ g$^{-1}$. The resulting strong optical line blanketing shifts the emission towards infrared bands \citep[the so called red kilonova,][]{Barnes:2013wka, Roberts:2011xz, Kasen:2013xka, Tanaka:2013ana}. Whether or not these elements are produced by the r-process nucleosynthesis mainly depends on the electron fraction $Y_e$ of the ejecta, for the low entropy and fast expansion conditions expected in the ejecta from binary NS mergers. If $Y_e$ $\lesssim$ 0.25, then the ejecta will be lanthanide-rich. If $Y_e$ $\gtrsim$ 0.25, then r-process nucleosynthesis runs out of free neutrons before lanthanides can be produced \citep{1997ApJ...482..951H,Lippuner:2015gwa}.
    
    In our model, we adopt grey opacity ranging from $1.0$ cm$^2$ g$^{-1}$ to $10.0$ cm$^2$ g$^{-1}$, which we take to be a function of the initial $Y_e$. Our choice is motivated by the study of \citet{Tanaka:2017lxb} which showed that bolometric light curves computed assuming grey opacity in this range are in good agreement with those obtained with wavelength-dependent radiation transfer results. A similar range is adopted in \citep{Villar:2017wcc} to fit AT2017gfo, although their lower bound is smaller. We use the following formula to set the opacity:
    \begin{equation}
    	\label{opacity_ye}
    	\kappa = 1 + \frac{9}{1+(4Y_e)^{12}}~~ \mathrm{ [cm^2 g^{-1}]}.
    \end{equation}
    This smoothly transits from $1.0$ cm$^2$ g$^{-1}$ to $10.0$ cm$^2$ g$^{-1}$. Accordingly, the opacity corresponding to $Y_e = 0.25$ is $5.5$ cm$^2$ g$^{-1}$. This formula reproduces the expected rapid change in opacity at around $Y_e \simeq 0.25$. We explore the impact of the slope of the transition at $Y_e \simeq 0.25$ in Appendix \ref{appendix_impact_of_opacity_formula}). There we show that the light curves are mostly insensitive to it.
    \begin{figure}
        \centering
        \includegraphics[scale=0.5]{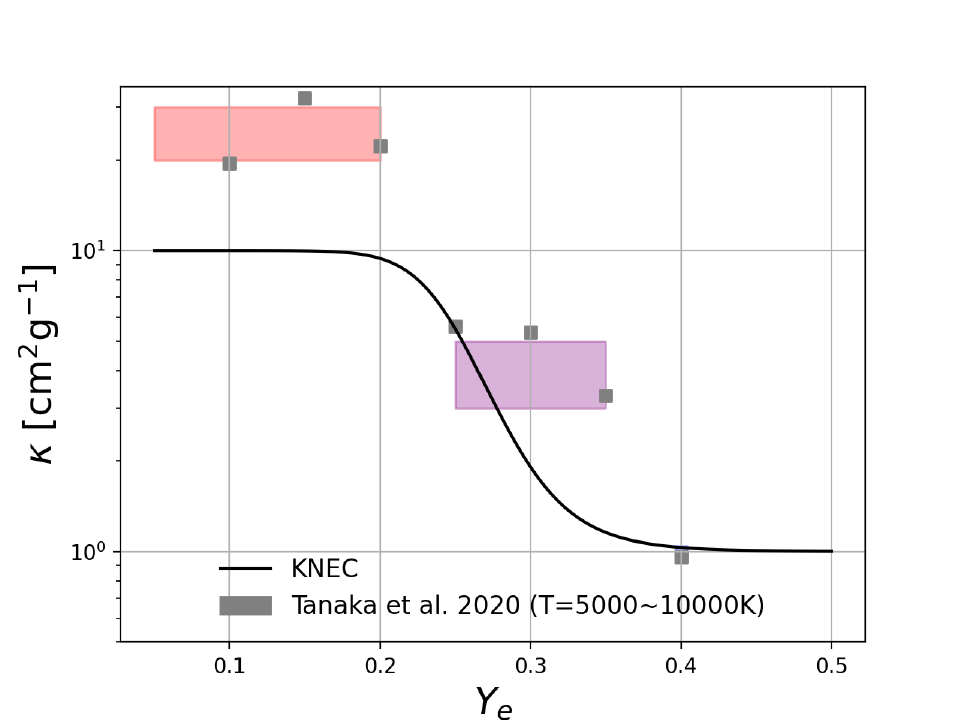}
        \caption{The solid line is opacity as a function of $Y_e$ in our model  (Equation \ref{opacity_ye}). The small grey squares show data from \citet{Tanaka:2019iqp}, and the large rectangles are the suggested opacity ranges in their paper at 5000 $\sim$ 10000 K. Note that the opacities from \citet{Tanaka:2019iqp} decrease steeply at lower temperature. The opacities used in our calculations are somewhat smaller, since we take $10$ cm$^2$ g$^{-1}$ as their maximum value.}
        \label{opacity_Ye}
    \end{figure}

    Figure \ref{opacity_Ye} shows the comparison between our opacity model with the results of \citet{Tanaka:2019iqp}. We remark that our model does not account for changes in the opacities, for example due to recombination, which are instead kept constant throughout our simulations. On the other hand, we emphasize that such treatment is consistent with the way these effective gray opacities have been constructed \citep{Kasen:2013xka, Tanaka:2013ana, Tanaka:2019iqp}. To ease the comparison with previous works, we also restrict the maximum opacity to $10\ {\rm cm}^2\ {\rm g}^{-1}$ \citep{Kasen:2013xka, Tanaka:2013ana, Perego:2017wtu, Villar:2017wcc}.
    
    \subsection{Heating rates}
    \label{subsection_heating}
    At the times relevant for kilonovae, the dominant source of heating is constituted by the decays of the heavy elements produced in the r-process nucleosynthesis. This energy release is described in terms of a heating rate which can be computed by evolving the abundances of the numerous characteristic nuclides in time while accounting for their mutual interactions and decays. Nuclear heating simulations are highly dependent on the dynamical and thermodynamical conditions of the ejecta, and in particular on the entropy, electron fraction and expansion timescale at the freeze-out from nuclear statistical equilibrium \citep[NSE, see, e.g.,][]{1997ApJ...482..951H,Lippuner:2015gwa}. In addition, simulations also depend on the nuclear physics inputs: distinct theoretical nuclear mass models, reaction rates or fission fragment distributions can lead to significantly different heating rates. This sensitivity is particularly strong at low electron fractions and the nuclear physics uncertainties can lead to changes in the predicted heating rates of about an order of magnitude \citep{Rosswog:2016dhy,Zhu_2021}.
    
\begin{figure*}
    \centering
    \includegraphics[width=.95\linewidth]{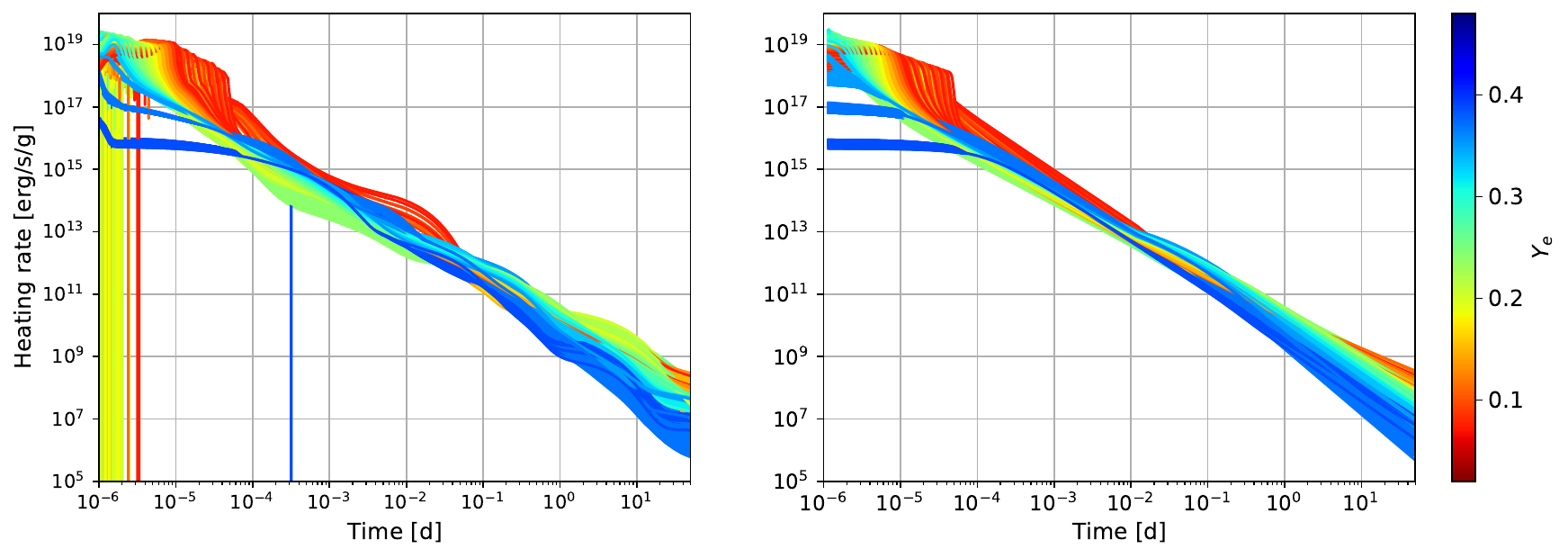}
    \caption{Heating rate trajectories for a grid of thermodynamic variables $0.05\leq Y_e\leq0.4$, $3$ $\mathrm{k_B/baryon}$ $\leq s\leq50$ $\mathrm{k_B/baryon}$ and $1$ ms $\leq\tau\leq30$ ms, corresponding (for visual clarity) to a subset of grid used in this work, as obtained by \texttt{SkyNet} (left) and as result of the fit discussed in the text (right). Trajectories are color-coded according to the initial electron fractions. The vertical lines visible for some of the \texttt{SkyNet} trajectories correspond to sudden endoenergetic changes in the nuclear composition, possibly occurring during the $r$-process nucleosynthesis, which are averaged out in the fit procedure and do not significantly affect the heating rate at later times. The fitted heating rates agree well with the \texttt{SkyNet} calculations.}
    \label{heatrates}
    \end{figure*}

    Here, we consider the time-dependent heating rates resulting from the broad nucleosynthesis calculations reported in \citet{Perego:2020evn}. %where more details can be found. 
    In that work, the nuclear abundance evolution of Lagrangian fluid elements was performed using the nuclear reaction network \texttt{SkyNet} \citep{Lippuner:2017tyn} with the finite-range droplet macroscopic model \citep[FRDM,][]{Moller:2015fba} for the nuclear masses. Each \texttt{SkyNet} run was initialized from the electron fraction $Y_e$, entropy $s$, and expansion timescale $\tau$ at a temperature of $6$ GK in NSE conditions. 
    More details about these nucleosynthesis calculations can be found in \citet{Perego:2020evn}. 
    The heating rates used in this work were computed over a comprehensive grid of $11700$ distinct trajectories with $0.01\leq Y_e\leq0.48$ linearly spaced, $1.5$ $k_{\rm B}~{\rm baryon^{-1}}$ $\leq s\leq 200$ $k_{\rm B}~{\rm baryon^{-1}}$ and $0.5$ ms $\leq\tau\leq200$ ms log-spaced. These intervals are expected to bracket the properties of the ejecta from BNS and NSBH mergers. In the left panel of Fig. \ref{heatrates} we report the heating rates obtained for the most representative conditions expected in the ejecta from NSNS mergers.
    
    In order to derive the heating rate for arbitrary initial conditions, we construct fits to the trajectories obtained with \texttt{SkyNet}. The fits describe the heating rate over a time interval ranging from $0.1$ seconds to $50$ days after the merger. The fitting function distinguishes between two regimes. At early times, $t\lesssim0.1$ days, we use the analytic fitting formula proposed by \citet{Korobkin:2012uy}, which was also derived from detailed nucleosynthesis calculations:
    \begin{equation}
    \label{eqKorfit}
        \dot{\epsilon}_{\mathrm{r}}(t)=\epsilon_0\left[\frac{1}{2}-\frac{1}{\pi}\arctan{\left(\frac{t-t_0}{\sigma}\right)}\right]^{\alpha},
    \end{equation}
    where $\epsilon_0$, $\alpha$, $t_0$ and $\sigma$ are fitting parameters. At later times, $t\gtrsim0.1$ days, we use a power-law fit, thus the fitting formula becomes:
    \begin{equation}
    \label{eqpowfit}
    \dot{\epsilon}_{\mathrm{r}}(t)=\epsilon_0't^{-\alpha'},
    \end{equation}
    where $\epsilon_0'$ and $\alpha'$ are additional fit parameters. The heating rate fits, as obtained from Equations \ref{eqKorfit} and \ref{eqpowfit}, are then joint together by a log-scaled smoothing procedure applied on the time interval $1\times10^3$ s $\leq t\leq4\times10^4$ s, centered on $t\sim0.1$ days in log-scale. The right panel of Fig. \ref{heatrates} shows the fitted version of the heating rate trajectories presented in the left panel.\\
    The quality of a single fit is evaluated using a mean fractional log error as employed in \cite{Lippuner:2015gwa}, defined as:
    \begin{equation}
    \label{eqlogerror}
    \Delta(\dot{\epsilon}_{\mathrm{r}})=\left<\frac{|\ln(\dot{\epsilon}_{\mathrm{r}}^o(t))-\ln(\dot{\epsilon}_{\mathrm{r}}(t))|}{\ln(\dot{\epsilon}_{\mathrm{r}}^o(t))}\right>,
    \end{equation}
    where $\dot{\epsilon}_{\mathrm{r}}^o(t)$ is the original \texttt{SkyNet} heating rate trajectory, while the mean is performed over the entire time window $0.1$ s $\leq t\leq50$ days without weighing over the time steps, in order to account for the original \texttt{SkyNet} resolution. For most trajectories we find relative errors smaller than $\sim1\%$. The largest errors are found at the boundary of the \texttt{SkyNet} grid, where the relative error can be as large as $\sim 5\%$.

    \begin{figure}
    \centering
    \includegraphics[scale=0.45]{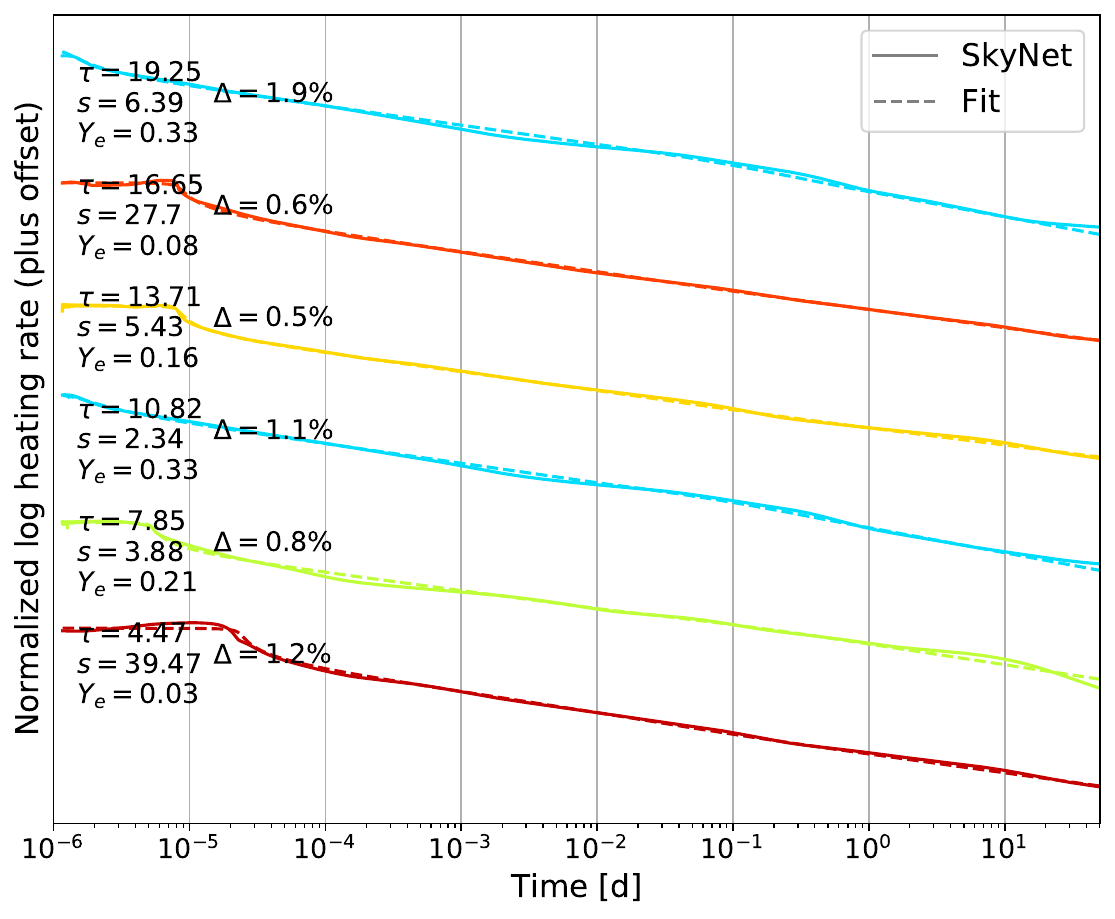}
    \caption{Comparison of the heating rates computed with \texttt{SkyNet} and with our fit for five different representative sets of off-grid thermodynamic variables. We also report the relative errors in the fits computed using Eq.~\eqref{eqlogerror}. The typical errors in the heating rate due to the fitting procedure are of at most a few percent.}
    \label{fiterror}
    \end{figure}
    
    The fitting coefficients are usually smooth functions of the thermodynamic variables, in particular for $Y_e\leq0.36$, $s\leq90$ $k_{\rm B}~{\rm baryon^{-1}}$ and $\tau\leq30$ ms. Isolated points or boundary regions for which the continuity of the fitting coefficients was poor were removed from the fit. Since the regions where the parameters evolve smoothly are the most relevant for our calculations, we adopt a trilinear interpolation of the fitting coefficients as a function of $Y_e$, $s$, and $\tau$. We validate this procedure by computing the error in the heating rate due to the fitting procedure for new \texttt{SkyNet} trajectories generated with input thermodynamic variables distinct from those used to construct the fit. The results for a subset of these trajectories are shown in Fig.~\ref{fiterror}. We find that the relative error of the fitting procedure is less than ${\sim}1\%$, well below the expected nuclear physics uncertainties.
    
  \subsection{Initial and boundary conditions} 
  \label{subsection_initial_boundary}
   
   We consider two kinds of ejecta profiles: (i) analytic wind profiles for code validation and parameter comparison, and (ii) realistic profiles extracted from NR simulations of merging neutron stars obtained with the \texttt{WhiskyTHC} code \citep{Radice:2012cu, Radice:2013hxh, Radice:2013xpa, Radice:2015nva, Radice:2016dwd, Radice:2017zta, Radice:2018pdn}. Both types of profile correspond to spherically symmetric outflows for which radius, temperature, density, velocity, initial $Y_e$, initial entropy, and expansion timescale are given as a function of the enclosed mass. The initial entropy and expansion timescale are new quantities that we have introduced and that are used to compute the heating rates and the opacities as discussed above. \texttt{SNEC} already tracks the electron fraction of the material, however our calculations only use the initial $Y_e$ of the matter. While this is consistent with our treatment of heating rates and opacity, which depends on the initial $Y_e$, this introduces an error at the level of the EOS, since we do not correctly account for the pressure contribution from free electrons. We leave the mean degree of ionization as a free parameter instead. As shown in \S \ref{subsection_other_differences}, our tests indicate that this is a negligible effect, since matter is still dominated by the radiation pressure when homologous expansion sets in.

   \begin{figure}
   \centering
   \includegraphics[scale=0.4]{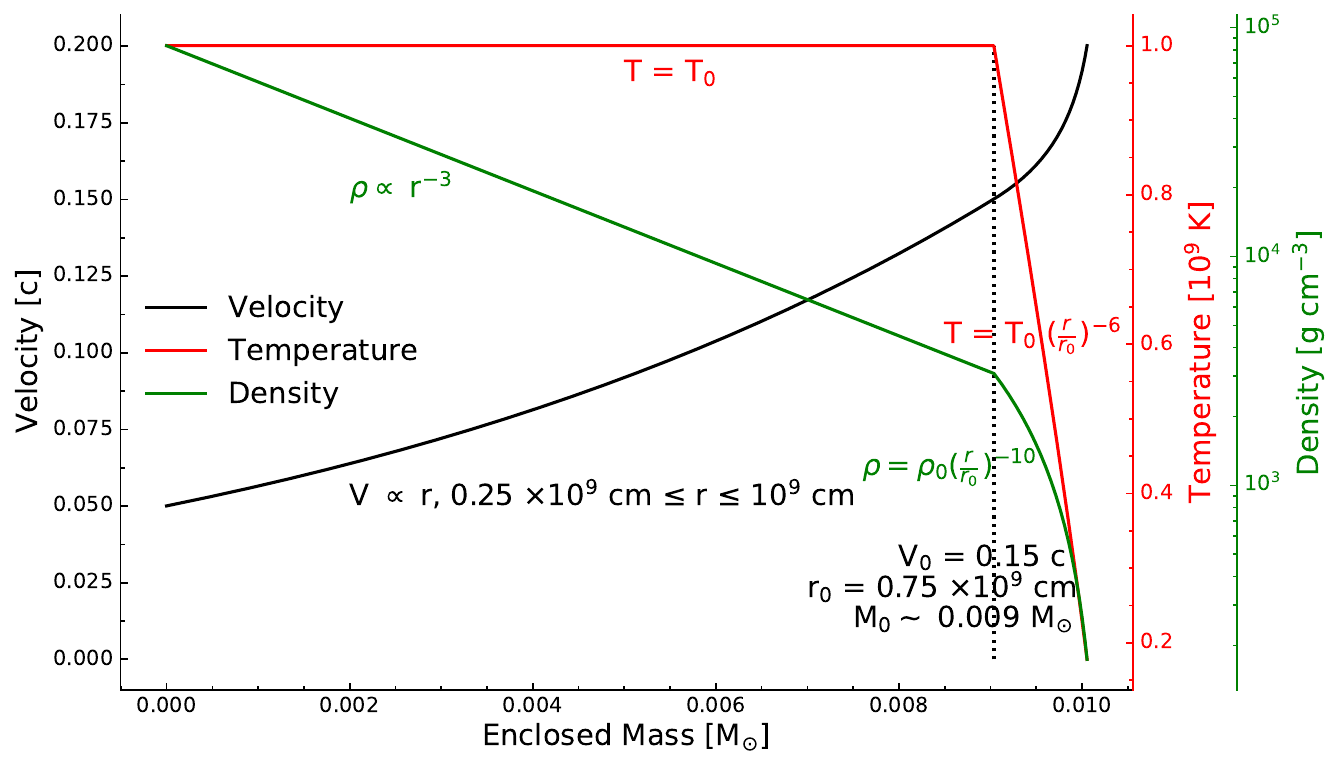}
   \caption{wind310T6 (optimal wind) profile: velocity, temperature, and density as a function of mass. The velocity is proportional to the radius. The maximum velocity and maximum radius are set to 0.2~c and $10^9$~cm, respectively. The density decays with radius with a power law exponent of 3, in the interior of the outflows, and of 10, in the outer regions, hence the name \textit{wind310}. The turning point between the two power laws $r_0$ is 0.75 $\times 10^9$~cm. The profile is designed with two power law factors in order to fit homologous expansion. Outside $r = r_0$, the temperature drops with radius with a power law factor of 6. This temperature drop reduces the otherwise large pressure gradients that would otherwise be present at the outer boundary producing very large expansion velocities.}
   \label{wind310T6_profile}
   \end{figure}
  
   We design analytic wind profiles similar to \cite{Metzger:2010sy} and \cite{Tanaka:2013ana}. The velocity is proportional to the radius and ranges between 0.05 c and 0.2 c. The maximum radius is set to $10^{9}$ cm and the total mass is 0.01 M$_{\odot}$ by default. The initial electron fraction $Y_e$, entropy $s$, and expansion timescale $\tau$ are uniform in the ejecta. We set $s$ to 10 $k_{\text{B}}~{\rm baryon^{-1}}$ and $\tau$ to 10 ms. In fact, the heating rates are relatively insensitive to $s$ and $\tau$. Inspired by \cite{Ishizaki:2021qne}, we use two power laws to describe the density as a function of $r$: 
    \begin{equation}
    \left\{\begin{array}{rl}
     \rho \propto r^{- k_1} & {\rm for~} r \le r_0 \, ,\\
     \rho \propto r^{- k_2} & {\rm for~} r \ge r_0 \, ,
    \end{array}\right.
    \end{equation}
    where $k_1$ is set to 3, and $k_2$ should be larger to represent a steep drop in density near the outer boundary. We experiment with various $k_2$ and find that $k_2 \gtrsim$ 10 produces results for which there is good agreement between the full radiation-hydrodynamics calculations and calculations assuming homologous expansion (see \S \ref{subsection_hydrodynamics}). $r_0$ is set to 0.75 $\times 10^{9}$ cm by default. We also use piecewise functions for the temperature $T$:
    \begin{equation}
    \label{wind_initial_teperature}
    \left\{\begin{array}{rl}
     T = T_0 = 10^9 K &  {\rm for~} r \le r_0,\\
     T  \propto r^{- \alpha} &  {\rm for~} r \ge r_0 \, .
    \end{array}\right.
    \end{equation}
    We find that the use of a power law decay for $T$ prevents the appearance of large pressure gradients at the outer boundary of our Lagrangean grid, which can otherwise generate unphysically large velocities for this type of artificial wind profiles.
    We find that $\alpha \gtrsim 6$ is enough to avoid this artifact. We denote the profile with ($k_1 = 3, k_2 = 10, \alpha = 6$) as \textit{wind310T6} profile, or \emph{optimal wind} profile (Figure \ref{wind310T6_profile}). More details regarding the boundary velocity problem with analytical wind profiles are discussed in Appendix \ref{appendix_boundary_velocity}.

   \begin{figure}
   \centering
   \includegraphics[scale=0.4]{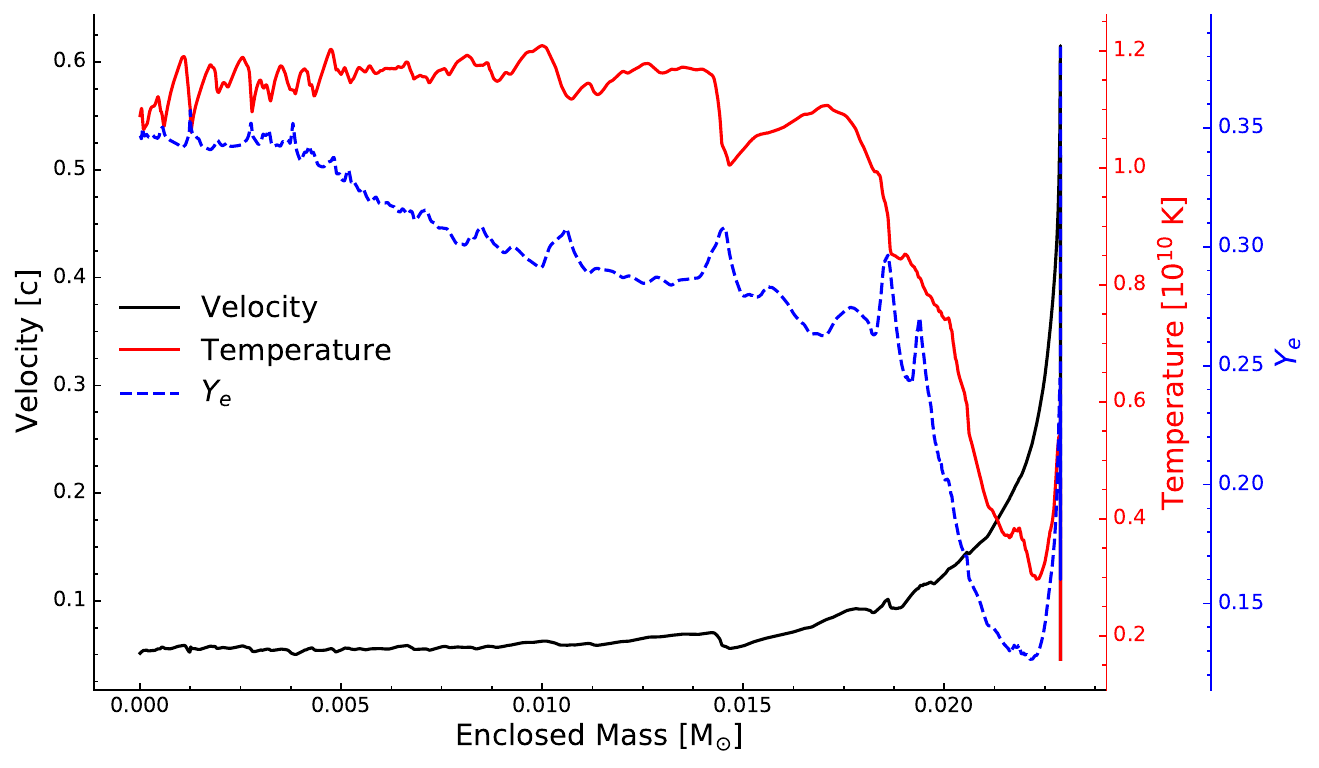}
   \caption{BLh profile: velocity, temperature, and $Y_e$ as a function of mass. The profile is taken from \texttt{WhiskyTHC} simulation of binary NS merger (1.4 and 1.2 $M_\odot$, BLh EOS) at ${\sim}$ 0.11 s after merger. The velocity is almost constant, but rises sharply to $\sim$ 0.6 c near the outer boundary. The low-$Y_e$ component near the outer boundary is often referred to as ``lanthanide curtain''. However, there exists a high-$Y_e$ component at the outermost tail of the ejecta.}
   \label{blh_profile}
   \end{figure}
   
   \begin{figure}
   \centering
   \includegraphics[scale=0.4]{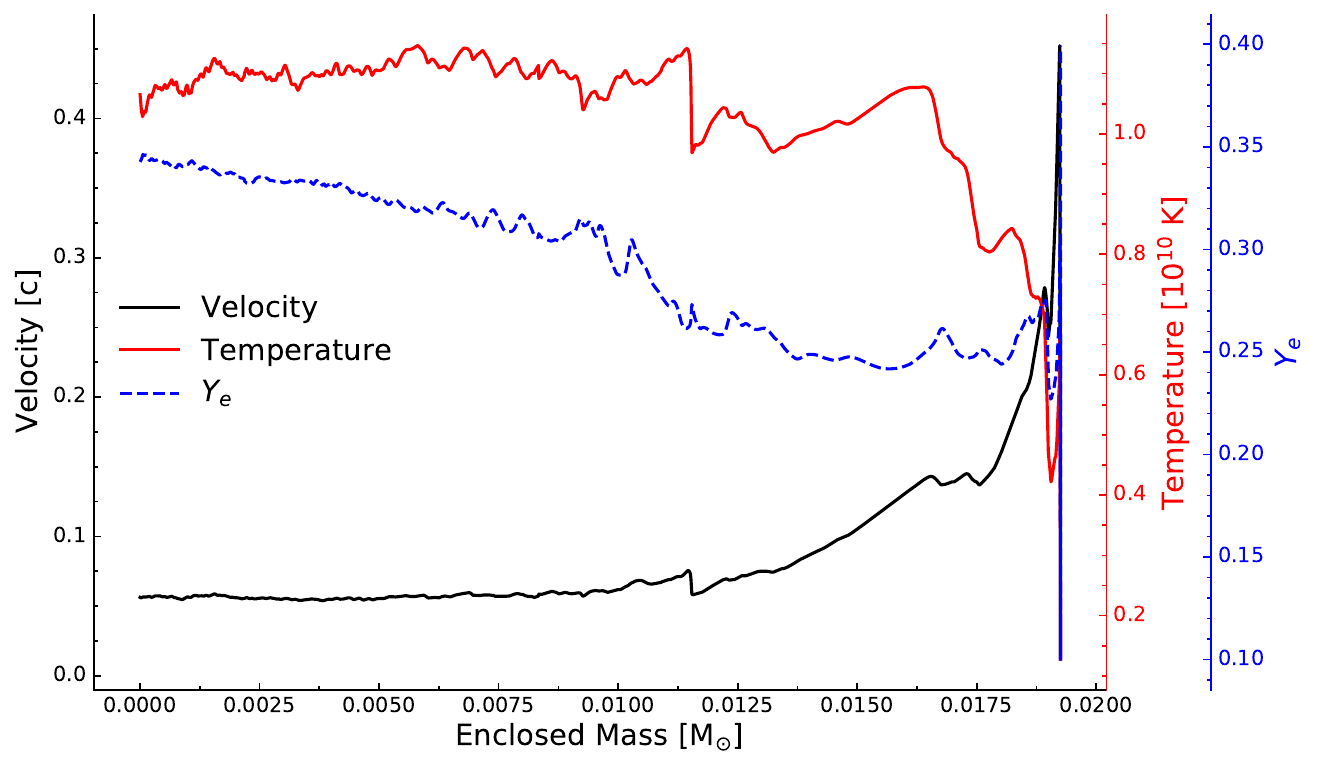}
   \caption{DD2 profile: velocity, temperature, and $Y_e$ as a function of mass. The profile is taken from \texttt{WhiskyTHC} simulation of binary NS merger (1.36 and 1.36 solar mass, DD2 EOS) at ${\sim}$ 0.11 s. Most part of the DD2 profile has a $Y_e$ larger than $0.25$. This is due to lack of the low-$Y_e$ tidal component, because the neutron stars are of equal mass here.}
   \label{DD2_profile}
   \end{figure}
    
   The NR profiles are constructed from outflow data recorded on fixed coordinate spheres as a function of time. In particular, we record the properties of matter crossing a sphere of radius $r \simeq 295\ {\rm km}$. We only consider matter that is unbound according to the Bernoulli criterion, that is with $h u_t < -1$, $h$ being the enthalpy and $u_t$ the covariant time component of the fluid four velocity. Thermodynamic properties of the material, including $Y_e$, are then converted to spherical symmetry using a mass-weighted average and tabulated as a function of the enclosed ejecta mass, $m$. Since, the $Y_e$ depends sensitively on the polar angle \citep{Perego:2017wtu}, this procedure introduces a systematic error in the computed light curves. We plan to address this in the future by performing isotropic-equivalent calculations that consider polar and equatorial ejecta separately. Since \texttt{SNEC} needs initial data at a fixed time and not inner boundary data as a function of time, we transform the data assuming homologous expansion. In particular, we compute $r(m)$ from the requirement that
   \[
        m(r) = 4 \pi \int_0^r \rho\, r^2\, \mathrm{d}r. 
   \]
   For this study we consider the following three binaries.
   \begin{enumerate}
       \item A $1.4\ M_\odot - 1.2\ M_\odot$ binary simulated with the BLh EOS \citep{Logoteta:2020yxf, Bombaci:2018ksa, Bernuzzi:2020txg} and evolved until $106\ {\rm ms}$ after merger. This binary produced a long-lived remnant. It is discussed in detail in \citet{Prakash:2021wpz}.
       \item A $1.364\ M_\odot - 1.364\ M_\odot$ binary targeted to GW170817 and simulated with the DD2 EOS  \citep{Typel:2009sy, Hempel:2009mc} until $113\, {\rm ms}$ after the merger. This binary produced a long-lived remnant. This system is discussed in detail in \citet{Nedora:2019jhl, Nedora:2020hxc}.
       \item A $1.4\ M_\odot - 1.2\ M_\odot$ binary simulated with the SFHo EOS \citep{Steiner:2012rk} and evolved until $32\ {\rm ms}$ after the merger. This binary produced a short-lived remnant. It is discussed in detail in \citet{Radice:2018pdn}.
   \end{enumerate}
   All three simulations modeled neutrino emission and re-absorption using the M0 scheme of \citet{Radice:2016dwd}. The DD2 binary also included a treatment of viscous angular momentum transport using the GRLES formalism \citep{Radice:2017zta, Radice:2020ids}.  We will refer to the three profiles generated from these simulations as being the DD2, SFHo, and BLh profiles, respectively.
   
   Velocity, temperature, and initial $Y_e$ for the BLh and DD2 profiles are shown in Fig. \ref{blh_profile} and \ref{DD2_profile}. An important difference between these two models is that the BLh ejecta has an outer shell of low-$Y_e$ material ($m \gtrsim 0.019$ M$_{\odot}$) ejected due to the tidal interaction between the two stars shortly prior to merger. This shell is absent for the equal mass DD2 model for which the outflows are driven by shocks and viscous and hydrodynamic torques on the postmerger disk. This ``lanthanide curtain'' leads to very different behaviors between light curves of BLh and DD2 profiles (\S \ref{subsection_general_features}). Both profiles also include a fast expanding moderate $Y_e$ outer shell of material. This mildly-relativistic component of the outflow is accelerated by shocks after the merger, when the remnant bounces back \citep{Radice:2018pdn, Nedora:2021eoj}. The SFHo profile is not shown, but it is qualitatively similar to the BLh profile. It also includes a lanthanide curtain. However, it has a smaller overall amount of ejecta, because black hole formation terminates the spiral-wave driven wind, which is the main mechanism driving the outflows in the first few tens of milliseconds after the merger \citep{Nedora:2020hxc}. Additional outflow is expected on longer timescales due to viscous and nuclear processes in the disk \citep{Shibata:2019wef}, but it is still not possible to simulate the binary over these longer timescales in full 3D numerical relativity. Additionally, we have performed calculations in which we extrapolate the outflow rates from the simulation as a function of time, as discussed in \S \ref{subsection_extrapolation_of_NR_informed_models} and Appendix~\ref{appendix_method_of_blh_extrapolation}.

   At the inner boundary, we keep the velocity constant, i.e. $v_{1} (t) = v_1(t=0)$. Other boundary conditions are the same as in the original \texttt{SNEC} code. Luminosity is zero at the inner boundary ($L_1 = 0$). The artificial viscosity, density, specific internal energy, temperature, and pressure all vanish at the outer boundary, while the luminosity is extrapolated at first order ($Q_{\text{imax}} = 0$, $\rho_{\text{imax}+1/2} = 0$, $\epsilon_{\text{imax}+1/2} = 0$, $T_{\text{imax}+1/2} = 0$, $p_{\text{imax}+1/2} = 0$, $L_{\text{imax}} = L_{\text{imax}-1}$).
   
   \subsection{Other differences from \texttt{SNEC}}
    \label{subsection_other_differences}
     
   \paragraph*{Composition and EOS.}
   The \texttt{SNEC} code computes the electron number density $n_e$ and mean degree of ionization $\bar{y}$ by solving Saha equations. Due to the complexity of the ejecta compositions and the lack of detailed knowledge of ionization energies for r-process elements, we are not able to solve the Saha equations here. Instead, we take $\bar{y}$ to be a free parameter in our code. We also provide another free parameter, the mean molecular weight $\mu$, such that $n_e = \frac{\bar{y} \rho}{m_p \mu}$, where $m_p$ is the mass of the proton.
   In the calculations presented here, we fix the mean degree of ionization $\bar{y}$ to 2 and mean molecular weight $\mu$ is set to 100.  In the calculation of the electron contribution to the pressure, we also fix the electron fraction to be 0.4. We remark that this electron fraction value is different from the initial $Y_e$ used for the opacity and heating rates calculations. It roughly corresponds to the electron fraction of matter at the end of the nucleosynthesis. We have checked that our results are insensitive to these choices by performing test calculations with $\bar{y}$ varying between 1 and 50 and $\mu$ varying between 50 and 150\footnote{Changes in the post-nucleosynthesis $Y_e$ are degenerate with $\bar{y}$ and $\mu$.}. We found that these parameters have a negligible impact on light curves. This is expected, since matter is radiation pressure dominated during the early phases of the expansion when pressure gradients drive the evolution of the outflows. Moreover, we neglect ionization correction terms in the specific internal energy and the partial derivative terms as shown below.
     
   The ejecta EOS we use is basically the same as the Paczynski EOS in the original \texttt{SNEC} code, but the ionization correction terms are omitted. It is useful to go through the detailed calculations in the note on \texttt{SNEC}'s website, and compare them with our expressions shown below. In fact, \texttt{SNEC}'s notes are based on \cite{paczynski1983models} with the addition of corrections due to partial ionization \citep{weiss2004cox}. The total pressure contains the contributions from ions, electrons and radiation 
   \begin{equation}
       p = P_{\text {ion}}+P_{\mathrm{e}}+P_{\mathrm{rad}} \, .
   \end{equation}
   In the original \texttt{SNEC} codes, the specific internal energy $\epsilon$ is expressed as 
   \begin{equation}
   \label{specific_internal_energy-snec}
      \epsilon=\frac{3}{2} N k_{\mathrm{B}} T+\frac{1}{f-1} \frac{P_{\mathrm{e}}}{\rho}+\frac{a T^{4}}{\rho}+N\left\{\sum_{k} \nu_{k}\left[\sum_{s} y_{s}^{k}\left(\sum_{m=1}^{s} \chi_{m-1}^{k}\right)\right]\right\}
   \end{equation}
   where $N$ is the number of ions per unit mass. $\nu_k$ is the number abundance of $k$-th element and $y_s^k$ is the degree of $s$-th ionization of the $k$-th element. $\chi_{m-1}^k$ is the ionization energy for the ionization process $(m-1)$-th state $ \rightarrow$ $m$-th state of $k$-th element. Since we do not have this information, we ignore the ionization correction term and use a simplified expression instead:
   \begin{equation}
       \epsilon=\frac{3}{2} N k_{\mathrm{B}} T+\frac{1}{f-1} \frac{P_{\mathrm{e}}}{\rho}+\frac{a T^{4}}{\rho}
   \end{equation}
   For the same reason, the partial derivative terms are simplified to: 
   \begin{equation}
	 \left(\frac{\partial \epsilon}{\partial T}\right)_{\rho}=\frac{3}{2} N k_{\mathrm{B}}+\frac{4 a T^{3}}{\rho}+\frac{1}{f-1} \frac{P_{\mathrm{end}}^{2}}{P_{\mathrm{e}} \rho T}
   \end{equation} 
   \begin{equation}
      \left(\frac{\partial p}{\partial T}\right)_{\rho}=N k_{\mathrm{B}} \rho+\frac{4 a T^{3}}{3}+\frac{P_{\mathrm{end}}^{2}}{P_{\mathrm{e}} T}
      \end{equation}
   \begin{equation}
   \label{partial_p_rho}
      \left(\frac{\partial p}{\partial \rho}\right)_{T}=N k_{\mathrm{B}} T+\frac{1}{P_{\mathrm{e}}}\left(\frac{P_{\mathrm{end}}^{2}}{\rho}+f \frac{P_{\mathrm{ed}}^{2}}{\rho}\right)
   \end{equation}
    where $P_{\mathrm{end}}$ and $P_{\mathrm{ed}}$ denote the pressure of a non-degenerate and degenerate electron gas, respectively. The $f$ in Eqs.~\eqref{specific_internal_energy-snec} to \eqref{partial_p_rho} is $f=\frac{d \ln P_{\mathrm{ed}}}{d \ln \rho}=\frac{5}{3}\left(\frac{P_{\mathrm{ed}}}{P_{\mathrm{ednr}}}\right)^{2}+\frac{4}{3}\left(\frac{P_{\mathrm{ed}}}{P_{\mathrm{edr}}}\right)^{2}
$, and $P_{\mathrm{ednr}}$ and $P_{\mathrm{edr}}$ correspond to the non-relativistic and relativistic cases for degenerate electron gas.

  	\paragraph*{Explosion setup.}
    \texttt{SNEC} provides two effective ways to explode the progenitor star of the supernova: thermal bomb and piston explosion. However, the designed analytic wind profiles and realistic profiles from \texttt{WhiskyTHC} already contain full initial conditions, so there is no need to set up explosions additionally. Thus we simply set the explosion type to thermal bomb and set the energy input to 0. We use the thermal bomb module only when we study the impact of shock cooling on kilonovae (see \S \ref{subsection_shock_cooling}).
    
    The \texttt{SNEC} code also implements a module called boxcar to smooth the compositional profile in the initial data. This tool mimics the mixing of ejecta during a supernova explosion. The boxcar has a given width, which is 0.4 $M_{\odot}$ by default. For each isotope, it sums up the isotope's mass within the width, and distributes the total isotopic mass to each shell equally. The boxcar moves from the inner to the outer boundary, and then this procedure is repeated until smoothness is achieved. We do not use the boxcar in our calculations, because we do not expect large scale mixing on the kilonova timescale.
    
    \paragraph*{Central remnant.}
    The mass of the inner remnant is also a parameter in the calculations as its gravitational pull can affect the evolution of the ejecta. We have fixed this inner remnant mass to be $M_{\mathrm{remnant}} = 3\ M_\odot$ in all calculations presented in this work.

    \subsection{Bolometric luminosities and Multicolor luminosities} 
    \label{subsection_lightcurve}
    
    Blackbody radiation assumption for kilonovae was commonly used in previous research, such as the single-temperature model in \citet{Li:1998bw}, and multi-component models in \citep{Villar:2017wcc,Perego:2017wtu}. The spectra of AT2017gfo were close to blackbody in the first ${\sim}2$ days \citep{Pian:2017gtc, Nicholl:2017ahq}. Non-thermal radiation is negligible at $T \sim$ 5000 $K$, although it may become important at late times when the ejecta becomes transparent \citep{Kasen:2013xka, Tanaka:2013ana}.
    
    We compute the emergent radiation from the photosphere and from all the mass shells above it using a multi-temperature blackbody model. In particular, we estimate the bolometric light curve as:
    \begin{equation}
        L_{\rm bol} = L_{\rm ph} + \int_{r_{\rm ph}}^{r_{\rm max}} \dot{\epsilon}\, \mathrm{d}m,
    \end{equation}
    where $L_{\mathrm{ph}}$ is the luminosity at the photosphere, $r_{\mathrm{ph}}$ is the photospheric radius, $r_{\rm max}$ is the outer boundary in our simulation, and $\dot{\epsilon}$ is the effective heating rate per unit mass. $\dot{\epsilon} = \epsilon_{\mathrm{th}} \dot{\epsilon_{\mathrm{r}}}$, $\dot{\epsilon_{\mathrm{r}}}$ is the heating rate introduced in \S \ref{subsection_heating}, and $\epsilon_{\mathrm{th}}$ is the thermalization efficiency, which is set to 0.5 by default.
    
    The observed flux density at frequency $\nu$ is 
    \begin{equation}
        f_{\nu} = \frac{1}{4 \pi D^2} \left( \frac{\pi L_{\rm ph}}{\sigma T_{\rm ph}^4}\, B_\nu(T_{\rm ph}) +
            \int_{r_{\rm ph}}^{r_{\rm max}} \frac{\pi \dot{\epsilon}}{\sigma T^4}\, B_\nu(T)\, \mathrm{d}m \right),
     %  	f_{\nu,obs} = \frac{L_{\mathrm{ph}}}{4 \pi D^2} \frac{\frac{2h \nu^3}{c^2} \frac{1}{e^{h \nu /k_{\mathrm{B}} T_{\mathrm{ph}}}-1}}{\sigma T_{\mathrm{ph}}^4 / \pi}  + \sum_i \frac{\dot{ \epsilon_i} \Delta m_i}{4 \pi D^2} \frac{\frac{2h \nu^3}{c^2} \frac{1}{e^{h \nu /k_{\mathrm{B}} T_i}-1}}{\sigma T_i^4 / \pi} 
    \end{equation}
    where $\sigma$ is the Stefan-Boltzmann constant, $B_{\nu}$ is the blackbody function, and $D$ is the luminosity distance to the source. Throughout this work we fix $D$ to 40 Mpc, the approximate distance to AT2017gfo \citep{Hjorth:2017yza}. Unlike the original \texttt{SNEC} code, we do not set a temperature floor here. We report our results using the AB magnitude system:
	\begin{equation}
    m_{\mathrm{AB}} = -2.5 \log _{10}\left(\frac{\int f_{\nu}(h \nu)^{-1} e(\nu) \mathrm{d} \nu}{\int 3631 \mathrm{Jy}(h \nu)^{-1} e(\nu) \mathrm{d} \nu}\right)
	\end{equation}
	We compute light curves in different bands using filter functions $e(\nu)$ downloaded from the SVO Filter Profile Service\footnote{\label{foot:filters}\url{http://svo2.cab.inta-csic.es/theory/fps/}} \citep{2012ivoa.rept.1015R, 2020sea..confE.182R}. We primarily use CTIO and Gemini bands.

% ===========================================================================
\section{Code validation}
\label{section_code_validation}
% ===========================================================================

\subsection{Energy conservation}
\label{subsection_energy_conservation}
The equation of energy conservation for the whole system is:
 \begin{equation}
 \label{conservation_1}
 \begin{split}
 	 	\frac{\mathrm{d}}{\mathrm{d} t} \int_{\Omega} \rho\left(\epsilon+\frac{1}{2} |\boldsymbol{v}|^{2}\right) \mathrm{d} V & =\int_{\Omega} \rho \boldsymbol{f_b} \cdot \boldsymbol{v} \mathrm{~d} V -\int_{\partial \Omega} p \boldsymbol{v} \cdot \mathrm{d} \boldsymbol{S} \\&-\int_{\partial \Omega} \boldsymbol{f_s} \cdot \mathrm{d} \boldsymbol{S}+ \dot{Q} \, ,
 \end{split}
 \end{equation}
 where $\Omega$ is a material volume (a region moving with the fluid), $\rho$ is the matter density, $\epsilon$ is the specific internal energy of the fluid (including the radiation), $v$ is the fluid velocity, and $p$ is the pressure. ${\rm d}/{\rm d}t$ is the total time derivative. In our simulations, the surface force $\boldsymbol{f_s}$ is zero, while the body force $\boldsymbol{f_b}$ is the gravitational force. $\dot{Q} = H - L_{\rm bol}$ is the net cooling/heating due to nuclear decays $H$ and emission $L_{\rm bol}$. This last term also includes the energy deposited into the outflows by the GRB jet, discussed in \S \ref{subsection_shock_cooling}.
 
We can rewrite the energy conservation equation as:
\begin{equation}
\label{conservation_2}
	\frac{d}{dt}(E_{\rm int} + E_{\rm kin}+E_{\rm grav}) = \dot Q - \int p \boldsymbol{v} \cdot d \boldsymbol{S},
\end{equation}
where
\begin{align}
    & E_{\rm int} = 4 \pi \int_{r_1}^{r_{\max}} \rho\, \epsilon\, r^2\, \mathrm{d}r, \\
    & E_{\rm kin} = \frac{4 \pi}{2} \int_{r_1}^{r_{\max}} \rho v^2\, r^2\, \mathrm{d}r, \\
    & E_{\rm grav} = - 4\pi \int_{r_1}^{r_{\max}} \rho\, \frac{G \mathcal{M}}{r}\, r^2\, \mathrm{d}r,
\end{align}
$r_1$ and $r_{\max}$ are the inner and outer radius of the ejecta and $\mathcal{M} = m(r) + M_{\rm remnant}$ is the enclosed mass including the central remnant. Since the outer boundary condition is $p_{\rm imax}=0$, the $p{\rm d}V$ term only includes a contribution from the inner boundary:
\[
-\int_{r_1}^{r_{\max}} p \boldsymbol{v} \cdot d \boldsymbol{S} = 4 \pi p_{1} v_1 r_1^2,
\]
where $p_1, v_1$ and $r_1$ are pressure, radial velocity, and radius at the inner boundary, respectively. The gravitational energy $E_{\rm grav}$ is dominated by the contribution of the gravitational attraction to the central remnant.

\begin{figure}
\centering
\includegraphics[scale=0.5]{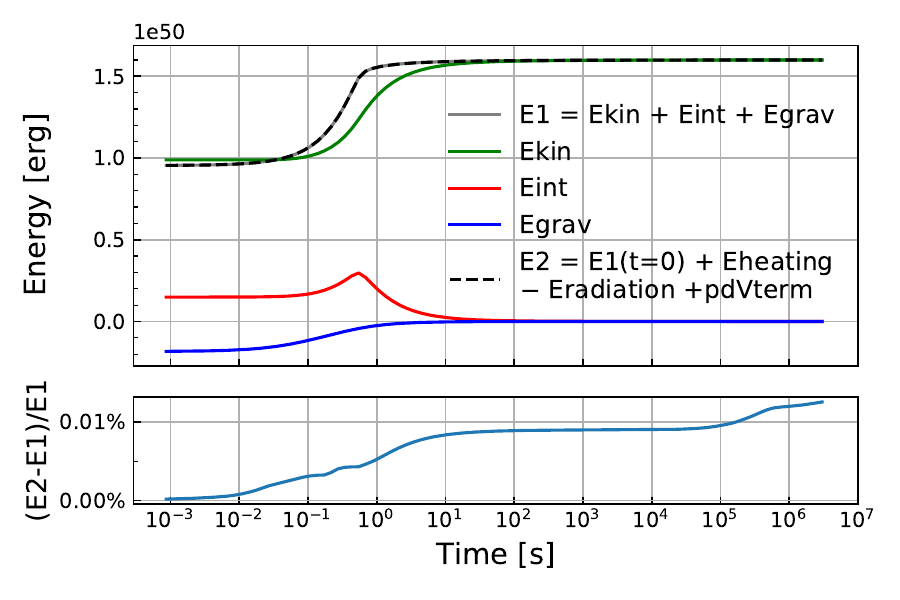}
\caption{Energy Conservation of the optimal wind with $Y_e$ = 0.1. E1 is the total energy of the ejecta (gravitational + kinetic + internal) as a function of time. E2 is the initial total energy of the ejecta plus the net cumulative energy injected/released by r-process heating, $p{\rm d}V$ work at boundary, and radiation emission. Perfect energy conservation would imply ${\rm E1} = {\rm E2}$. Their maximum relative difference here is around 0.01\%, indicating that our simulation well conserve energy.}
\label{E1E2_wind310T6_ye0.1}
\end{figure} 

To test how well energy is conserved in our calculations, we integrate Eq. \eqref{conservation_2} to obtain an overall energy balance. Here, we discuss energy conservation in the context of the optimal wind profile with initial $Y_e$ = 0.1, which is a representative case. In Fig.~\ref{E1E2_wind310T6_ye0.1}, E1 is the total energy of the ejecta including internal, kinetic, and gravitational energy. E2 is the initial total energy of the ejecta plus the net cumulative energy injected/released by r-process heating, $p{\rm d}V$ work at boundary, and radiation emission. Eheating (t) ($= \int_0^t H\, \mathrm{d}t$) is the r-process heating, and Eradiation (t) ($= \int_0^t L_{\mathrm{bol}} \mathrm{d}t$) is the energy loss due to kilonova emission. If energy were perfectly conserved, then E1 and E2 would be identical. Since energy is not perfectly conserved in our simulation, we monitor $|{\rm E1} - {\rm E2}|$ to check the level of violation of energy conservation. That said, we find that \texttt{SNEC} conserves energy with a high degree of precision. In the case of the optimal wind profile, the maximum relative difference between E1 and E2 is $\sim$ 0.01\%. In the case of the BLh profile the dynamics is more complex, but energy is also conserved to better than one percent (see Appendix \ref{appendix_energy_conservation_blh}). 

Figure \ref{E1E2_wind310T6_ye0.1} also shows the relative importance of the different forms of energy in the outflows. Overall, most of the energy is in the form of kinetic energy. Internal energy roughly balances gravity at very early times $t \lesssim 0.1\ {\rm s}$ and peaks at a $t \lesssim 1\ {\rm s}$. At a time of about one second, r-process heating peaks and the internal energy now provides a significant contribution to the energy budget and can play a role in the dynamics of the outflows. This is consistent with the findings of \citet{Rosswog:2013kqa}, who reported that the inclusion/omission of r-process heating lead to appreciable differences in the structure of the outflows after about one second. \citet{Foucart:2021ikp} also discusses the importance of heating in the context of neutron star binary merger simulations.

\begin{figure}
\centering
\includegraphics[scale=0.5]{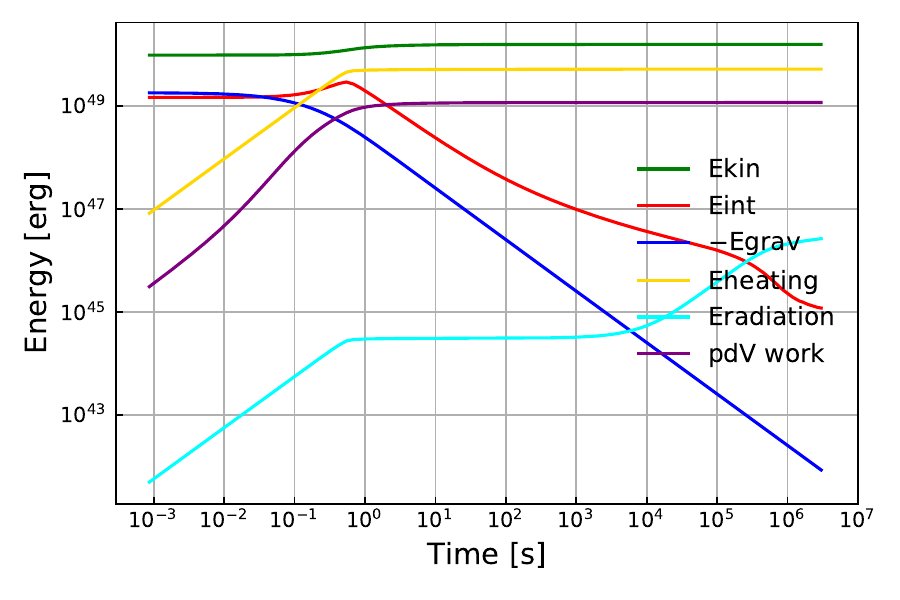}
\caption{Energy budget for the optimal wind with $Y_e$ = 0.1. The total energy budget is dominated by the kinetic energy. Only a small fraction of the energy is radiated. The internal energy and gravitational energy become less important as the ejecta cools and moves away from the central engine. The $p{\rm d}V$ term and the energy from heating are most important at around 1 second, when the bulk of the neutron captures are taking place in the r-process.}
\label{logEnergy_wind310T6_ye0.1}
\end{figure} 

Figure \ref{logEnergy_wind310T6_ye0.1} shows the energy balance in logarithmic scale. We find that the heating and $p{\rm d}V$ work at inner boundary are important when r-process nucleosynthesis is taking place. Only a small fraction of the overall energy of the ejecta is radiated. 
 
\subsection{Comparison with analytic models}
We compare the \texttt{SNEC} calculations with two alternative semi-analytic models: SADS (Semi-Analytic Diffusion Solver) and Arnett-Chatzopoulos-Villar's single component semi-analytic model (ACV). SADS implements a semi-analytic formula for the kilonova luminosity as proposed by \cite{Wollaeger:2017ahm}. The model considers an homogeneous sphere with constant density, temperature, and opacity, which expands homologously starting from a few hours after merger. We model the radioactive heating in the ejecta using the heating rates described in \S \ref{subsection_heating}. A semi-analytic solution of the radiative transfer equations is obtained under the assumption that matter is optically thick. The opacity is calculated starting from the input $Y_e$ by means of Equation \ref{opacity_ye}. Along with the thermodynamical ejecta properties defining heating rates, that is $Y_e$, $s$ and $\tau$, the model considers the ejecta mass $M_{\mathrm{ej}}$ and its maximum expansion velocity $v_{\mathrm{max}}$ as input variables, while it assumes a fixed value of $T_0=10^4$ K for the temperature of the homogeneous sphere at the starting time $t_0=10^4$ s.

ACV \citep{Arnett:1982ioj, Chatzopoulos:2011vj, Villar:2017wcc} is based on an analytic solution originally proposed by \cite{Arnett:1982ioj} for light curves of Type II supernovae with $^{56}$Ni heating only, and later generalized to any given heating function by \cite{Chatzopoulos:2011vj}. The model treats a radiation-dominated gas in spherical symmetry with an homologous expansion law. The luminosity is obtained starting from the first law of thermodynamics for the expanding envelope and by invoking the diffusion approximation. A constant grey opacity is employed, and the input energy generation rate is provided by the radioactive heating rate in order to adapt the energy source to kilonovae. \cite{Villar:2017wcc} has used three ejecta components to obtain excellent agreement with data from GW170817. Here we return to one-component spherically-symmetric ejecta. Both opacity and heating rate models are the same as those employed by \texttt{SNEC} and SADS.

For this comparison, \texttt{SNEC} is prepared using the initial and boundary conditions described in \S \ref{subsection_initial_boundary} and \S \ref{subsection_other_differences}. In particular, we initialise the simulations using the analytic \textit{wind310T6} profile, which, as discussed in \S \ref{subsection_hydrodynamics}, is found to provide a good agreement between \texttt{SNEC} calculations performed in full radiation-hydrodynamics and those which instead assume homologous expansion.

\begin{figure*}
\centering
\includegraphics[width=.95\linewidth]{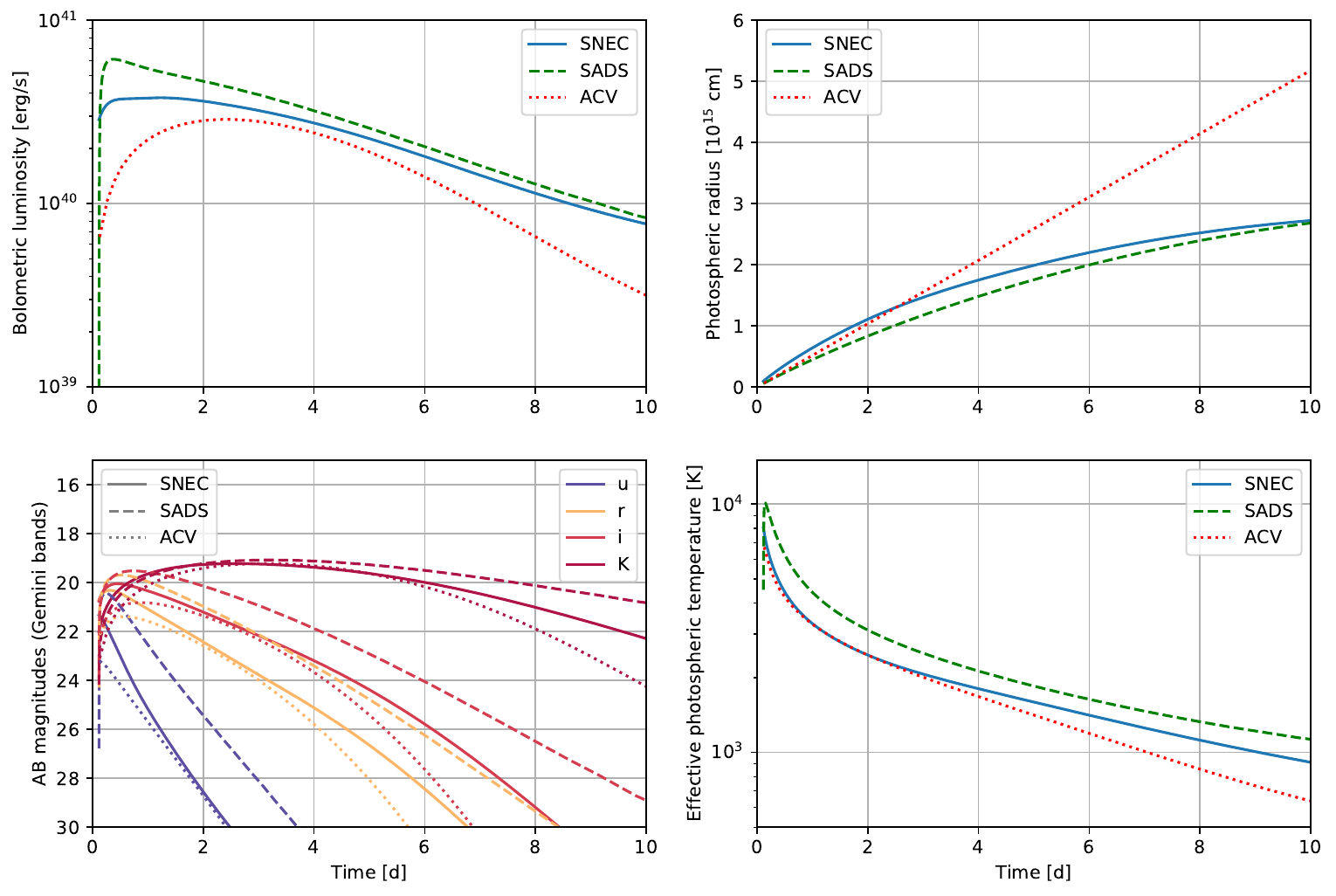}
\caption{\texttt{SNEC} results for the bolometric luminosity and AB magnitudes using the optimal wind profile in correspondence of the input quantities $M_{\mathrm{ej}}=0.01$ $M_{\odot}$, $v_{\mathrm{max}}=0.2$~c, $Y_e=0.1$, $s=10$ $\mathrm{k_B/baryon}$ and $\tau=10$ ms, compared to the same results obtained with SADS and ACV models. Photospheric radius and effective temperature are shown for illustration.}
\label{lum_Rph_mod_var_ye0.1}
\end{figure*}

Figure \ref{lum_Rph_mod_var_ye0.1} shows bolometric luminosity, AB magnitudes in a few different Gemini bands, photospheric radius, and effective photospheric temperature obtained from \texttt{SNEC}, SADS and ACV models. All calculations assumed fiducial values of $M_{\mathrm{ej}}=0.01$ $M_{\odot}$, $v_{\mathrm{max}}=0.2$~c, $Y_e=0.1$, $s=10$ $k_\mathrm{B} {\rm baryon^{-1}}$ and $\tau=10$ ms. We emphasize that all three calculations have adopted the same heating rates, effective gray opacities, and heating efficiencies. The three models show good overall agreement in their prediction for the bolometric luminosity, especially on a timescale of a few days from the merger. The agreement is somewhat worse at early and late times. SADS model tends to overestimate the luminosity at early times, since it assumes that all radioactive decay energy is immediately radiated as blackbody emission. ACV underestimates the bolometric luminosity and overestimates the photospheric radius at late times. This is due to the fact that in this model the photospheric radius is assumed to coincide with the average ejecta radius $r_{\mathrm{avg}}=v_{\mathrm{avg}}t$, which increases indefinitely and eventually becomes unphysical. In addition, ACV does not account for any luminosity contribution from the optically thin region outside the photosphere. ACV can avoid the unphysical photospheric radius expansion by applying a temperature floor, as done by \cite{Villar:2017wcc} when comparing the three component model with data from GW170817.
On the other hand, ACV shows a better agreement with \texttt{SNEC} than SADS in the color light curves, especially in the blue and optical bands. SADS prediction of a bluer spectrum is caused by its systematic overestimation of the effective photospheric temperature. This effect arises both because the bolometric luminosity is typically overestimated and because the photospheric radius, which is computed independently, is slightly underestimated. In SADS the latter is found analytically by imposing a homologous density profile \citep{Wollaeger:2017ahm} in the condition $\tau(r_{\mathrm{ph}})=2/3$, where $\tau(r)$ is the optical depth of the material at a certain radius $r$. This solution typically includes a first increase of the radius up to a maximum value, after which the latter decreases again back to zero. All models agree well in the infrared bands at a timescale of a few days. This is not too surprising since hydrodynamic effects (\S \ref{subsection_hydrodynamics}) and the details of the radiative transfer in the ejecta become less important at these times.

\begin{figure}
\centering
\includegraphics[scale=0.5]{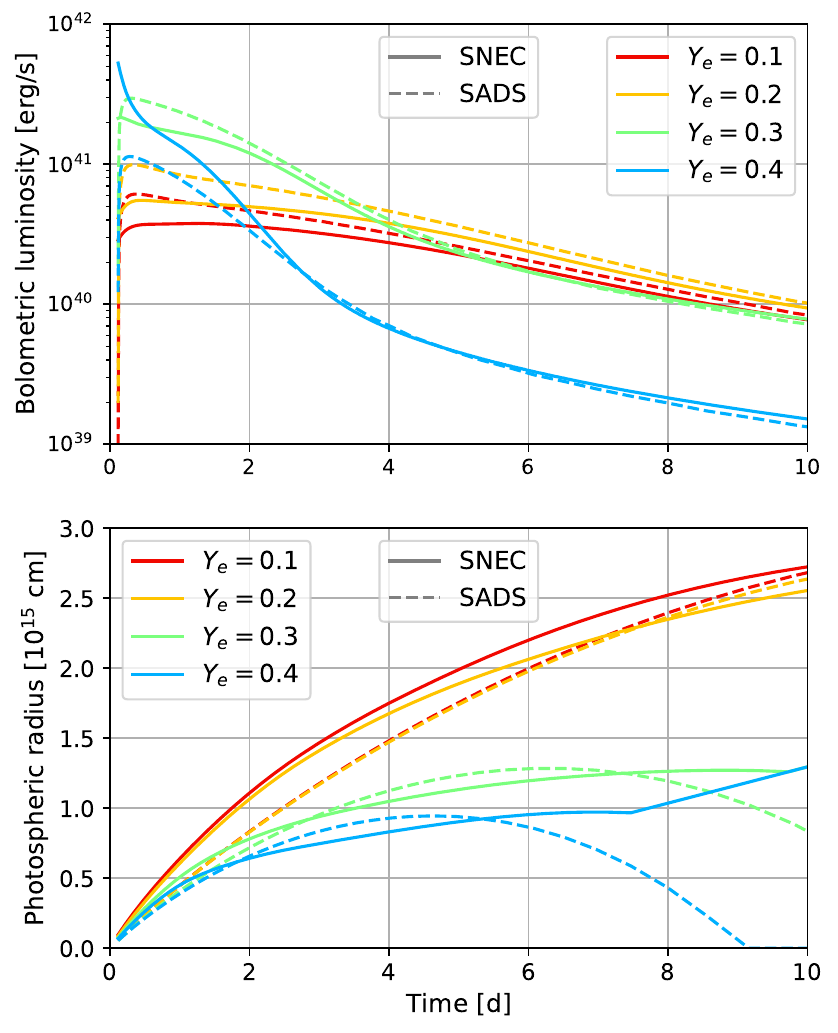}
\caption{Bolometric luminosities and photospheric radii obtained with \texttt{SNEC} and SADS for different initial electron fractions at $M_{\mathrm{ej}}=0.01$ $M_{\odot}$, $v_{\mathrm{max}}=0.2$~c, $s=10$ $\mathrm{k_B/baryon}$ and $\tau=10$ ms.}
\label{lum_Rph_Ye_var_SADS}
\end{figure}

Figure \ref{lum_Rph_Ye_var_SADS} shows a comparison between \texttt{SNEC} and SADS for different values of the ejecta $Y_e$. We find that for most values of $Y_e$ SADS overestimates the bolometric luminosity, as it was the case in the previous comparison for $Y_e=0.1$. However, for large values of $Y_e \gtrsim 0.4$ the situation is reversed and SADS underestimates the bolometric luminosity. The reason is that, for such values of $Y_e$, the heating rate is dominated by the decay of a relatively small number of nuclear species, so it peaks at earlier times and then exponentially decay. This early energy release is not captured by SADS, since the SADS calculations only start $\sim3$ hours after merger. On the other hand, the \texttt{SNEC} simulations also track the emission and thermalization of this energy and its subsequent release at later times.

Figure \ref{lum_Rph_Ye_var_SADS} also shows some general trends in the light curve of kilonovae. In particular, it can be seen that the maximum bolometric luminosities for the optimal wind profiles with $M_{\text{ej}} = 0.01 M_{\odot}$ and initial $v_{\text{max}} = 0.2$~c range between $10^{40}$ and $10^{42}\ {\rm erg}\ {\rm s}^{-1}$. Kilonova light curves produced by wind profiles with $Y_e \gtrsim 0.25$ have larger peak luminosity and evolve more rapidly than those produced by more neutron rich outflows. In fact, even if the total amount of heating produced by the $r$-process is larger for smaller $Y_e$, the overall radiated energy as well as its distribution in time depends on both the radioactive heating and the material opacity. Indeed, a small opacity is expected to cause the emission to peak earlier and the peak luminosity to be brighter. For a kilonova at a distance of 40 Mpc, the $Y_e = 0.1$ wind model with \texttt{SNEC} predicts a peak luminosity of about 21 magnitudes in the $u$-band and of 19 magnitudes in the $K_s$-band. The latter is reached at around 3 days after merger. The $Y_e = 0.4$ wind model with \texttt{SNEC} predicts a similar peak luminosity in the Ks-band, but the peak is reached one day earlier. Moreover, the $Y_e = 0.4$ wind is much brighter in the $u$-band and peaks at around $18.5$ magnitudes. Ten days after the merger, the $K_s$-band magnitude has dropped to about 21 magnitudes for the $Y_e = 0.1$ wind and to about 27 magnitudes for the $Y_e = 0.4$ wind. These trends are consistent with the expectations \citep{Metzger:2019zeh}.

% ===========================================================================
\section{First applications of \texttt{SNEC}}
\label{section_first_applications_of_SNEC}
% ===========================================================================
\subsection{General features}
\label{subsection_general_features}

\begin{figure}
\centering
\includegraphics[scale=0.5]{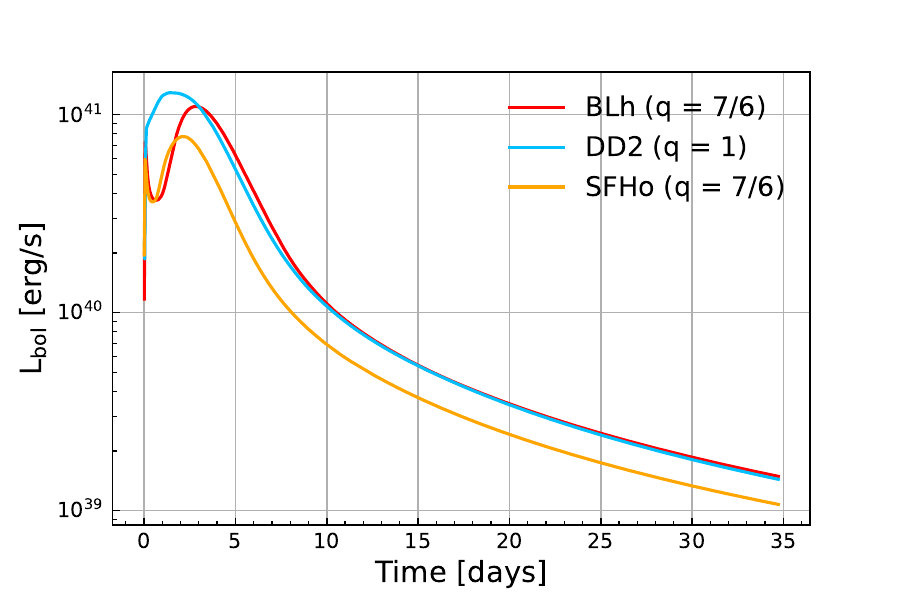}
\caption{Bolometric light curves for the numerical relativity outflow profiles BLh, DD2, and SFHo (\S \ref{subsection_initial_boundary}). The BLh and SFHo profiles have the same mass ratio $q = M_1/M_2 = 7/6$, and the DD2 profile is from an equal-mass neutron star binary. The BLh and DD2 outflows have a comparable mass and brightness after ${\sim}3$ days. The SFHo outflow is less massive and produces a dimmer kilonova. Both the BLh and SFHo kilonovae have double peaked light curves due to the lanthanide curtain effect, while the DD2 model does not.}

\label{LC_blh_DD2_sfho}
\end{figure} 

We use \texttt{SNEC} to generate synthetic light curves using profiles from numerical relativity simulations of merging neutron stars. Figure \ref{LC_blh_DD2_sfho} shows the bolometric luminosities of the BLh, SFHo, and DD2 profiles. In the following discussion, we take these light curves as a baseline for comparison as we study the impact of uncertainties in the heating rates, we consider time-extrapolated outflow rates from the simulations, and we study the impact of the thermal energy deposition due to a GRB jet and cocoon breaking through the ejecta. Among these outflow profiles, the SFHo profile has the smallest amount of ejecta (${\sim}9.2 \times 10^{-3}\ M_\odot$), because the associated merger simulation was discontinued after black hole formation, when the outflow rate due to the spiral-wave wind dropped to zero. Additional mass ejection would have been driven by viscous and nuclear processes in the disk over a timescale of a few second, but these cannot yet be modeled in full-3D numerical relativity simulations. For these reasons, it is not surprising that the SFHo profile gives rise to the faintest kilonova among the considered models. The BLh and DD2 profiles have a similar amount of mass: $2.29 \times 10^{-2}\ M_\odot$ and $1.93 \times 10^{-2}\ M_\odot$, respectively. For this reason they produce kilonovae that have very similar brightness after the first few days and both are brighter than the SFHo outflow. Interestingly, both the BLh and SFHo light curves have a double peak, while the DD2 light curve has a single peak. This is due to the presence of a low-$Y_e$ component of the outflow for BLh and SFHo, which is absent in the DD2 profile (see Figs.~\ref{blh_profile} and \ref{DD2_profile}). This outflow component is due to the partial tidal desruption of the secondary star prior to merger. It is absent for the DD2 profile which is associated to an equal mass merger.

\begin{figure*}
\centering
\includegraphics[width=.95\linewidth]{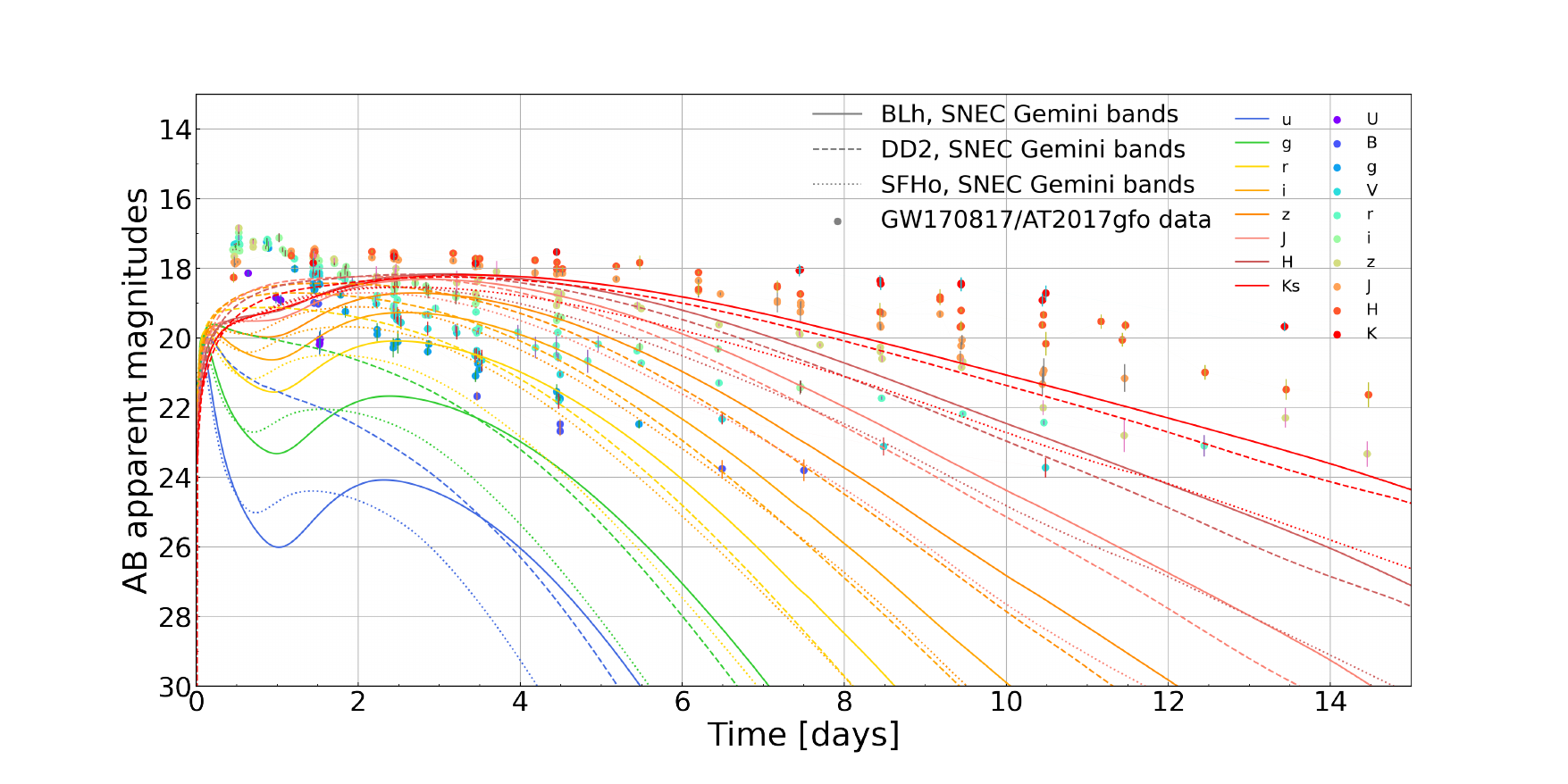}
\caption{GW170817/AT2017gfo data (dots with error bars) and \texttt{SNEC}'s AB apparent magnitudes of BLh (solid line), DD2 (dashed line), and SFHo (dotted line) at 40 Mpc. The observation data covers U to K band for various telescopes. We adopt Gemini bands from u to Ks for \texttt{SNEC} results. This comparison shows that the current NR informed models including BLh, DD2, and SFHo, which have an ejecta mass of 0.023 $M_{\odot}$, 0.019 $M_{\odot}$, and 0.009 $M_{\odot}$, do not match the observation. This indicates that a larger ejecta mass, or additional factors contributing to light curves should be considered to fit the observation.}
\label{ABmags_blh_DD2_sfho_GW170817}
\end{figure*}   

These trends are reflected in Fig.~\ref{ABmags_blh_DD2_sfho_GW170817} which shows the AB magnitudes of the kilonova emerging from these three profiles assuming a distance of 40~Mpc in different bands. The difference between the DD2 equal mass model and the others is even more apparent in the blue bands at early times. Our calculations suggest that high cadence observations of kilonovae could constrain the presence/absence of a lanthanide curtain, which in turn would constrain the mass ratio of the binary. However, we caution the reader that the impact of the presence of a massive tidal tail on the light curve is likely exaggerated by the assumption of spherical symmetry used in our calculations. In reality, we expect that this effect will only be prominent for edge-on binaries.

Figure \ref{ABmags_blh_DD2_sfho_GW170817} also show the photometric data for AT2017gfo. The observation data is collected from kilonova.space\footnote{\url{https://kilonova.space}} \citep{Villar:2017wcc}. The \texttt{SNEC} results use Gemini filters, and we also calculate CTIO bands, while the observation data is from various instruments. The differences in filters have little influence in the comparison. AT2017gfo is significantly brighter than any of our models. This is not unexpected given the approximations in our models, most notably the fact that our merger simulations cannot yet self-consistently compute the full evolution of the postmerger disk due to the long timescales involved and the assumption of spherical symmetry \citep{Perego:2014fma}. In particular, the works of \citet{Perego:2017wtu, Korobkin:2020spe} and \citet{Heinzel:2020qlt} showed that multidimensional effects and viewing angle, which we cannot take into account with \texttt{SNEC}, have a strong impact on the color light curves from kilonovae. It is also worth mentioning that \citet{Breschi:2021tbm} performed a Bayesian selection analysis of the AT2017gfo and ruled out spherically symmetric kilonova models with high confidence.  That said, fitting the observation is not the purpose of this paper, and we leave it to our future work.

\begin{figure}
\centering
\includegraphics[scale=0.5]{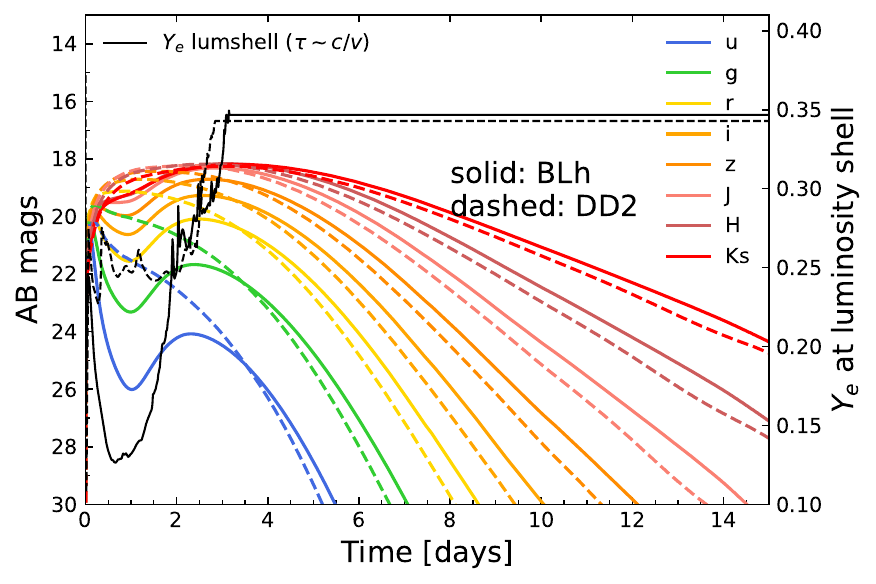}
\caption{BLh and DD2 profile: AB apparent magnitudes and $Y_e$ at luminosity shell.  Luminosity shell locates at the point whose optical depth $\tau$ and velocity $v$ satisfy $\tau = c/v$. When the ejecta becomes transparent enough, the luminosity shell falls onto the inner boundary. This figure shows that the first peak of BLh color light curves is related to the outermost fast high-$Y_e$ component of the ejecta. The gap between BLh's double peaks is due to the low-$Y_e$ lanthanide curtain.}
\label{ABmags_Yelumshell_blh_DD2}
\end{figure} 

The multicolor light curves properties depend most directly on the initial $Y_e$ at the luminosity shell of the ejecta. The latter is defined as the shell at which radiation diffusion and expansion timescales become comparable, that is when the optical depth $\tau \sim c/v$. \texttt{SNEC} locates the luminosity shell by sweeping through the ejecta. It starts from the exterior, where $\tau = 0$, and moves towards the interior until $\tau$ becomes equal to $c/v$. At early times the luminosity shell is close to the surface of the outflows, but at later times the shell is found at increasingly large depth into the outflows, as the material expands and becomes transparent. Eventually, the luminosity shell becomes the inner boundary of the ejecta. Figure \ref{ABmags_Yelumshell_blh_DD2} combines AB magnitudes in different bands for the BLh and DD2 profiles and the $Y_e$ at the location of the luminosity shell, both as a function of time. Both profiles have an outer shell of rapidly expanding, high-$Y_e$ material launched when the remnant bounces back after merger. In both cases, the kilonova is blue in the very first few hours after merger. The u-band magnitude for the BLh model reaches ${\sim} 20$ magnitudes in the first hours of the merger, before dropping rapidly. In the BLh case the kilonova becomes fainter and redder very quickly as the luminosity shell passes through the tidal tail, which is very neutron rich. The kilonova becomes bright again when the luminosity shell reaches the inner part of the ejecta which has higher $Y_e$ due to the combined effects of shock heating and neutrino irradiation from the central remnant \citep{Radice:2016dwd}.

\subsection{Hydrodynamics}
\label{subsection_hydrodynamics}
Most of the previous models, ranging from analytic and semi-analytic to Monte Carlo radiative transfer, assume homologous expansion and neglect the effects of pressure work (e.g. \citealt{Tanaka:2013ana, wollaeger2013radiation, bulla2019possis}). There are some attempts to combine hydrodynamics and radiative transfer \citep{gittings2008rage, Roth:2014wda}, but the study of the effects of hydrodynamics on kilonovae is very limited. \citet{Ishizaki:2021qne} includes hydrodynamics to study fallback accretion, but does not include radiative transfer. Our work is one of the first radiation-hydrodynamics study of kilonovae. Radiation-hydrodynamics simulations are performed right after the merger until $\sim$ 35 days. The hydrodynamics can also be turned off in our code, so that the velocity is frozen and the ejecta undergoes free expansion ($v(t) = v(t=0), r = vt$).

\begin{figure}
\centering
\includegraphics[scale=0.5]{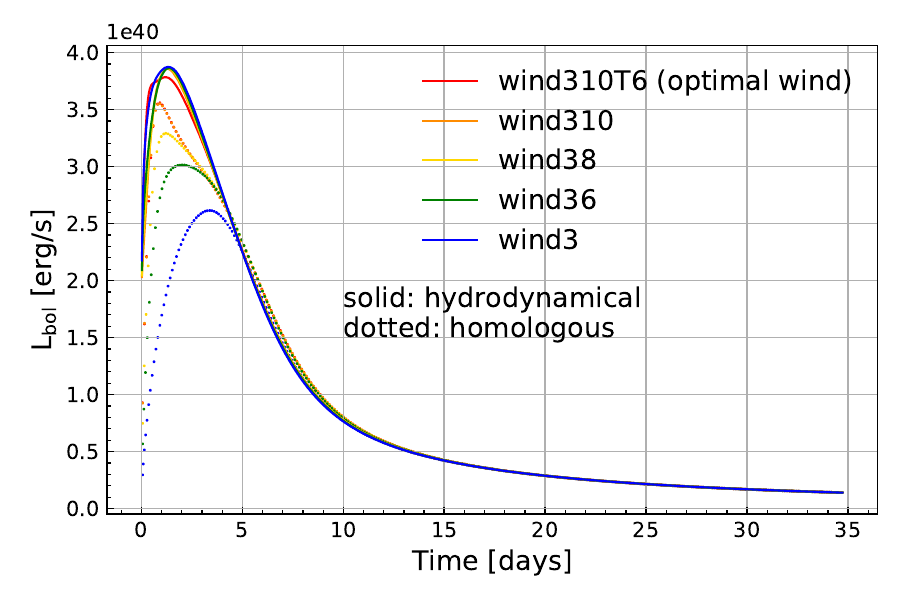}
\caption{ Bolometric light curves computed for different wind profiles assuming homologous expansion, or with full hydrodynamics. For all models, initial $Y_e$ = 0.1, $s$ = 10 $\mathrm{k_B/baryon}$, $\tau$ = 10 ms. The solid lines show the hydrodynamical results from \texttt{SNEC} using the wind profiles. In general the light curves coincide with each other. The dotted lines show homologous expansion results, i.e. hydrodynamics is turned off in \texttt{SNEC}. As we increase the second powerlaw index of density, the light curves get closer to hydrodynamical results. }
\label{CodeValidation-Hydro_wind3X_ye0.1}
\end{figure} 

\begin{figure}
\centering
\includegraphics[scale=0.5]{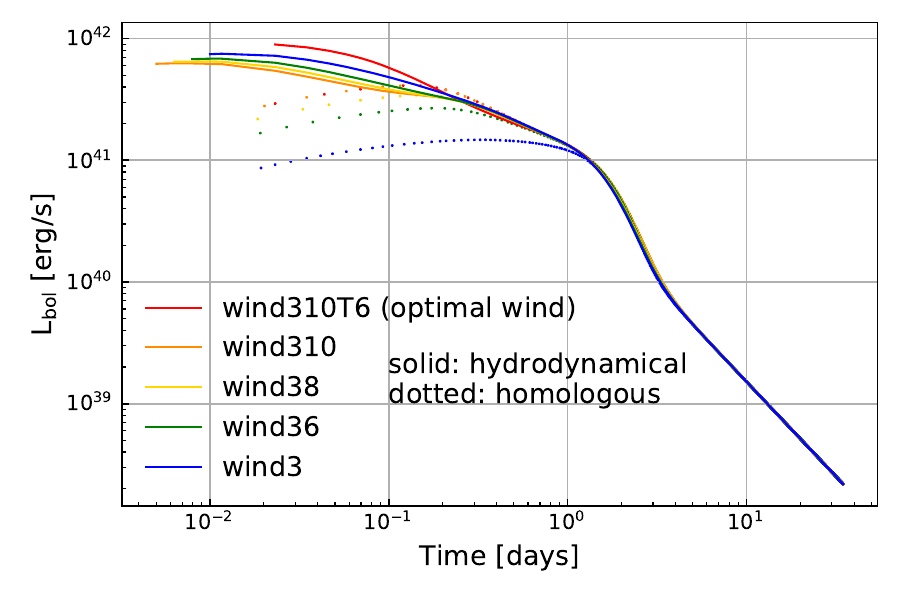}
\caption{The same as Figure \ref{CodeValidation-Hydro_wind3X_ye0.1} except $Y_e$ = 0.4. After about 0.2 day, the hydrodynamical results for the optimal wind profile (wind310T6) agree with homologous expansion, while for other profiles it takes longer time for the agreement.}
\label{CodeValidation-Hydro_wind3X_ye0.4}
\end{figure} 

Figure~\ref{CodeValidation-Hydro_wind3X_ye0.1} shows the bolometric light curves computed with and without the assumption of homologous expansions and for different wind profiles. The initial $Y_e$ is set to 0.1 in all calculations. We remind the reader that wind3 profile refers to $\rho \propto r^{-3}$, while wind36, wind38, wind310 use 2 powerlaws for density ($k_1 = 3$ and $k_2$ = 6, 8, 10 respectivelly). Their temperature is uniformly $10^9$ K. Wind310T6 is the optimal wind profile introduced in \S \ref{subsection_initial_boundary}. The bolometric luminosity from the hydrodynamic calculations (solid lines) are insensitive to the initial profiles, because the hydrodynamical evolutions at the beginning of the simulation smooths the differences in the ejecta structures. On the other hand, the homologous expansion results vary by a factor of $\sim$ 2 depending on profiles. When increasing the second powerlaw factor for the density we find better agreement between the homologous expansion results and those obtained with the hydrodynamics calculations. So if the density profile includes a sharp drop near the outer boundary, which is reasonable as seen from NR simulations, homologous expansion is a good assumption. Fig.~\ref{CodeValidation-Hydro_wind3X_ye0.4} shows the corresponding results for initial $Y_e$ = 0.4. Also in this case, we find that homologous expansion calculations are very sensitive to the details of the outflow profiles. These tests suggest that wind profiles similar to the optimal wind profile introduced here should be employed for radiative transfer calculations that assume homologous expansion.

\begin{figure}
\centering
\includegraphics[scale=0.5]{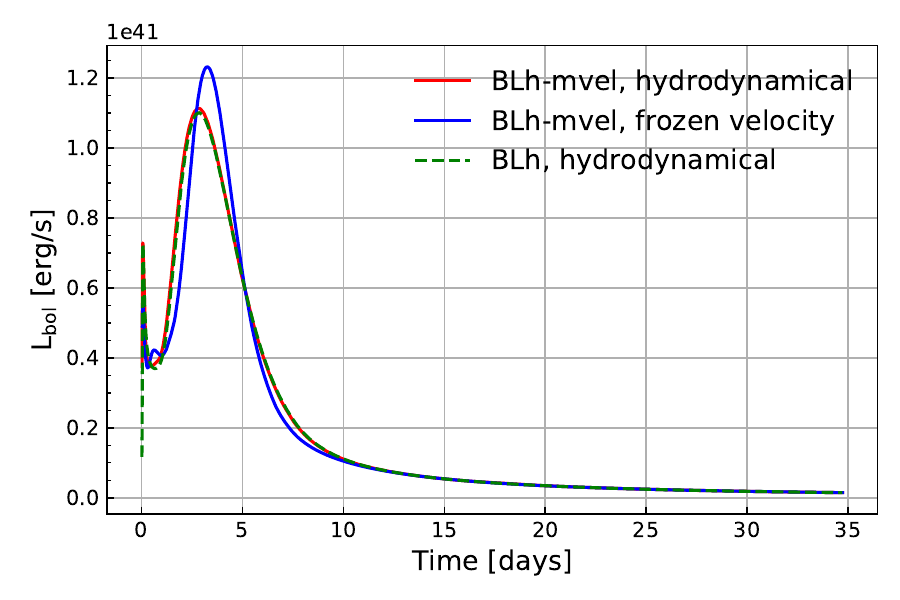}
\caption{Effect of the hydrodynamics on the bolometric light curves for the BLh profile with modified velocity (BLh-mvel profile). The red line shows the results from the radiation-hydrodynamic calculations while the blue line shows the results obtained with hydrodynamics turned off and frozen velocity. As a comparison, the green dashed line shows the light curve for the original BLh profile. Hydrodynamic models predict faster expansion driven by pressure forces in the outflows and more rapidly evolving light curves.} 
\label{CodeValidation-Hydro_blh-mvel2}
\end{figure} 

\begin{figure}
\centering
\includegraphics[scale=0.5]{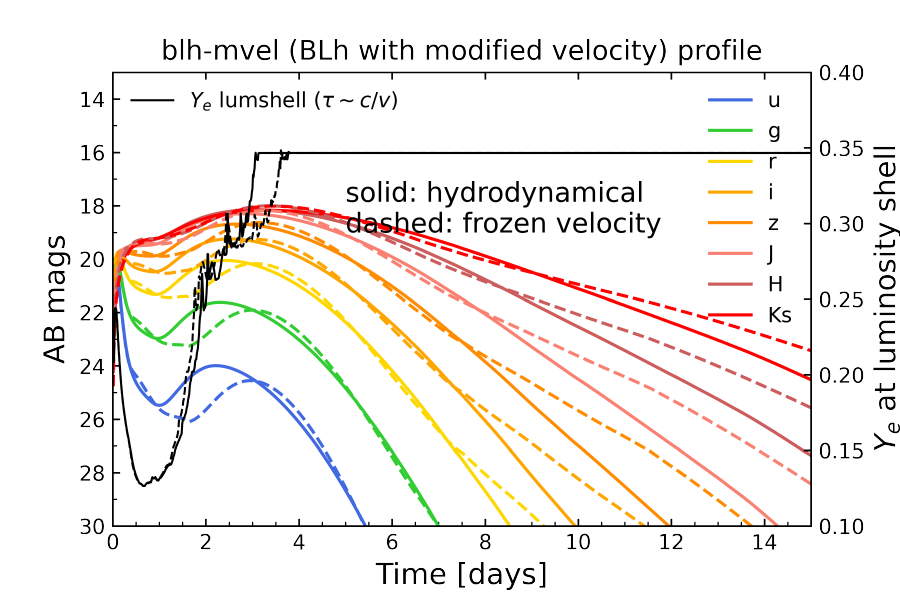}
\caption{Effect of the hydrodynamics on the multicolor light curves for the BLh with modified velocity (BLh-mvel profile). The solid lines show the results from the radiation-hydrodynamic calculations, while dashed lines show the result obtained with hydrodynamics turned off and frozen velocity. The impact of hydrodynamics is particularly evident at early times in the blue bands.}
\label{CodeValidation-Hydro_ABmags_blh-mvel2}
\end{figure} 

In addition to considering the impact of homologous expansion in the case of idealized wind profiles, we also consider its impact for the BLh profile, which we take as representative of a realistic profiles from a NR simulations. Because there are fluctuations in the initial velocity distribution, we cannot directly turn off the hydrodynamics and froze the velocity in this case. Indeed, the velocity $v$ must increase monotonically with the radius $r$ (or enclosed mass $m$) to avoid shell crossing. To achieve this, we replace the velocity in the BLh with a fit constructed using a monotonically increasing function (see Fig.~\ref{blh-mvel_profile} in Appendix \ref{appendix_boundary_velocity}). We use this BLh-with-modified-velocity (BLh-mvel) profile for this test. The comparison of bolometric luminosity between BLh-mvel profile with and without hydrodynamics is shown in Fig.~\ref{CodeValidation-Hydro_blh-mvel2}. In general, the two are consistent. The comparison between the multi-color light curves is shown in Fig.~\ref{CodeValidation-Hydro_ABmags_blh-mvel2}. We find that the inclusion of hydrodynamic effect shifts the second peak of the light curve by one day, from ${\sim} 3$ days after merger in the homologous expansion calculations to ${\sim} 2$ days after merger in the radiation-hydrodynamics calculation. This difference is explained by the more rapid expansion of the lanthanide curtain driven by pressure forces in the hydrodynamics model. As a consequence of this fast expansion, the optical depth drops more rapidly and light from the lanthanide-poor inner part of the ejecta escapes at earlier times, so the kilonova peaks sooner. This effect can be seen in Fig.~\ref{CodeValidation-Hydro_ABmags_blh-mvel2}, where we also show the $Y_e$ at the luminosity shell as a function of time. The faster expansion of the ejecta in the hydrodynamic models also lead to faster cooling for the hydrodynamics simulation compared to the homologous expansion simulation. This results in a more rapid drop in the color light curves for the former after ${\sim}9$ days.

\subsection{Impact of uncertainties in the heating rates}

\begin{figure}
\centering
\includegraphics[scale=0.5]{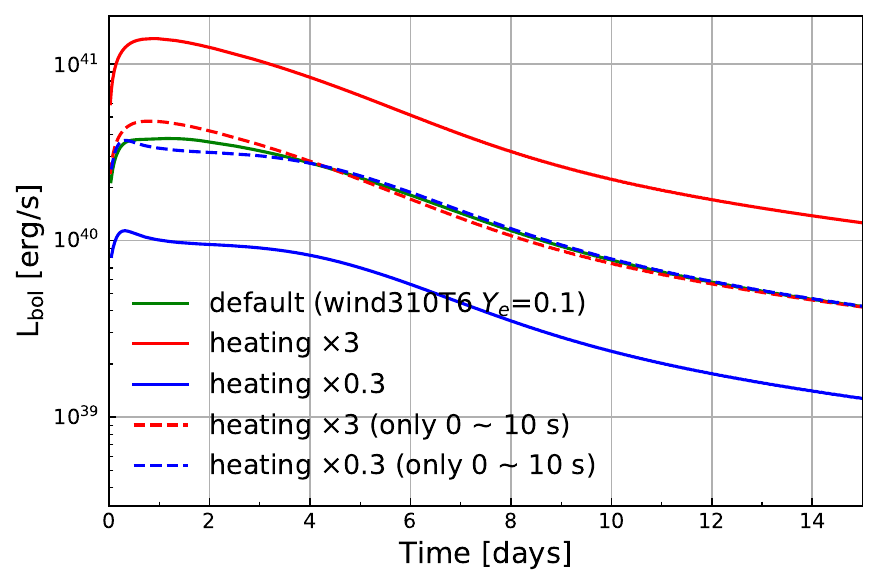}
\caption{Bolometric light curves with different heating rates and the wind310T6 (optimal wind) profile with $Y_e = 0.1$, $s = 10$~$\mathrm{k_B/baryon}$, $\tau = 10$~ms. The green line represents the default heating rates introduced in \S \ref{subsection_heating}, with thermalization efficiency $\epsilon_{\mathrm{th}} = 0.5$. The red and blue solid lines display the light curves when the heating rate is multiplied by 3 and 0.3, respectively. The dashed light curves are obtained by changing the heating rates from the baseline only in the first ten seconds.}
\label{LC_wind310T6_ye0.1_heating}
\end{figure} 

The energy released by nuclear decays and its thermalization efficiency are affected by systematic nuclear physics uncertainties \citep{Zhu:2020eyk, Barnes:2020nfi}. These uncertainties span about an order of magnitude in the heating rate. To quantify their impact in our calculations, we perform simulations in which we vary the heating rates by a factor of $3$ or $0.3$. Fig.~\ref{LC_wind310T6_ye0.1_heating} shows the impact of changes in the heating rate in the case of the optimal wind profile with $Y_e = 0.1$. Unsurprisingly, the bolometric luminosity increases/decreases proportionally to the heating rate when we change the heating rates throughout the entire calculation. Interestingly, we find modest, but measurable differences in the bolometric luminosity even if we change the heating rates only in the first 10 seconds of the calculations, that is during the time the actual r-process is actually taking place. These difference persist for the first few days after merger.

\begin{figure}
\centering
\includegraphics[scale=0.5]{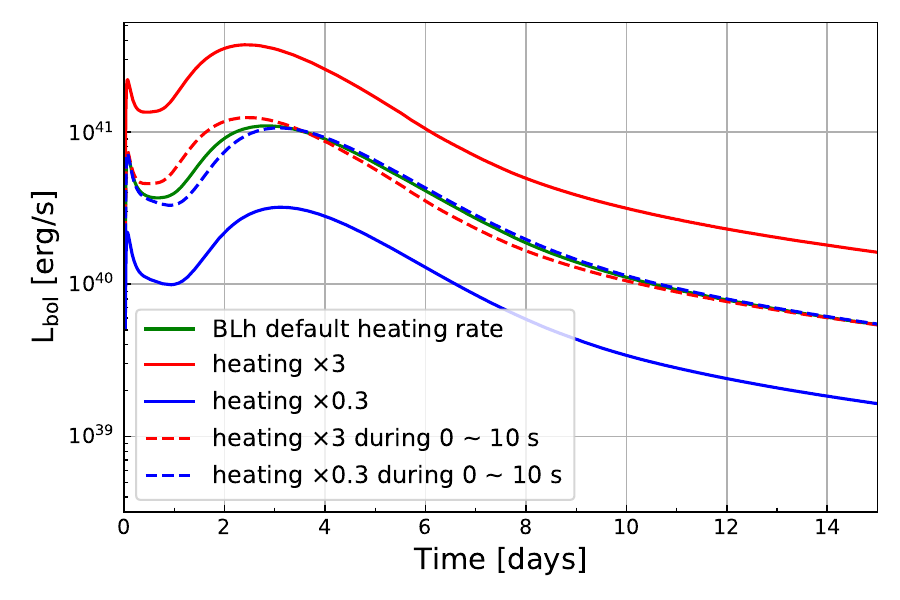}
\caption{Bolometric light curves obtained with different heating rates and the BLh profile. The uncertainty of heating rates changes the light curves by an order of magnitude. The green line represents the default heating rates introduced in \S \ref{subsection_heating}, with thermalization efficiency $\epsilon_{\mathrm{th}} = 0.5$. The red and blue solid lines display the light curves when the heating rate is multiplied by 3 and 0.3, respectively. The dashed light curves are obtained by changing the heating rates from the baseline only in the first ten seconds.}
\label{LC_blh_heating}
\end{figure} 

\begin{figure}
\centering
\includegraphics[scale=0.5]{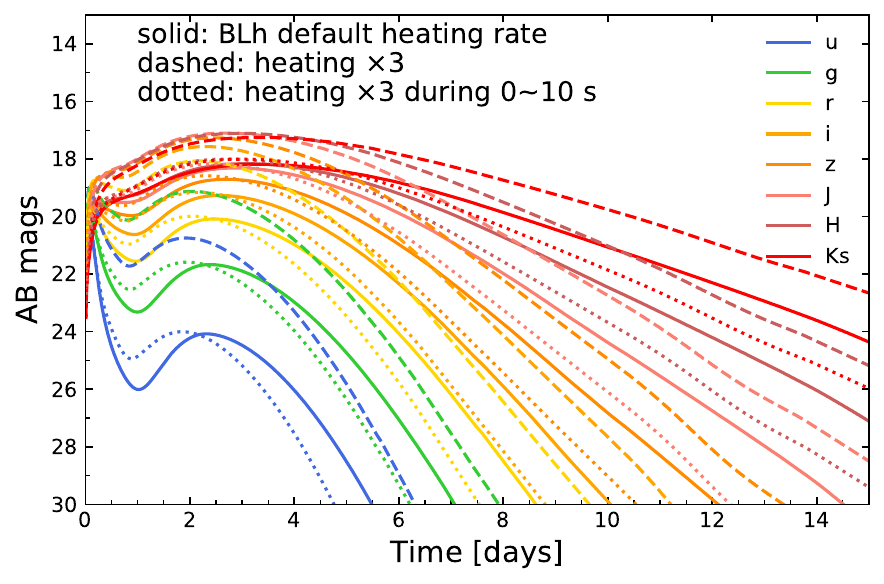}
\caption{Multicolor light curves obtained with different heating rates and the BLh profile. The solid lines represent the Gemini band AB magnitudes predicted with the default heating rates. The dashed lines show the light curves with heating rates multiplied by 3. The dotted lines show the light curves obtained by multiplying the heating rates by 3 in the first 10 seconds only.}
\label{ABmags_blh_heatingx3}
\end{figure} 

The same trend is also seen in Fig.~\ref{LC_blh_heating} for the BLh profile. In this case, the uncertainties in the heating rates in the first ten seconds result in a shift of the peak time by about a day. The multicolor light curves corresponding to the models with baseline and increased heating rates are shown in Fig.~\ref{ABmags_blh_heatingx3}. To investigate the origin of these difference we have repeated the BLh calculation with the hydrodynamics turned off (assuming homologous expansion). We find that when the assumption of homologous expansion is used, the heating rate in the first ten seconds has no impact on the light curve. We conclude that these changes in the light curve are the result of changes in the structure of the outflows. When the heating is increased in the first ten seconds, this leads to higher temperatures and, consequently, higher pressures and, as a result, the expansion of the ejecta is slightly accelerated. The lanthanide curtain is lifted at earlier time and the light curve peaks sooner. These results are consistent with those of \citet{Klion:2021jzr}, who investigated the impact of r-process heating in the first 60 second of the outflows. They also find that enhanced heating at early times can produce slightly brighter light curves that peak at earlier times. However, both in our calculations and those of \citet{Klion:2021jzr}, these effects are modest and possibly degenerate with other properties of the ejecta.

\subsection{Extrapolation of NR informed models}
\label{subsection_extrapolation_of_NR_informed_models}
The realistic profiles from our numerical relativity simulations only capture the amount of ejecta that has crossed a coordinate sphere with radius $r = 295\ {\rm km}$ by the time we terminate our calculations. Here, we estimate the contribution of material ejected at later times by extrapolating the outflow rate in time. This is clearly a crude estimate, considering that the flow is expected to change in a qualitative way once the accretion rate onto the central object drops below a critical value \citep{Beloborodov:2008nx, De:2020jdt}. However, this approach let us test the sensitivity of our models to the length of the numerical relativity simulations without the need to introduce additional parameters.

\begin{figure}
\centering
\includegraphics[scale=0.5]{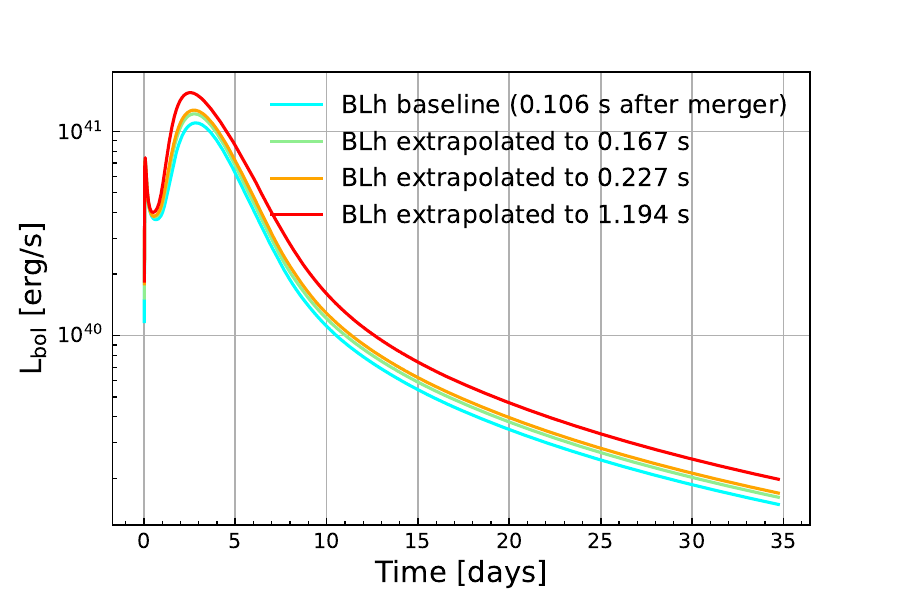}
\caption{Bolometric light curves for the time-extrapolated BLh models. The original BLh profile, or BLh baseline, is extracted from \texttt{WhiskyTHC} simulation until 0.106 sec after merger. We extrapolate the profile to 1.5, 2, and 10 times the total \texttt{WhiskyTHC} simulation time, which corresponds to 0.167, 0.227, and 1.194 sec after merger, respectively. }
\label{LC_blh_extrapolation}
\end{figure} 

\begin{figure}
\centering
\includegraphics[scale=0.5]{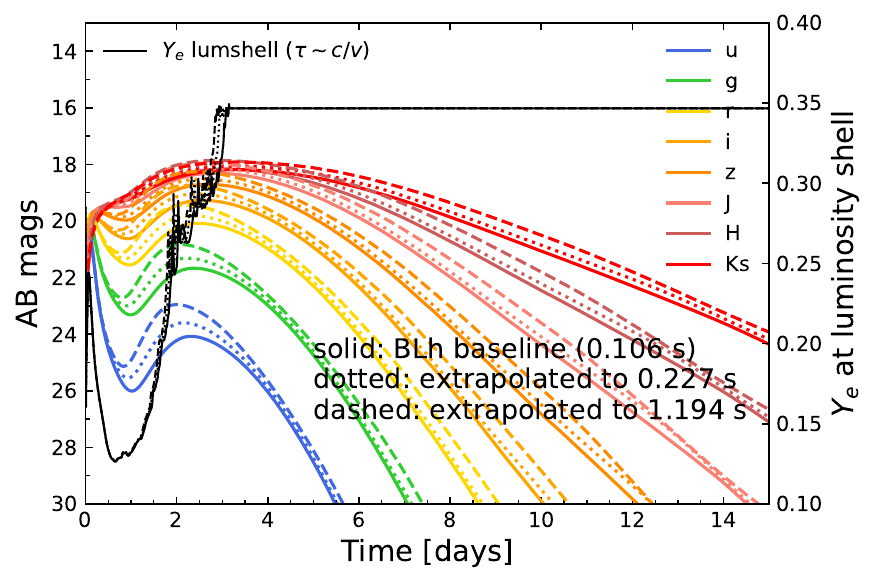}
\caption{AB magnitudes and $Y_e$ at the luminosity shell for the time-extrapolated BLh models. The late time ejecta predicted by the extrapolation procedure is not very neutron rich and only contributes to increase in the second peak of BLh multicolor light curves.}
\label{ABmags_Yelumshell_blh_extrapolation}
\end{figure} 

We extrapolate the BLh outflow rate in time to 1.5, 2, and 10 times the total \texttt{WhiskyTHC} simulation duration, i.e. 0.167, 0.227, and 1.194 sec after merger, respectively. The details of the extrapolation method are documented in Appendix \ref{appendix_method_of_blh_extrapolation}. Since the outflow rate is decaying, the overall ejecta mass increases only by a small factor, even when extrapolating to very late times (see Fig.~\ref{FluxMass_blh_extrapolation}). Consequently, the kilonova is only slightly brighter for the time-extrapolated profiles, as shown in Fig.~\ref{LC_blh_extrapolation}. It should be noted that only the second peak of the BLh light curves is enhanced, which can be seen more clearly in multi-band magnitudes (Fig.~ \ref{ABmags_Yelumshell_blh_extrapolation}). This is expected, since during the first peak the the luminosity shell is still localized at the outer surface of the ejecta, which are unaffected by the extrapolation. However, the kilonova becomes bluer at about 2 days, and the influence of lanthanide curtain on blue bands is weakened. This is also not surprising, since the material added to the profile by the extrapolation procedure has a high electron fraction, because the $Y_e$ increases towards the interior of the ejecta (see Fig.~\ref{blh_profile}).

\subsection{Impact of shock cooling}
\label{subsection_shock_cooling}
Although the r-process heating can explain the general features of the GW170817/AT2017gfo kilonova, the nature of the emission in the first ${\sim}1$ day is still unclear. \citet{Piro:2017ayh} suggested that this early signal might be due to the radiative cooling of shock-heated material. The shock might have originated from the interaction between the GRB jet and the ejecta. When the jet propagates through the ejecta, it forms a hot cocoon around it and generates a shock structure including a reverse shock. The shock deposits energy as it propagates and heats the ejecta, although the way of energy deposition is not clear \citep{Piro:2017ayh, Gottlieb:2017mqv, Nakar:2016cih, Lazzati:2016yxl, lazzati2021two}. According to \citep{Gottlieb:2017mqv, gottlieb2018cocoonshock, duffell2018jet, Nativi:2020moj, lundman2021first}, the jet energy ranges between $10^{48}$~erg and $10^{51}$~erg, while the plausible cocoon energy is between $5\times 10^{45}$~erg and $5\times 10^{49}$~erg. The jet break out time was of 1.7~s in GW170817 \citep{LIGOScientific:2017ync}.

In this section, we use the BLh profile to explore the impact of shock cooling on kilonova emission. We use the ``thermal bomb'' routine in \texttt{SNEC} to inject a shock with energy $E_{\rm shock}$ at the base of the ejecta. This routines injects energy with an exponential time dependency between the start time $t_{\text{start}}^{\text{b}}$ and the end time $t_{\text{end}}^{\text{b}}$ of the bomb: 
\begin{equation}
P^{\mathrm{b}}(t)=d^{\prime} e^{-c^{\prime} t}
\end{equation}
where $P^{\mathrm{b}}(t)$ is the injected bomb energy per unit time. The ratio $P^{\mathrm{b}}(t_{\mathrm{start}}^{\mathrm{b}})/P^{\mathrm{b}}(t_{\mathrm{end}}^{\mathrm{b}}) = R_t$ is set to 100 by default in \texttt{SNEC}. Therefore, 
\begin{equation}
c^{\prime}=\frac{\ln R_{t}}{\left(t_{\text {end }}^{\mathrm{b}}-t_{\text {start }}^{\mathrm{b}}\right)}, \quad d^{\prime}=\frac{c^{\prime} E_{\text {shock }}}{e^{-c^{\prime} t_{\text {start }}^{\mathrm{b}}}-e^{-c^{\prime} t_{\text {end }}^{\mathrm{b}}}}
\end{equation}
Similarly, at each time, the energy is spread exponentially between the start point $m_{\text{start}}$ and the end point $m_{\text{end}}$:
\begin{equation}
P_{\mathrm{m}, i}^{\mathrm{b}}\left(m_{i}\right)=b^{\prime} e^{-a^{\prime} m_{i}}
\end{equation}
The ratio $ 
P_{\mathrm{m}, i}^{\mathrm{b}}\left(m_{\mathrm{start}}^{\mathrm{b}}\right) / P_{\mathrm{m}, i}^{\mathrm{b}}\left(m_{\mathrm{end}}^{\mathrm{b}}\right)=R_{m}$ is also set to 100, then we obtain
\begin{equation}
a^{\prime}=\frac{\ln R_{m}}{m_{\mathrm{end}}^{\mathrm{b}}-m_{\mathrm{start}}^{\mathrm{b}}}, \quad b^{\prime}=\frac{d^{\prime} e^{-c^{\prime} t}}{\sum_{i} e^{-a^{\prime} m_{i}} \Delta m_{i+1 / 2}}
\end{equation}

We test different configurations of these parameters and find that the results are not very sensitive to the time interval, which we vary between ($0 - 50\ {\rm ms}$), ($0 - 100\ {\rm ms}$), ($50\ {\rm ms} - 100 {\rm ms}$) and ($0 - 1\ {\rm s}$), and to the choice of the spatial region in which the energy is injected, which we vary between ($0 - 0$; i.e., only deposited at the inner boundary) and ($0 - 0.01\ M_{\odot}$). The results are instead sensitive to the overall injected energy.

We find that shocks with $E_{\text{shock}} < 10^{49}$~erg have negligible impact on the kilonova light curve. This is not too surprising given that the initial kinetic energy in the ejecta is ${\sim}10^{50}$~erg. However, it is important to remark that our calculations assume spherical symmetry, while the cocoon is expected to be asymmetric at the time of breakout. A very rough estimate of the impact of anisotropy can be obtained by using the isotropic equivalent energy of the shock, instead of its actual energy. In so doing, we effectively assume that the fluid elements are only weakly coupled in the angular direction. Accordingly, we interpret the \texttt{SNEC} calculations as describing the evolution of a portion of the outflow subtended by a fixed solid angle. A better treatment would require performing 2D axisymmetric or 3D simulations \citep{Gottlieb:2017mqv, gottlieb2018cocoonshock, duffell2018jet, Nativi:2020moj, lundman2021first}.

\begin{figure}
\centering
\includegraphics[scale=0.5]{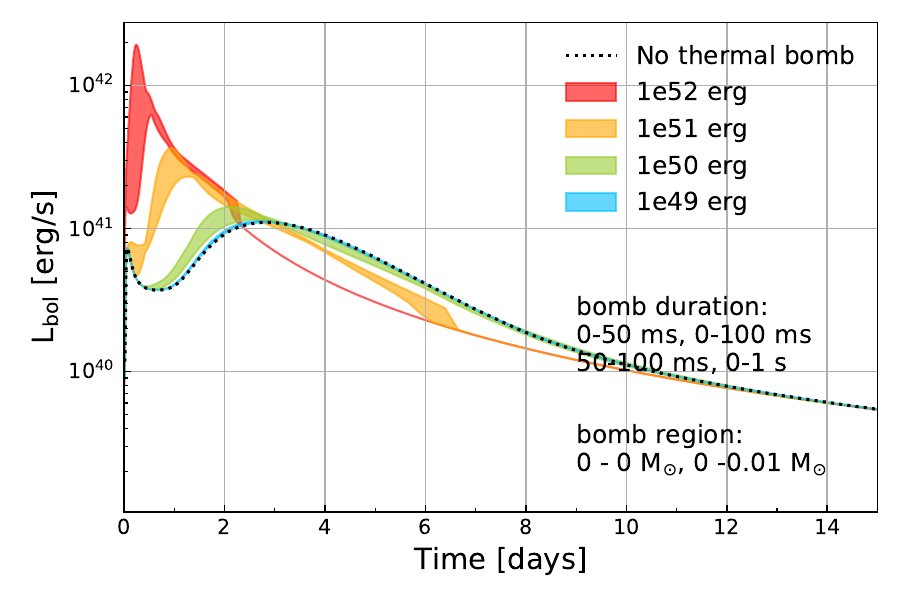}
\caption{Bolometric light curve for the BLh profile with thermal bomb shock heating. The dotted line shows the baseline (no thermal bomb). The different colors represent different amounts of energy carried by the injected shock, ranging from $10^{49}$~erg to $10^{52}$~erg (isotropic-equivalent). For each energy, the band spans results from different bomb configurations: time extent in $\big\{(0 - 50\ {\rm ms}), (0 - 100\ {\rm ms}), (50\ {\rm ms} - 100\ {\rm ms}), (0 - 1\ {\rm s})\big\}$, spatial extent $\big\{ (0, 0) , (0 - 0.01 M_{\odot})\big\}$.}
\label{LC_blh_TB_1e49-1e52}
\end{figure}

We vary $E_{\text{shock}}$ from $10^{49}$~erg to $10^{52}$~erg. Figure \ref{LC_blh_TB_1e49-1e52} shows the bolometric light curves with shock injection. We find that, if the shock energy is large enough, it can increase the bolometric luminosity by up to an order of magnitude. The shock can also alter the morphology of the light curve, suppressing the minimum on the light curve at $t \simeq 1\ {\rm day}$ and thus hiding the lanthanide curtain. These changes are in part due to the radiative cooling of the shock heating material. However, the main effect of the shock is to accelerate the expansion of the ejecta which, as a result, becomes transparent at earlier times.

\begin{figure}
\centering
\includegraphics[scale=0.5]{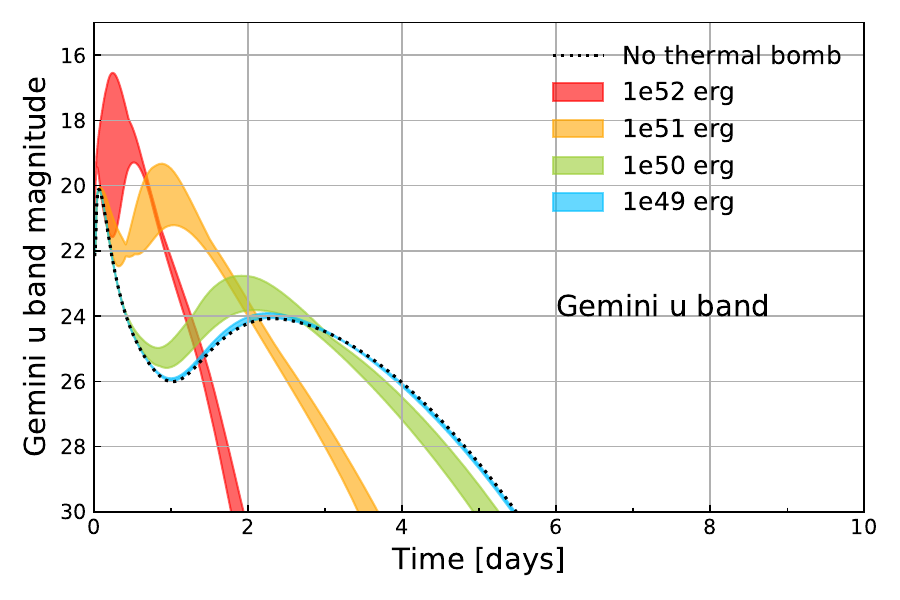}
\caption{u-band light curve for BLh models with thermal bomb. A sufficiently strong shock can significantly accelerate the expansion of the ejecta. On the one hand, faster expansion and radiation from the shock cooling suppress the lanthanide curtain effect and boost the luminosity at early times. On the other hand, the fast expansion causes the material to become optically thin at early times, so the kilonova light curve evolves on shorter timescales.}
\label{uGemini_blh_TB_1e49-1e52}
\end{figure}

\begin{figure}
\centering
\includegraphics[scale=0.5]{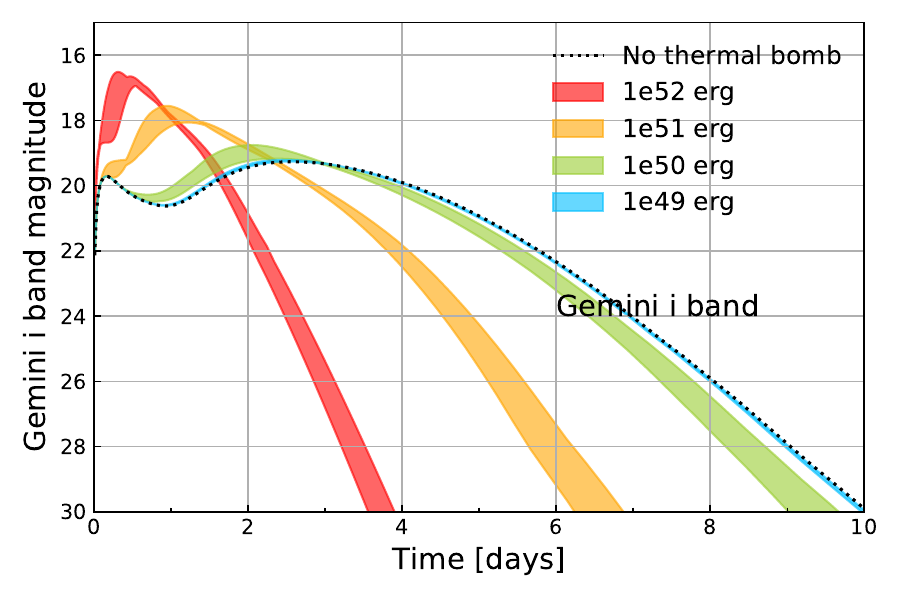}
\caption{i-band light curve for BLh models with thermal bomb. This figure is to be contrasted with Fig.~\ref{uGemini_blh_TB_1e49-1e52} which shows the u-band emission for the same models. The impact of a shock injected at the base of the outflow on the i-band light curve is similar, but somewhat less pronounced, than that on the u-band light curve.}
\label{iGemini_blh_TB_1e49-1e52}
\end{figure}

\begin{figure}
\centering
\includegraphics[scale=0.5]{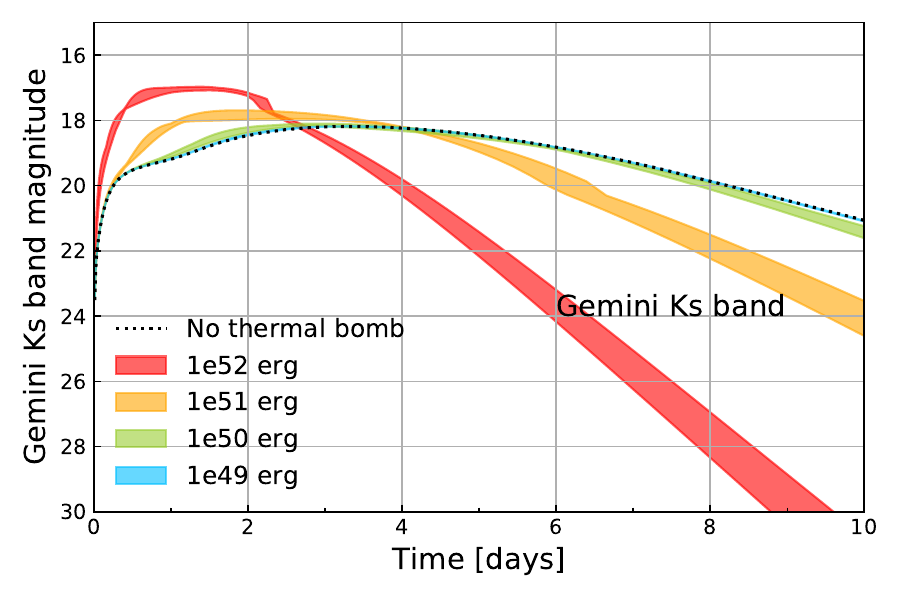}
\caption{Ks-band light curve for BLh models with thermal bomb. This figure is to be contrasted with Figs.~\ref{uGemini_blh_TB_1e49-1e52} and \ref{iGemini_blh_TB_1e49-1e52} which show the u-band and i-band emission for the same models. The more rapid expansion of the outflows caused by the shock also influences this band. With large thermal bomb energies the kilonova is brighter and evolves on faster time scales.}
\label{KsGemini_blh_TB_1e49-1e52}
\end{figure}

Figures \ref{uGemini_blh_TB_1e49-1e52}, \ref{iGemini_blh_TB_1e49-1e52}, and \ref{KsGemini_blh_TB_1e49-1e52} show the Gemini u-band, i-band, and Ks-band of the results. The blue/optical bands are more significantly influenced by the shock, which can boost the luminosity of the kilonova by up to 4 magnitudes in these bands. The impact on the peak luminosity in the red/infrared bands is more modest, but we still find that an energetic shock can boost the luminosity by about 1 magnitude even in these bands. In all cases we find that a shock at the base of the outflow can significantly accelerate the kilonova: making it peak at earlier times and fade more rapidly. Overall our results motivate the need for further investigation of the impact of the jet on the kilonova emission using multidimensional models.

% ===========================================================================
\section{Conclusion and Discussion}
\label{section_conclusion_and_discussion}
% ===========================================================================
We studied the kilonova emission from the ejecta of BNS mergers by means of radiation-hydrodynamical simulations. We considered both analytic wind profiles and ejecta profiles from numerical relativity simulations, and then employed the \texttt{SNEC} code to compute the associated color light curves. To this aim, we developed new modules for the \texttt{SNEC} code, including for the calculation of r-process heating rates and opacities, and specialized the built-in Paczynsky EOS in \texttt{SNEC} to the case of merger outflows. 
%% DR: I am not sure that this is really a conclusion to the study.
%% Although the velocity at the outer boundary may diverge, this does not affect kilonova light curves, and the problem can be alleviated by smoothing the initial velocity profile or constructing a smaller temperature gradient near the outer boundary.
We validated our approach by carefully checking energy conservation and comparing our results with those obtained from simpler semi-analytic models. As first applications of the code, we computed self-consistent kilonova light curves from a set of merger simulations;  we studied the impact of pressure forces and hydrodynamics, of nuclear physics uncertainties, and of shock cooling on the kilonova light curves.

We considered three merger simulations employing three EOS (BLh, DD2, and SFHo) and two different mass ratios. The DD2 binary considers an equal mass binary ($1.365\ M_\odot - 1.365\ M_\odot$), while the BLh and SFHo consider binaries with a mass asymmetry ($1.4\ M_\odot - 1.2\ M_\odot$). All the corresponding ejecta profiles show an outer fast component with high $Y_e$, but the bulk of the outflows has a moderate neutron richness. Additionally, the BLh and SFHo outflows have a very neutron rich tidal component between the outer high $Y_e$ outflow and the bulk of the outflow. The combined presence of a fast high-$Y_e$ outer shell and of a lanthanide curtain results in a double peaked morphology of the light curve. This is a new feature revealed by our calculations.

It is not our goal to fit observational data, but when comparing our models to AT2017gfo we found them to be underluminous, especially in the first few days. This remains true even when considering outflow rates from the merger simulations extrapolated to late times. This may suggest that GW170817 ejected more mass than predicted by our models, or that the adopted heating rates and opacities are underestimated or overestimated, respectively. Shock cooling and, more in general, the interaction between the ejecta and the GRB jet might also alleviate this disagreement. That said our results should be considered as provisional given our assumption of spherical symmetry in \texttt{SNEC} \citep{Perego:2017wtu}.

We studied the impact of hydrodynamic effects by comparing light curves produced with and without the assumption of homologous expansion. We found that hydrodynamics can have a substantial impact on the light curve, especially when considering idealized wind profiles. However, these effects are substantially smaller for more realistic wind profiles for which there are smaller pressure gradients close to the surface of the ejecta. The impact of hydrodynamics is also relatively small when considering ejecta from simulations.

We studied the impact of nuclear physics uncertainties on heating rates on the kilonova light curves. As expected, we find that the bolometric luminosities are directly proportional to the heating rate. Surprisingly, however, we also found that changes to the heating rate in the first 10 seconds can result in small, but appreciable differences in the kilonova properties. These differences arise due to changes in the structure of the outflows resulting from the increased/decreased pressure.

Finally, we studied the impact of the interaction between the dynamical ejecta and the GRB jet. To this aim, we injected shocks at the base of the ejecta using the thermal bomb module of \texttt{SNEC} with different total energies and with different bomb parameters. We found that the shock has a substantial impact on the kilonova light curve when the energy of the shock is comparable to or larger than the initial kinetic energy of the ejecta (${\sim}10^{50}\ {\rm erg}$). The shock accelerates the ejecta which, as a result, become transparent at earlier times. The resulting kilonova light curves evolve more rapidly and are bluer. The shock injection impacts predominantly the UV/optical bands in the first ${\sim}2$~days of the merger.

The approach we have developed here is complementary to other efforts that employ wavelength-dependent Monte Carlo radiative transfer, but neglect hydrodynamic effects. We have made a number of approximations that need to be improved to be able to compute reliable, realistic synthetic kilonova light curves from numerical relativity. Among these, the most serious one is the assumption of spherical symmetry. We plan to go beyond this approximation by porting the routines we have developed and tested with \texttt{SNEC} into the \texttt{Athena++} code \citep{2020ApJS..249....4S} and use a technique similar to that introduced by \citet{2021arXiv210903899H} to track the expansion of the ejecta over a timescale of several weeks. \texttt{SNEC} simulations could be post processed using Monte Carlo radiative transfer codes to compute improved color light curve and to compute synthetic spectra. Other possible future avenue of research include coupling \texttt{SNEC} with a nuclear reaction network like \texttt{SkyNet} and adopting time dependent thermalization efficiencies and improved opacities. 

\section*{Data Availability}
Simulation data produced as part of this work will be made available upon request to the corresponding author.

\section*{Acknowledgement}
We thank Viktoriya Morozova for help with the \texttt{SNEC} code.
David Radice acknowledges funding from the U.S. Department of Energy, Office of
Science, Division of Nuclear Physics under Award Number(s)
DE-SC0021177 and from the National Science Foundation under Grants No. 
PHY-2011725, PHY-2020275, PHY-2116686, and AST-2108467.

\bibliographystyle{mnras}
\bibliography{references}

\begin{thebibliography}{}
\makeatletter
\relax
\def\mn@urlcharsother{\let\do\@makeother \do\$\do\&\do\#\do\^\do\_\do\%\do\~}
\def\mn@doi{\begingroup\mn@urlcharsother \@ifnextchar [ {\mn@doi@}
  {\mn@doi@[]}}
\def\mn@doi@[#1]#2{\def\@tempa{#1}\ifx\@tempa\@empty \href
  {http://dx.doi.org/#2} {doi:#2}\else \href {http://dx.doi.org/#2} {#1}\fi
  \endgroup}
\def\mn@eprint#1#2{\mn@eprint@#1:#2::\@nil}
\def\mn@eprint@arXiv#1{\href {http://arxiv.org/abs/#1} {{\tt arXiv:#1}}}
\def\mn@eprint@dblp#1{\href {http://dblp.uni-trier.de/rec/bibtex/#1.xml}
  {dblp:#1}}
\def\mn@eprint@#1:#2:#3:#4\@nil{\def\@tempa {#1}\def\@tempb {#2}\def\@tempc
  {#3}\ifx \@tempc \@empty \let \@tempc \@tempb \let \@tempb \@tempa \fi \ifx
  \@tempb \@empty \def\@tempb {arXiv}\fi \@ifundefined
  {mn@eprint@\@tempb}{\@tempb:\@tempc}{\expandafter \expandafter \csname
  mn@eprint@\@tempb\endcsname \expandafter{\@tempc}}}

\bibitem[\protect\citeauthoryear{Abbott et~al.,}{Abbott
  et~al.}{2017}]{LIGOScientific:2017ync}
Abbott B.,  et~al., 2017, \mn@doi [Astrophys.J.Lett.]
  {10.3847/2041-8213/aa91c9}, 848, L12

\bibitem[\protect\citeauthoryear{Abbott et~al.,}{Abbott
  et~al.}{2020a}]{KAGRA:2013rdx}
Abbott B.,  et~al., 2020a, \mn@doi [Living Rev.Rel.]
  {10.1007/s41114-020-00026-9}, 23, 3

\bibitem[\protect\citeauthoryear{Abbott et~al.,}{Abbott
  et~al.}{2020b}]{LIGOScientific:2020aai}
Abbott B.,  et~al., 2020b, \mn@doi [Astrophys.J.Lett.]
  {10.3847/2041-8213/ab75f5}, 892, L3

\bibitem[\protect\citeauthoryear{Abbott et~al.,}{Abbott
  et~al.}{2021}]{LIGOScientific:2021qlt}
Abbott R.,  et~al., 2021, \mn@doi [Astrophys.J.Lett.]
  {10.3847/2041-8213/ac082e}, 915, L5

\bibitem[\protect\citeauthoryear{{Arcavi} et~al.,}{{Arcavi}
  et~al.}{2017}]{2017ApJ...848L..33A}
{Arcavi} I.,  et~al., 2017, \mn@doi [\apjl] {10.3847/2041-8213/aa910f}, \href
  {https://ui.adsabs.harvard.edu/abs/2017ApJ...848L..33A} {848, L33}

\bibitem[\protect\citeauthoryear{{Arnett}}{{Arnett}}{1980}]{1980ApJ...237..541A}
{Arnett} W.~D.,  1980, \mn@doi [\apj] {10.1086/157898}, \href
  {https://ui.adsabs.harvard.edu/abs/1980ApJ...237..541A} {237, 541}

\bibitem[\protect\citeauthoryear{Arnett}{Arnett}{1982}]{Arnett:1982ioj}
Arnett W.~D.,  1982, \mn@doi [Astrophys. J.] {10.1086/159681}, 253, 785

\bibitem[\protect\citeauthoryear{Barnes \& Kasen}{Barnes \&
  Kasen}{2013}]{Barnes:2013wka}
Barnes J.,  Kasen D.,  2013, \mn@doi [Astrophys.J.]
  {10.1088/0004-637X/775/1/18}, 775, 18

\bibitem[\protect\citeauthoryear{Barnes, Kasen, Wu  \& Martínez-Pinedo}{Barnes
  et~al.}{2016}]{Barnes:2016umi}
Barnes J.,  Kasen D.,  Wu M.-R.,   Martínez-Pinedo G.,  2016, \mn@doi
  [Astrophys.J.] {10.3847/0004-637X/829/2/110}, 829, 110

\bibitem[\protect\citeauthoryear{Barnes, Zhu, Lund, Sprouse, Vassh, McLaughlin,
  Mumpower  \& Surman}{Barnes et~al.}{2021}]{Barnes:2020nfi}
Barnes J.,  Zhu Y.,  Lund K.,  Sprouse T.,  Vassh N.,  McLaughlin G.,  Mumpower
  M.,   Surman R.,  2021, \mn@doi [Astrophys.J.] {10.3847/1538-4357/ac0aec},
  918, 44

\bibitem[\protect\citeauthoryear{Bauswein, Goriely  \& Janka}{Bauswein
  et~al.}{2013}]{Bauswein:2013yna}
Bauswein A.,  Goriely S.,   Janka H.-T.,  2013, \mn@doi [Astrophys.J.]
  {10.1088/0004-637X/773/1/78}, 773, 78

\bibitem[\protect\citeauthoryear{Beloborodov}{Beloborodov}{2008}]{Beloborodov:2008nx}
Beloborodov A.~M.,  2008, \mn@doi [AIP Conf.Proc.] {10.1063/1.3002509}, 1054,
  51

\bibitem[\protect\citeauthoryear{Berger}{Berger}{2014}]{Berger:2013jza}
Berger E.,  2014, \mn@doi [Ann.Rev.Astron.Astrophys.]
  {10.1146/annurev-astro-081913-035926}, 52, 43

\bibitem[\protect\citeauthoryear{Berger, Fong  \& Chornock}{Berger
  et~al.}{2013}]{Berger:2013wna}
Berger E.,  Fong W.,   Chornock R.,  2013, \mn@doi [Astrophys.J.Lett.]
  {10.1088/2041-8205/774/2/L23}, 774, L23

\bibitem[\protect\citeauthoryear{Bernuzzi et~al.,}{Bernuzzi
  et~al.}{2020}]{Bernuzzi:2020txg}
Bernuzzi S.,  et~al., 2020, \mn@doi [Mon.Not.Roy.Astron.Soc.]
  {10.1093/mnras/staa1860}, 497, 1488

\bibitem[\protect\citeauthoryear{Bombaci \& Logoteta}{Bombaci \&
  Logoteta}{2018}]{Bombaci:2018ksa}
Bombaci I.,  Logoteta D.,  2018, \mn@doi [Astron.Astrophys.]
  {10.1051/0004-6361/201731604}, 609, A128

\bibitem[\protect\citeauthoryear{Bovard, Martin, Guercilena, Arcones, Rezzolla
  \& Korobkin}{Bovard et~al.}{2017}]{Bovard:2017mvn}
Bovard L.,  Martin D.,  Guercilena F.,  Arcones A.,  Rezzolla L.,   Korobkin
  O.,  2017, \mn@doi [Phys.Rev.D] {10.1103/PhysRevD.96.124005}, 96

\bibitem[\protect\citeauthoryear{Breschi, Perego, Bernuzzi, Del~Pozzo, Nedora,
  Radice  \& Vescovi}{Breschi et~al.}{2021}]{Breschi:2021tbm}
Breschi M.,  Perego A.,  Bernuzzi S.,  Del~Pozzo W.,  Nedora V.,  Radice D.,
  Vescovi D.,  2021, \mn@doi [Mon.Not.Roy.Astron.Soc.]
  {10.1093/mnras/stab1287}, 505, 1661

\bibitem[\protect\citeauthoryear{Bulla}{Bulla}{2019}]{bulla2019possis}
Bulla M.,  2019, Monthly Notices of the Royal Astronomical Society, 489, 5037

\bibitem[\protect\citeauthoryear{Bulla et~al.,}{Bulla
  et~al.}{2021}]{Bulla:2020jjr}
Bulla M.,  et~al., 2021, \mn@doi [Mon.Not.Roy.Astron.Soc.]
  {10.1093/mnras/staa3796}, 501, 1891

\bibitem[\protect\citeauthoryear{Chatzopoulos, Wheeler  \& Vinko}{Chatzopoulos
  et~al.}{2012}]{Chatzopoulos:2011vj}
Chatzopoulos E.,  Wheeler J.,   Vinko J.,  2012, \mn@doi [Astrophys.J.]
  {10.1088/0004-637X/746/2/121}, 746, 121

\bibitem[\protect\citeauthoryear{Chornock et~al.,}{Chornock
  et~al.}{2017}]{Chornock:2017sdf}
Chornock R.,  et~al., 2017, \mn@doi [Astrophys.J.Lett.]
  {10.3847/2041-8213/aa905c}, 848, L19

\bibitem[\protect\citeauthoryear{Ciolfi \& Kalinani}{Ciolfi \&
  Kalinani}{2020}]{Ciolfi:2020wfx}
Ciolfi R.,  Kalinani J.~V.,  2020, \mn@doi [Astrophys.J.Lett.]
  {10.3847/2041-8213/abb240}, 900, L35

\bibitem[\protect\citeauthoryear{Coughlin et~al.,}{Coughlin
  et~al.}{2018}]{Coughlin:2018miv}
Coughlin M.~W.,  et~al., 2018, \mn@doi [Mon.Not.Roy.Astron.Soc.]
  {10.1093/mnras/sty2174}, 480, 3871

\bibitem[\protect\citeauthoryear{Coulter et~al.,}{Coulter
  et~al.}{2017}]{Coulter:2017wya}
Coulter D.,  et~al., 2017, \mn@doi [Science] {10.1126/science.aap9811}, 358,
  1556

\bibitem[\protect\citeauthoryear{Cowan, Sneden, Lawler, Aprahamian, Wiescher,
  Langanke, Martínez-Pinedo  \& Thielemann}{Cowan
  et~al.}{2021}]{Cowan:2019pkx}
Cowan J.~J.,  Sneden C.,  Lawler J.~E.,  Aprahamian A.,  Wiescher M.,  Langanke
  K.,  Martínez-Pinedo G.,   Thielemann F.-K.,  2021, \mn@doi [Rev.Mod.Phys.]
  {10.1103/RevModPhys.93.015002}, 93

\bibitem[\protect\citeauthoryear{Cowperthwaite et~al.,}{Cowperthwaite
  et~al.}{2017}]{Cowperthwaite:2017dyu}
Cowperthwaite P.,  et~al., 2017, \mn@doi [Astrophys.J.Lett.]
  {10.3847/2041-8213/aa8fc7}, 848, L17

\bibitem[\protect\citeauthoryear{De \& Siegel}{De \& Siegel}{2020}]{De:2020jdt}
De S.,  Siegel D.,  2020, arXiv:2011.07176

\bibitem[\protect\citeauthoryear{Dean, Fernández  \& Metzger}{Dean
  et~al.}{2021}]{Dean:2021gpd}
Dean C.,  Fernández R.,   Metzger B.~D.,  2021, arXiv:2108.08311

\bibitem[\protect\citeauthoryear{Dessart, Ott, Burrows, Rosswog  \&
  Livne}{Dessart et~al.}{2009}]{Dessart:2008zd}
Dessart L.,  Ott C.,  Burrows A.,  Rosswog S.,   Livne E.,  2009, \mn@doi
  [Astrophys.J.] {10.1088/0004-637X/690/2/1681}, 690, 1681

\bibitem[\protect\citeauthoryear{Drout et~al.,}{Drout
  et~al.}{2017}]{Drout:2017ijr}
Drout M.,  et~al., 2017, \mn@doi [Science] {10.1126/science.aaq0049}, 358, 1570

\bibitem[\protect\citeauthoryear{Duffell, Quataert, Kasen  \& Klion}{Duffell
  et~al.}{2018}]{duffell2018jet}
Duffell P.~C.,  Quataert E.,  Kasen D.,   Klion H.,  2018, The Astrophysical
  Journal, 866, 3

\bibitem[\protect\citeauthoryear{Evans et~al.,}{Evans
  et~al.}{2017}]{Evans:2017mmy}
Evans P.,  et~al., 2017, \mn@doi [Science] {10.1126/science.aap9580}, 358, 1565

\bibitem[\protect\citeauthoryear{Fernández \& Metzger}{Fernández \&
  Metzger}{2013}]{Fernandez:2013tya}
Fernández R.,  Metzger B.~D.,  2013, \mn@doi [Mon.Not.Roy.Astron.Soc.]
  {10.1093/mnras/stt1312}, 435, 502

\bibitem[\protect\citeauthoryear{Fernández, Tchekhovskoy, Quataert, Foucart
  \& Kasen}{Fernández et~al.}{2019}]{Fernandez:2018kax}
Fernández R.,  Tchekhovskoy A.,  Quataert E.,  Foucart F.,   Kasen D.,  2019,
  \mn@doi [Mon.Not.Roy.Astron.Soc.] {10.1093/mnras/sty2932}, 482, 3373

\bibitem[\protect\citeauthoryear{Fong et~al.,}{Fong
  et~al.}{2014}]{Fong:2013lba}
Fong W.,  et~al., 2014, \mn@doi [Astrophys.J.] {10.1088/0004-637X/780/2/118},
  780, 118

\bibitem[\protect\citeauthoryear{Fontes, Fryer, Hungerford, Wollaeger  \&
  Korobkin}{Fontes et~al.}{2020}]{Fontes:2019tlk}
Fontes C.,  Fryer C.,  Hungerford A.,  Wollaeger R.,   Korobkin O.,  2020,
  \mn@doi [Mon.Not.Roy.Astron.Soc.] {10.1093/mnras/staa485}, 493, 4143

\bibitem[\protect\citeauthoryear{Foucart, O'Connor, Roberts, Kidder, Pfeiffer
  \& Scheel}{Foucart et~al.}{2016}]{Foucart:2016rxm}
Foucart F.,  O'Connor E.,  Roberts L.,  Kidder L.~E.,  Pfeiffer H.~P.,   Scheel
  M.~A.,  2016, \mn@doi [Phys.Rev.D] {10.1103/PhysRevD.94.123016}, 94

\bibitem[\protect\citeauthoryear{Foucart, Duez, Hebert, Kidder, Pfeiffer  \&
  Scheel}{Foucart et~al.}{2020}]{Foucart:2020qjb}
Foucart F.,  Duez M.~D.,  Hebert F.,  Kidder L.~E.,  Pfeiffer H.~P.,   Scheel
  M.~A.,  2020, \mn@doi [Astrophys.J.Lett.] {10.3847/2041-8213/abbb87}, 902,
  L27

\bibitem[\protect\citeauthoryear{Foucart, Moesta, Ramirez, Wright, Darbha  \&
  Kasen}{Foucart et~al.}{2021}]{Foucart:2021ikp}
Foucart F.,  Moesta P.,  Ramirez T.,  Wright A.~J.,  Darbha S.,   Kasen D.,
  2021, arXiv:2109.00565

\bibitem[\protect\citeauthoryear{Fujibayashi, Kiuchi, Nishimura, Sekiguchi  \&
  Shibata}{Fujibayashi et~al.}{2018}]{Fujibayashi:2017puw}
Fujibayashi S.,  Kiuchi K.,  Nishimura N.,  Sekiguchi Y.,   Shibata M.,  2018,
  \mn@doi [Astrophys.J.] {10.3847/1538-4357/aabafd}, 860, 64

\bibitem[\protect\citeauthoryear{Fujibayashi, Wanajo, Kiuchi, Kyutoku,
  Sekiguchi  \& Shibata}{Fujibayashi et~al.}{2020}]{Fujibayashi:2020dvr}
Fujibayashi S.,  Wanajo S.,  Kiuchi K.,  Kyutoku K.,  Sekiguchi Y.,   Shibata
  M.,  2020, \mn@doi [Astrophys.J.] {10.3847/1538-4357/abafc2}, 901, 122

\bibitem[\protect\citeauthoryear{Gittings et~al.,}{Gittings
  et~al.}{2008}]{gittings2008rage}
Gittings M.,  et~al., 2008, Computational Science \& Discovery, 1, 015005

\bibitem[\protect\citeauthoryear{Gottlieb, Nakar  \& Piran}{Gottlieb
  et~al.}{2018a}]{Gottlieb:2017mqv}
Gottlieb O.,  Nakar E.,   Piran T.,  2018a, \mn@doi [Mon. Not. Roy. Astron.
  Soc.] {10.1093/mnras/stx2357}, 473, 576

\bibitem[\protect\citeauthoryear{Gottlieb, Nakar, Piran  \&
  Hotokezaka}{Gottlieb et~al.}{2018b}]{gottlieb2018cocoonshock}
Gottlieb O.,  Nakar E.,  Piran T.,   Hotokezaka K.,  2018b, Monthly Notices of
  the Royal Astronomical Society, 479, 588

\bibitem[\protect\citeauthoryear{Grossman, Korobkin, Rosswog  \&
  Piran}{Grossman et~al.}{2014}]{Grossman:2013lqa}
Grossman D.,  Korobkin O.,  Rosswog S.,   Piran T.,  2014, \mn@doi
  [Mon.Not.Roy.Astron.Soc.] {10.1093/mnras/stt2503}, 439, 757

\bibitem[\protect\citeauthoryear{{Habegger} \& {Heitsch}}{{Habegger} \&
  {Heitsch}}{2021}]{2021arXiv210903899H}
{Habegger} R.,  {Heitsch} F.,  2021, arXiv e-prints, \href
  {https://ui.adsabs.harvard.edu/abs/2021arXiv210903899H} {p. arXiv:2109.03899}

\bibitem[\protect\citeauthoryear{Hallinan et~al.,}{Hallinan
  et~al.}{2017}]{Hallinan:2017woc}
Hallinan G.,  et~al., 2017, \mn@doi [Science] {10.1126/science.aap9855}, 358,
  1579

\bibitem[\protect\citeauthoryear{Heinzel et~al.,}{Heinzel
  et~al.}{2021}]{Heinzel:2020qlt}
Heinzel J.,  et~al., 2021, \mn@doi [Mon. Not. Roy. Astron. Soc.]
  {10.1093/mnras/stab221}, 502, 3057

\bibitem[\protect\citeauthoryear{Hempel \& Schaffner-Bielich}{Hempel \&
  Schaffner-Bielich}{2010}]{Hempel:2009mc}
Hempel M.,  Schaffner-Bielich J.,  2010, \mn@doi [Nucl.Phys.A]
  {10.1016/j.nuclphysa.2010.02.010}, 837, 210

\bibitem[\protect\citeauthoryear{Hjorth et~al.,}{Hjorth
  et~al.}{2017}]{Hjorth:2017yza}
Hjorth J.,  et~al., 2017, \mn@doi [Astrophys.J.Lett.]
  {10.3847/2041-8213/aa9110}, 848, L31

\bibitem[\protect\citeauthoryear{{Hoffman}, {Woosley}  \& {Qian}}{{Hoffman}
  et~al.}{1997}]{1997ApJ...482..951H}
{Hoffman} R.~D.,  {Woosley} S.~E.,   {Qian} Y.~Z.,  1997, \mn@doi [\apj]
  {10.1086/304181}, \href
  {https://ui.adsabs.harvard.edu/abs/1997ApJ...482..951H} {482, 951}

\bibitem[\protect\citeauthoryear{Hotokezaka \& Nakar}{Hotokezaka \&
  Nakar}{2019}]{Hotokezaka:2019uwo}
Hotokezaka K.,  Nakar E.,  2019, \mn@doi [arXiv:1909.02581]
  {10.3847/1538-4357/ab6a98}

\bibitem[\protect\citeauthoryear{Hotokezaka, Kyutoku, Tanaka, Kiuchi,
  Sekiguchi, Shibata  \& Wanajo}{Hotokezaka et~al.}{2013}]{Hotokezaka:2013kza}
Hotokezaka K.,  Kyutoku K.,  Tanaka M.,  Kiuchi K.,  Sekiguchi Y.,  Shibata M.,
    Wanajo S.,  2013, \mn@doi [Astrophys.J.Lett.]
  {10.1088/2041-8205/778/1/L16}, 778, L16

\bibitem[\protect\citeauthoryear{Hotokezaka, Wanajo, Tanaka, Bamba, Terada  \&
  Piran}{Hotokezaka et~al.}{2016}]{Hotokezaka:2015cma}
Hotokezaka K.,  Wanajo S.,  Tanaka M.,  Bamba A.,  Terada Y.,   Piran T.,
  2016, \mn@doi [Mon.Not.Roy.Astron.Soc.] {10.1093/mnras/stw404}, 459, 35

\bibitem[\protect\citeauthoryear{Hotokezaka, Sari  \& Piran}{Hotokezaka
  et~al.}{2017}]{Hotokezaka:2017dbk}
Hotokezaka K.,  Sari R.,   Piran T.,  2017, \mn@doi [Mon.Not.Roy.Astron.Soc.]
  {10.1093/mnras/stx411}, 468, 91

\bibitem[\protect\citeauthoryear{Hotokezaka, Kiuchi, Shibata, Nakar  \&
  Piran}{Hotokezaka et~al.}{2018}]{Hotokezaka:2018gmo}
Hotokezaka K.,  Kiuchi K.,  Shibata M.,  Nakar E.,   Piran T.,  2018, \mn@doi
  [Astrophys.J.] {10.3847/1538-4357/aadf92}, 867, 95

\bibitem[\protect\citeauthoryear{Hotokezaka, Tanaka, Kato  \&
  Gaigalas}{Hotokezaka et~al.}{2021}]{Hotokezaka:2021ofe}
Hotokezaka K.,  Tanaka M.,  Kato D.,   Gaigalas G.,  2021, \mn@doi
  [arXiv:2102.07879] {10.1093/mnras/stab1975}

\bibitem[\protect\citeauthoryear{Iglesias \& Rogers}{Iglesias \&
  Rogers}{1996}]{Iglesias:1996bh}
Iglesias C.~A.,  Rogers F.~J.,  1996, \mn@doi [Astrophys. J.] {10.1086/177381},
  464, 943

\bibitem[\protect\citeauthoryear{Ishizaki, Kiuchi, Ioka  \& Wanajo}{Ishizaki
  et~al.}{2021}]{Ishizaki:2021qne}
Ishizaki W.,  Kiuchi K.,  Ioka K.,   Wanajo S.,  2021, arXiv:2104.04708

\bibitem[\protect\citeauthoryear{Jin, Xu, Fan, Wu  \& Wei}{Jin
  et~al.}{2013}]{Jin:2013jca}
Jin Z.-P.,  Xu D.,  Fan Y.-Z.,  Wu X.-F.,   Wei D.-M.,  2013, \mn@doi [The
  Astrophysical Journal Letter] {10.1088/2041-8205/775/1/L19}, 775, L19

\bibitem[\protect\citeauthoryear{Jin et~al.,}{Jin et~al.}{2016}]{Jin:2016pnm}
Jin Z.-P.,  et~al., 2016, \mn@doi [Nature Commun.] {10.1038/ncomms12898}, 7

\bibitem[\protect\citeauthoryear{Jin, Covino, Liao, Li, D'Avanzo, Fan  \&
  Wei}{Jin et~al.}{2020}]{Jin:2019uqr}
Jin Z.-P.,  Covino S.,  Liao N.-H.,  Li X.,  D'Avanzo P.,  Fan Y.-Z.,   Wei
  D.-M.,  2020, Nature Astron., 4, 77

\bibitem[\protect\citeauthoryear{Just, Bauswein, Pulpillo, Goriely  \&
  Janka}{Just et~al.}{2015}]{Just:2014fka}
Just O.,  Bauswein A.,  Pulpillo R.~A.,  Goriely S.,   Janka H.~T.,  2015,
  \mn@doi [Mon.Not.Roy.Astron.Soc.] {10.1093/mnras/stv009}, 448, 541

\bibitem[\protect\citeauthoryear{Just, Kullmann, Goriely, Bauswein, Janka  \&
  Collins}{Just et~al.}{2021b}]{Just:2021vzy}
Just O.,  Kullmann I.,  Goriely S.,  Bauswein A.,  Janka H.-T.,   Collins
  C.~E.,  2021b, arXiv:2109.14617

\bibitem[\protect\citeauthoryear{Just, Goriely, Janka, Nagataki  \&
  Bauswein}{Just et~al.}{2021a}]{Just:2021cls}
Just O.,  Goriely S.,  Janka H.-T.,  Nagataki S.,   Bauswein A.,  2021a,
  arXiv:2102.08387

\bibitem[\protect\citeauthoryear{Kasen \& Barnes}{Kasen \&
  Barnes}{2019}]{Kasen:2018drm}
Kasen D.,  Barnes J.,  2019, \mn@doi [Astrophys.J.] {10.3847/1538-4357/ab06c2},
  876, 128

\bibitem[\protect\citeauthoryear{Kasen, Badnell  \& Barnes}{Kasen
  et~al.}{2013}]{Kasen:2013xka}
Kasen D.,  Badnell N.,   Barnes J.,  2013, \mn@doi [Astrophys.J.]
  {10.1088/0004-637X/774/1/25}, 774, 25

\bibitem[\protect\citeauthoryear{Kashyap, Raman  \& Ajith}{Kashyap
  et~al.}{2019}]{Kashyap:2019ypm}
Kashyap R.,  Raman G.,   Ajith P.,  2019, \mn@doi [Astrophys.J.Lett.]
  {10.3847/2041-8213/ab543f}, 886, L19

\bibitem[\protect\citeauthoryear{Kasliwal et~al.,}{Kasliwal
  et~al.}{2017}]{Kasliwal:2017ngb}
Kasliwal M.,  et~al., 2017, \mn@doi [Science] {10.1126/science.aap9455}, 358,
  1559

\bibitem[\protect\citeauthoryear{Kasliwal et~al.,}{Kasliwal
  et~al.}{2018}]{Kasliwal:2018fwk}
Kasliwal M.~M.,  et~al., 2018, \mn@doi [arXiv:1812.08708]
  {10.1093/mnrasl/slz007}

\bibitem[\protect\citeauthoryear{Kastaun \& Galeazzi}{Kastaun \&
  Galeazzi}{2015}]{Kastaun:2014fna}
Kastaun W.,  Galeazzi F.,  2015, \mn@doi [Phys. Rev. D]
  {10.1103/PhysRevD.91.064027}, 91, 064027

\bibitem[\protect\citeauthoryear{Kawaguchi, Shibata  \& Tanaka}{Kawaguchi
  et~al.}{2018}]{Kawaguchi:2018ptg}
Kawaguchi K.,  Shibata M.,   Tanaka M.,  2018, \mn@doi [Astrophys.J.Lett.]
  {10.3847/2041-8213/aade02}, 865, L21

\bibitem[\protect\citeauthoryear{Kawaguchi, Shibata  \& Tanaka}{Kawaguchi
  et~al.}{2019}]{Kawaguchi:2019nju}
Kawaguchi K.,  Shibata M.,   Tanaka M.,  2019, \mn@doi [arXiv:1908.05815]
  {10.3847/1538-4357/ab61f6}

\bibitem[\protect\citeauthoryear{Kawaguchi, Fujibayashi, Shibata, Tanaka  \&
  Wanajo}{Kawaguchi et~al.}{2021}]{Kawaguchi:2020vbf}
Kawaguchi K.,  Fujibayashi S.,  Shibata M.,  Tanaka M.,   Wanajo S.,  2021,
  \mn@doi [Astrophys. J.] {10.3847/1538-4357/abf3bc}, 913, 100

\bibitem[\protect\citeauthoryear{Klion, Tchekhovskoy, Kasen, Kathirgamaraju,
  Quataert  \& Fernández}{Klion et~al.}{2021a}]{Klion:2021jzr}
Klion H.,  Tchekhovskoy A.,  Kasen D.,  Kathirgamaraju A.,  Quataert E.,
  Fernández R.,  2021a, arXiv:2108.04251

\bibitem[\protect\citeauthoryear{Klion, Duffell, Kasen  \& Quataert}{Klion
  et~al.}{2021b}]{Klion:2020efn}
Klion H.,  Duffell P.~C.,  Kasen D.,   Quataert E.,  2021b, \mn@doi
  [Mon.Not.Roy.Astron.Soc.] {10.1093/mnras/stab042}, 502, 865

\bibitem[\protect\citeauthoryear{Korobkin, Rosswog, Arcones  \&
  Winteler}{Korobkin et~al.}{2012}]{Korobkin:2012uy}
Korobkin O.,  Rosswog S.,  Arcones A.,   Winteler C.,  2012, \mn@doi [Mon. Not.
  Roy. Astron. Soc.] {10.1111/j.1365-2966.2012.21859.x}, 426, 1940

\bibitem[\protect\citeauthoryear{Korobkin et~al.,}{Korobkin
  et~al.}{2021}]{Korobkin:2020spe}
Korobkin O.,  et~al., 2021, \mn@doi [Astrophys.J.] {10.3847/1538-4357/abe1b5},
  910, 116

\bibitem[\protect\citeauthoryear{Kulkarni}{Kulkarni}{2005}]{Kulkarni:2005jw}
Kulkarni S.,  2005, arXiv:astro-ph/0510256

\bibitem[\protect\citeauthoryear{Kullmann, Goriely, Just, Ardevol-Pulpillo,
  Bauswein  \& Janka}{Kullmann et~al.}{2021}]{Kullmann:2021gvo}
Kullmann I.,  Goriely S.,  Just O.,  Ardevol-Pulpillo R.,  Bauswein A.,   Janka
  H.-T.,  2021, arXiv:2109.02509

\bibitem[\protect\citeauthoryear{Lamb et~al.,}{Lamb
  et~al.}{2019}]{Lamb:2019lao}
Lamb G.,  et~al., 2019, \mn@doi [arXiv:1905.02159] {10.3847/1538-4357/ab38bb}

\bibitem[\protect\citeauthoryear{Lazzati, Deich, Morsony  \& Workman}{Lazzati
  et~al.}{2017}]{Lazzati:2016yxl}
Lazzati D.,  Deich A.,  Morsony B.~J.,   Workman J.~C.,  2017, \mn@doi [Mon.
  Not. Roy. Astron. Soc.] {10.1093/mnras/stx1683}, 471, 1652

\bibitem[\protect\citeauthoryear{Lazzati, Perna, Ciolfi, Giacomazzo,
  L{\'o}pez-C{\'a}mara  \& Morsony}{Lazzati et~al.}{2021}]{lazzati2021two}
Lazzati D.,  Perna R.,  Ciolfi R.,  Giacomazzo B.,  L{\'o}pez-C{\'a}mara D.,
  Morsony B.,  2021, The Astrophysical Journal Letters, 918, L6

\bibitem[\protect\citeauthoryear{{Lee}, {Ramirez-Ruiz}  \&
  {L{\'o}pez-C{\'a}mara}}{{Lee} et~al.}{2009}]{2009ApJ...699L..93L}
{Lee} W.~H.,  {Ramirez-Ruiz} E.,   {L{\'o}pez-C{\'a}mara} D.,  2009, \mn@doi
  [\apjl] {10.1088/0004-637X/699/2/L93}, \href
  {https://ui.adsabs.harvard.edu/abs/2009ApJ...699L..93L} {699, L93}

\bibitem[\protect\citeauthoryear{Lehner, Liebling, Palenzuela, Caballero,
  O'Connor, Anderson  \& Neilsen}{Lehner et~al.}{2016}]{Lehner:2016lxy}
Lehner L.,  Liebling S.~L.,  Palenzuela C.,  Caballero O.,  O'Connor E.,
  Anderson M.,   Neilsen D.,  2016, \mn@doi [Class.Quant.Grav.]
  {10.1088/0264-9381/33/18/184002}, 33

\bibitem[\protect\citeauthoryear{Li \& Paczynski}{Li \&
  Paczynski}{1998}]{Li:1998bw}
Li L.-X.,  Paczynski B.,  1998, \mn@doi [Astrophys.J.Lett.] {10.1086/311680},
  507, L59

\bibitem[\protect\citeauthoryear{Lippuner \& Roberts}{Lippuner \&
  Roberts}{2015}]{Lippuner:2015gwa}
Lippuner J.,  Roberts L.~F.,  2015, \mn@doi [Astrophys.J.]
  {10.1088/0004-637X/815/2/82}, 815, 82

\bibitem[\protect\citeauthoryear{Lippuner \& Roberts}{Lippuner \&
  Roberts}{2017}]{Lippuner:2017tyn}
Lippuner J.,  Roberts L.~F.,  2017, \mn@doi [Astrophys. J. Suppl.]
  {10.3847/1538-4365/aa94cb}, 233, 18

\bibitem[\protect\citeauthoryear{Logoteta, Perego  \& Bombaci}{Logoteta
  et~al.}{2021}]{Logoteta:2020yxf}
Logoteta D.,  Perego A.,   Bombaci I.,  2021, \mn@doi [Astron.Astrophys.]
  {10.1051/0004-6361/202039457}, 646, A55

\bibitem[\protect\citeauthoryear{Lundman \& Beloborodov}{Lundman \&
  Beloborodov}{2021}]{lundman2021first}
Lundman C.,  Beloborodov A.~M.,  2021, The Astrophysical Journal Letters, 907,
  L13

\bibitem[\protect\citeauthoryear{Margutti \& Chornock}{Margutti \&
  Chornock}{2020}]{Margutti:2020xbo}
Margutti R.,  Chornock R.,  2020, arXiv:2012.04810

\bibitem[\protect\citeauthoryear{Martin, Perego, Arcones, Thielemann, Korobkin
  \& Rosswog}{Martin et~al.}{2015}]{Martin:2015hxa}
Martin D.,  Perego A.,  Arcones A.,  Thielemann F.-K.,  Korobkin O.,   Rosswog
  S.,  2015, \mn@doi [Astrophys.J.] {10.1088/0004-637X/813/1/2}, 813, 2

\bibitem[\protect\citeauthoryear{Metzger}{Metzger}{2020}]{Metzger:2019zeh}
Metzger B.~D.,  2020, \mn@doi [Living Rev.Rel.] {10.1007/s41114-019-0024-0},
  23, 1

\bibitem[\protect\citeauthoryear{Metzger \& Fernandez}{Metzger \&
  Fernandez}{2021}]{Metzger:2021grk}
Metzger B.~D.,  Fernandez R.,  2021, \mn@doi [Astrophys.J.Lett.]
  {10.3847/2041-8213/ac1169}, 916, L3

\bibitem[\protect\citeauthoryear{Metzger \& Fernández}{Metzger \&
  Fernández}{2014}]{Metzger:2014ila}
Metzger B.~D.,  Fernández R.,  2014, \mn@doi [Mon.Not.Roy.Astron.Soc.]
  {10.1093/mnras/stu802}, 441, 3444

\bibitem[\protect\citeauthoryear{Metzger, Piro  \& Quataert}{Metzger
  et~al.}{2008}]{Metzger:2008av}
Metzger B.,  Piro A.,   Quataert E.,  2008, \mn@doi [Mon.Not.Roy.Astron.Soc.]
  {10.1111/j.1365-2966.2008.13789.x}, 390, 781

\bibitem[\protect\citeauthoryear{Metzger, Piro  \& Quataert}{Metzger
  et~al.}{2009}]{Metzger:2008jt}
Metzger B.,  Piro A.,   Quataert E.,  2009, \mn@doi [Mon.Not.Roy.Astron.Soc.]
  {10.1111/j.1365-2966.2008.14380.x}, 396, 304

\bibitem[\protect\citeauthoryear{Metzger et~al.,}{Metzger
  et~al.}{2010}]{Metzger:2010sy}
Metzger B.,  et~al., 2010, \mn@doi [Mon.Not.Roy.Astron.Soc.]
  {10.1111/j.1365-2966.2010.16864.x}, 406, 2650

\bibitem[\protect\citeauthoryear{Metzger, Thompson  \& Quataert}{Metzger
  et~al.}{2018}]{Metzger:2018uni}
Metzger B.~D.,  Thompson T.~A.,   Quataert E.,  2018, \mn@doi [Astrophys.J.]
  {10.3847/1538-4357/aab095}, 856, 101

\bibitem[\protect\citeauthoryear{Miller, Sprouse, Fryer, Ryan, Dolence,
  Mumpower  \& Surman}{Miller et~al.}{2020}]{Miller:2019mfl}
Miller J.~M.,  Sprouse T.~M.,  Fryer C.~L.,  Ryan B.~R.,  Dolence J.~C.,
  Mumpower M.~R.,   Surman R.,  2020, \mn@doi [Astrophys.J.]
  {10.3847/1538-4357/abb4e3}, 902, 66

\bibitem[\protect\citeauthoryear{M\"oller, Sierk, Ichikawa  \& Sagawa}{M\"oller
  et~al.}{2016}]{Moller:2015fba}
M\"oller P.,  Sierk A.~J.,  Ichikawa T.,   Sagawa H.,  2016, \mn@doi [Atom.
  Data Nucl. Data Tabl.] {10.1016/j.adt.2015.10.002}, 109-110, 1

\bibitem[\protect\citeauthoryear{Morozova, Piro, Renzo, Ott, Clausen, Couch,
  Ellis  \& Roberts}{Morozova et~al.}{2015}]{Morozova:2015bla}
Morozova V.,  Piro A.,  Renzo M.,  Ott C.,  Clausen D.,  Couch S.,  Ellis J.,
  Roberts L.,  2015, \mn@doi [Astrophys.J.] {10.1088/0004-637X/814/1/63}, 814,
  63

\bibitem[\protect\citeauthoryear{Morozova, Piro, Renzo  \& Ott}{Morozova
  et~al.}{2016}]{Morozova:2016asf}
Morozova V.,  Piro A.~L.,  Renzo M.,   Ott C.~D.,  2016, \mn@doi [Astrophys.J.]
  {10.3847/0004-637X/829/2/109}, 829, 109

\bibitem[\protect\citeauthoryear{Morozova, Piro  \& Valenti}{Morozova
  et~al.}{2017}]{Morozova:2016efp}
Morozova V.,  Piro A.~L.,   Valenti S.,  2017, \mn@doi [Astrophys.J.]
  {10.3847/1538-4357/aa6251}, 838, 28

\bibitem[\protect\citeauthoryear{Morozova, Piro  \& Valenti}{Morozova
  et~al.}{2018}]{Morozova:2017hbk}
Morozova V.,  Piro A.~L.,   Valenti S.,  2018, \mn@doi [Astrophys.J.]
  {10.3847/1538-4357/aab9a6}, 858, 15

\bibitem[\protect\citeauthoryear{Morozova, Piro, Fuller  \& Van~Dyk}{Morozova
  et~al.}{2020}]{Morozova:2019hiu}
Morozova V.,  Piro A.~L.,  Fuller J.,   Van~Dyk S.~D.,  2020, \mn@doi
  [Astrophys.J.Lett.] {10.3847/2041-8213/ab77c8}, 891, L32

\bibitem[\protect\citeauthoryear{Murguia-Berthier et~al.,}{Murguia-Berthier
  et~al.}{2017}]{Murguia-Berthier:2017kkn}
Murguia-Berthier A.,  et~al., 2017, \mn@doi [Astrophys.J.Lett.]
  {10.3847/2041-8213/aa91b3}, 848, L34

\bibitem[\protect\citeauthoryear{Mösta, Radice, Haas, Schnetter  \&
  Bernuzzi}{Mösta et~al.}{2020}]{Mosta:2020hlh}
Mösta P.,  Radice D.,  Haas R.,  Schnetter E.,   Bernuzzi S.,  2020, \mn@doi
  [Astrophys.J.Lett.] {10.3847/2041-8213/abb6ef}, 901, L37

\bibitem[\protect\citeauthoryear{Nakar}{Nakar}{2007}]{Nakar:2007yr}
Nakar E.,  2007, \mn@doi [Phys.Rept.] {10.1016/j.physrep.2007.02.005}, 442, 166

\bibitem[\protect\citeauthoryear{Nakar \& Piran}{Nakar \&
  Piran}{2017}]{Nakar:2016cih}
Nakar E.,  Piran T.,  2017, \mn@doi [Astrophys. J.]
  {10.3847/1538-4357/834/1/28}, 834, 28

\bibitem[\protect\citeauthoryear{Nativi, Bulla, Rosswog, Lundman, Kowal, Gizzi,
  Lamb  \& Perego}{Nativi et~al.}{2020}]{Nativi:2020moj}
Nativi L.,  Bulla M.,  Rosswog S.,  Lundman C.,  Kowal G.,  Gizzi D.,  Lamb
  G.~P.,   Perego A.,  2020, \mn@doi [Mon. Not. Roy. Astron. Soc.]
  {10.1093/mnras/staa3337}, 500, 1772

\bibitem[\protect\citeauthoryear{Nedora, Bernuzzi, Radice, Perego, Endrizzi  \&
  Ortiz}{Nedora et~al.}{2019}]{Nedora:2019jhl}
Nedora V.,  Bernuzzi S.,  Radice D.,  Perego A.,  Endrizzi A.,   Ortiz N.,
  2019, \mn@doi [Astrophys.J.Lett.] {10.3847/2041-8213/ab5794}, 886, L30

\bibitem[\protect\citeauthoryear{Nedora et~al.,}{Nedora
  et~al.}{2020}]{Nedora:2020qtd}
Nedora V.,  et~al., 2020, arXiv:2011.11110

\bibitem[\protect\citeauthoryear{Nedora, Radice, Bernuzzi, Perego, Daszuta,
  Endrizzi, Prakash  \& Schianchi}{Nedora et~al.}{2021a}]{Nedora:2021eoj}
Nedora V.,  Radice D.,  Bernuzzi S.,  Perego A.,  Daszuta B.,  Endrizzi A.,
  Prakash A.,   Schianchi F.,  2021a, arXiv:2104.04537

\bibitem[\protect\citeauthoryear{Nedora et~al.,}{Nedora
  et~al.}{2021b}]{Nedora:2020hxc}
Nedora V.,  et~al., 2021b, \mn@doi [Astrophys.J.] {10.3847/1538-4357/abc9be},
  906, 98

\bibitem[\protect\citeauthoryear{Nicholl et~al.,}{Nicholl
  et~al.}{2017}]{Nicholl:2017ahq}
Nicholl M.,  et~al., 2017, \mn@doi [Astrophys.J.Lett.]
  {10.3847/2041-8213/aa9029}, 848, L18

\bibitem[\protect\citeauthoryear{Oechslin, Janka  \& Marek}{Oechslin
  et~al.}{2007}]{Oechslin:2006uk}
Oechslin R.,  Janka H.-T.,   Marek A.,  2007, \mn@doi [Astron.Astrophys.]
  {10.1051/0004-6361:20066682}, 467, 395

\bibitem[\protect\citeauthoryear{Paczynski}{Paczynski}{1983}]{paczynski1983models}
Paczynski B.,  1983, The Astrophysical Journal, 267, 315

\bibitem[\protect\citeauthoryear{Perego, Rosswog, Cabezón, Korobkin, Käppeli,
  Arcones  \& Liebendörfer}{Perego et~al.}{2014}]{Perego:2014fma}
Perego A.,  Rosswog S.,  Cabezón R.~M.,  Korobkin O.,  Käppeli R.,  Arcones
  A.,   Liebendörfer M.,  2014, \mn@doi [Mon.Not.Roy.Astron.Soc.]
  {10.1093/mnras/stu1352}, 443, 3134

\bibitem[\protect\citeauthoryear{Perego, Radice  \& Bernuzzi}{Perego
  et~al.}{2017}]{Perego:2017wtu}
Perego A.,  Radice D.,   Bernuzzi S.,  2017, \mn@doi [Astrophys.J.Lett.]
  {10.3847/2041-8213/aa9ab9}, 850, L37

\bibitem[\protect\citeauthoryear{Perego et~al.,}{Perego
  et~al.}{2020}]{Perego:2020evn}
Perego A.,  et~al., 2020, arXiv:2009.08988

\bibitem[\protect\citeauthoryear{Perego, Thielemann  \& Cescutti}{Perego
  et~al.}{2021}]{Perego:2021dpw}
Perego A.,  Thielemann F.-K.,   Cescutti G.,  2021, \mn@doi [arXiv:2109.09162]
  {10.1007/978-981-15-4702-7_13-1}

\bibitem[\protect\citeauthoryear{Pian et~al.,}{Pian
  et~al.}{2017}]{Pian:2017gtc}
Pian E.,  et~al., 2017, \mn@doi [Nature] {10.1038/nature24298}, 551, 67

\bibitem[\protect\citeauthoryear{Piro \& Kollmeier}{Piro \&
  Kollmeier}{2018}]{Piro:2017ayh}
Piro A.~L.,  Kollmeier J.~A.,  2018, \mn@doi [Astrophys. J.]
  {10.3847/1538-4357/aaaab3}, 855, 103

\bibitem[\protect\citeauthoryear{Piro \& Morozova}{Piro \&
  Morozova}{2016}]{Piro:2015kro}
Piro A.~L.,  Morozova V.~S.,  2016, \mn@doi [Astrophys.J.]
  {10.3847/0004-637X/826/1/96}, 826, 96

\bibitem[\protect\citeauthoryear{Prakash et~al.,}{Prakash
  et~al.}{2021}]{Prakash:2021wpz}
Prakash A.,  et~al., 2021, arXiv:2106.07885

\bibitem[\protect\citeauthoryear{Radice}{Radice}{2017}]{Radice:2017zta}
Radice D.,  2017, \mn@doi [Astrophys.J.Lett.] {10.3847/2041-8213/aa6483}, 838,
  L2

\bibitem[\protect\citeauthoryear{Radice}{Radice}{2020}]{Radice:2020ids}
Radice D.,  2020, \mn@doi [Symmetry] {10.3390/sym12081249}, 12, 1249

\bibitem[\protect\citeauthoryear{Radice \& Rezzolla}{Radice \&
  Rezzolla}{2012}]{Radice:2012cu}
Radice D.,  Rezzolla L.,  2012, \mn@doi [Astron.Astrophys.]
  {10.1051/0004-6361/201219735}, 547, A26

\bibitem[\protect\citeauthoryear{Radice, Rezzolla  \& Galeazzi}{Radice
  et~al.}{2014a}]{Radice:2013xpa}
Radice D.,  Rezzolla L.,   Galeazzi F.,  2014a, \mn@doi [Class.Quant.Grav.]
  {10.1088/0264-9381/31/7/075012}, 31

\bibitem[\protect\citeauthoryear{Radice, Rezzolla  \& Galeazzi}{Radice
  et~al.}{2014b}]{Radice:2013hxh}
Radice D.,  Rezzolla L.,   Galeazzi F.,  2014b, \mn@doi
  [Mon.Not.Roy.Astron.Soc.] {10.1093/mnrasl/slt137}, 437, L46

\bibitem[\protect\citeauthoryear{Radice, Rezzolla  \& Galeazzi}{Radice
  et~al.}{2015}]{Radice:2015nva}
Radice D.,  Rezzolla L.,   Galeazzi F.,  2015, ASP Conf.Ser., 498, 121

\bibitem[\protect\citeauthoryear{Radice, Galeazzi, Lippuner, Roberts, Ott  \&
  Rezzolla}{Radice et~al.}{2016}]{Radice:2016dwd}
Radice D.,  Galeazzi F.,  Lippuner J.,  Roberts L.~F.,  Ott C.~D.,   Rezzolla
  L.,  2016, \mn@doi [Mon.Not.Roy.Astron.Soc.] {10.1093/mnras/stw1227}, 460,
  3255

\bibitem[\protect\citeauthoryear{Radice, Perego, Hotokezaka, Fromm, Bernuzzi
  \& Roberts}{Radice et~al.}{2018}]{Radice:2018pdn}
Radice D.,  Perego A.,  Hotokezaka K.,  Fromm S.~A.,  Bernuzzi S.,   Roberts
  L.~F.,  2018, \mn@doi [Astrophys.J.] {10.3847/1538-4357/aaf054}, 869, 130

\bibitem[\protect\citeauthoryear{Radice, Bernuzzi  \& Perego}{Radice
  et~al.}{2020}]{Radice:2020ddv}
Radice D.,  Bernuzzi S.,   Perego A.,  2020, \mn@doi [Ann.Rev.Nucl.Part.Sci.]
  {10.1146/annurev-nucl-013120-114541}, 70, 95

\bibitem[\protect\citeauthoryear{Roberts, Kasen, Lee  \& Ramirez-Ruiz}{Roberts
  et~al.}{2011}]{Roberts:2011xz}
Roberts L.~F.,  Kasen D.,  Lee W.~H.,   Ramirez-Ruiz E.,  2011, \mn@doi
  [Astrophys.J.Lett.] {10.1088/2041-8205/736/1/L21}, 736, L21

\bibitem[\protect\citeauthoryear{{Rodrigo} \& {Solano}}{{Rodrigo} \&
  {Solano}}{2020}]{2020sea..confE.182R}
{Rodrigo} C.,  {Solano} E.,  2020, in XIV.0 Scientific Meeting (virtual) of the
  Spanish Astronomical Society. p.~182

\bibitem[\protect\citeauthoryear{{Rodrigo}, {Solano}  \& {Bayo}}{{Rodrigo}
  et~al.}{2012}]{2012ivoa.rept.1015R}
{Rodrigo} C.,  {Solano} E.,   {Bayo} A.,  2012, {SVO Filter Profile Service
  Version 1.0}, IVOA Working Draft 15 October 2012,
  \mn@doi{10.5479/ADS/bib/2012ivoa.rept.1015R}

\bibitem[\protect\citeauthoryear{Rossi et~al.,}{Rossi
  et~al.}{2020}]{Rossi:2019fnm}
Rossi A.,  et~al., 2020, \mn@doi [Mon.Not.Roy.Astron.Soc.]
  {10.1093/mnras/staa479}, 493, 3379

\bibitem[\protect\citeauthoryear{Rosswog \& Davies}{Rosswog \&
  Davies}{2002}]{Rosswog:2001fh}
Rosswog S.,  Davies M.,  2002, \mn@doi [Mon.Not.Roy.Astron.Soc.]
  {10.1046/j.1365-8711.2002.05409.x}, 334, 481

\bibitem[\protect\citeauthoryear{Rosswog \& Liebendoerfer}{Rosswog \&
  Liebendoerfer}{2003}]{Rosswog:2003rv}
Rosswog S.,  Liebendoerfer M.,  2003, \mn@doi [Mon.Not.Roy.Astron.Soc.]
  {10.1046/j.1365-8711.2003.06579.x}, 342, 673

\bibitem[\protect\citeauthoryear{Rosswog, Liebendoerfer, Thielemann, Davies,
  Benz  \& Piran}{Rosswog et~al.}{1999}]{Rosswog:1998hy}
Rosswog S.,  Liebendoerfer M.,  Thielemann F.,  Davies M.,  Benz W.,   Piran
  T.,  1999, Astron.Astrophys., 341, 499

\bibitem[\protect\citeauthoryear{Rosswog, Ramirez-Ruiz  \& Davies}{Rosswog
  et~al.}{2003}]{Rosswog:2003tn}
Rosswog S.,  Ramirez-Ruiz E.,   Davies M.~B.,  2003, \mn@doi
  [Mon.Not.Roy.Astron.Soc.] {10.1046/j.1365-2966.2003.07032.x}, 345, 1077

\bibitem[\protect\citeauthoryear{Rosswog, Piran  \& Nakar}{Rosswog
  et~al.}{2013}]{Rosswog:2012wb}
Rosswog S.,  Piran T.,   Nakar E.,  2013, \mn@doi [Mon.Not.Roy.Astron.Soc.]
  {10.1093/mnras/sts708}, 430, 2585

\bibitem[\protect\citeauthoryear{Rosswog, Korobkin, Arcones, Thielemann  \&
  Piran}{Rosswog et~al.}{2014}]{Rosswog:2013kqa}
Rosswog S.,  Korobkin O.,  Arcones A.,  Thielemann F.,   Piran T.,  2014,
  \mn@doi [Mon.Not.Roy.Astron.Soc.] {10.1093/mnras/stt2502}, 439, 744

\bibitem[\protect\citeauthoryear{Rosswog, Feindt, Korobkin, Wu, Sollerman,
  Goobar  \& Martinez-Pinedo}{Rosswog et~al.}{2017}]{Rosswog:2016dhy}
Rosswog S.,  Feindt U.,  Korobkin O.,  Wu M.~R.,  Sollerman J.,  Goobar A.,
  Martinez-Pinedo G.,  2017, \mn@doi [Class. Quant. Grav.]
  {10.1088/1361-6382/aa68a9}, 34, 104001

\bibitem[\protect\citeauthoryear{Rosswog, Sollerman, Feindt, Goobar, Korobkin,
  Wollaeger, Fremling  \& Kasliwal}{Rosswog et~al.}{2018}]{Rosswog:2017sdn}
Rosswog S.,  Sollerman J.,  Feindt U.,  Goobar A.,  Korobkin O.,  Wollaeger R.,
   Fremling C.,   Kasliwal M.,  2018, \mn@doi [Astron.Astrophys.]
  {10.1051/0004-6361/201732117}, 615, A132

\bibitem[\protect\citeauthoryear{Roth \& Kasen}{Roth \&
  Kasen}{2015}]{Roth:2014wda}
Roth N.,  Kasen D.,  2015, \mn@doi [Astrophys. J. Suppl.]
  {10.1088/0067-0049/217/1/9}, 217, 9

\bibitem[\protect\citeauthoryear{Ruffert, Janka  \& Schaefer}{Ruffert
  et~al.}{1996}]{Ruffert:1995fs}
Ruffert M.,  Janka H.,   Schaefer G.,  1996, Astron.Astrophys., 311, 532

\bibitem[\protect\citeauthoryear{Sekiguchi, Kiuchi, Kyutoku  \&
  Shibata}{Sekiguchi et~al.}{2011}]{Sekiguchi:2011zd}
Sekiguchi Y.,  Kiuchi K.,  Kyutoku K.,   Shibata M.,  2011, \mn@doi
  [Phys.Rev.Lett.] {10.1103/PhysRevLett.107.051102}, 107

\bibitem[\protect\citeauthoryear{Sekiguchi, Kiuchi, Kyutoku  \&
  Shibata}{Sekiguchi et~al.}{2015}]{Sekiguchi:2015dma}
Sekiguchi Y.,  Kiuchi K.,  Kyutoku K.,   Shibata M.,  2015, \mn@doi
  [Phys.Rev.D] {10.1103/PhysRevD.91.064059}, 91

\bibitem[\protect\citeauthoryear{Sekiguchi, Kiuchi, Kyutoku, Shibata  \&
  Taniguchi}{Sekiguchi et~al.}{2016}]{Sekiguchi:2016bjd}
Sekiguchi Y.,  Kiuchi K.,  Kyutoku K.,  Shibata M.,   Taniguchi K.,  2016,
  \mn@doi [Phys.Rev.D] {10.1103/PhysRevD.93.124046}, 93

\bibitem[\protect\citeauthoryear{Shibata \& Hotokezaka}{Shibata \&
  Hotokezaka}{2019}]{Shibata:2019wef}
Shibata M.,  Hotokezaka K.,  2019, \mn@doi [Ann.Rev.Nucl.Part.Sci.]
  {10.1146/annurev-nucl-101918-023625}, 69, 41

\bibitem[\protect\citeauthoryear{Shibata, Fujibayashi  \& Sekiguchi}{Shibata
  et~al.}{2021}]{Shibata:2021bbj}
Shibata M.,  Fujibayashi S.,   Sekiguchi Y.,  2021, \mn@doi [Phys.Rev.D]
  {10.1103/PhysRevD.103.043022}, 103

\bibitem[\protect\citeauthoryear{Siegel}{Siegel}{2019}]{Siegel:2019mlp}
Siegel D.~M.,  2019, \mn@doi [Eur.Phys.J.A] {10.1140/epja/i2019-12888-9}, 55,
  203

\bibitem[\protect\citeauthoryear{Siegel \& Metzger}{Siegel \&
  Metzger}{2018}]{Siegel:2017jug}
Siegel D.~M.,  Metzger B.~D.,  2018, \mn@doi [Astrophys.J.]
  {10.3847/1538-4357/aabaec}, 858, 52

\bibitem[\protect\citeauthoryear{Siegel, Ciolfi  \& Rezzolla}{Siegel
  et~al.}{2014}]{Siegel:2014ita}
Siegel D.~M.,  Ciolfi R.,   Rezzolla L.,  2014, \mn@doi [Astrophys.J.Lett.]
  {10.1088/2041-8205/785/1/L6}, 785, L6

\bibitem[\protect\citeauthoryear{Smartt et~al.,}{Smartt
  et~al.}{2017}]{Smartt:2017fuw}
Smartt S.,  et~al., 2017, \mn@doi [Nature] {10.1038/nature24303}, 551, 75

\bibitem[\protect\citeauthoryear{Soares-Santos et~al.,}{Soares-Santos
  et~al.}{2017}]{Soares-Santos:2017lru}
Soares-Santos M.,  et~al., 2017, \mn@doi [Astrophys.J.Lett.]
  {10.3847/2041-8213/aa9059}, 848, L16

\bibitem[\protect\citeauthoryear{Steiner, Hempel  \& Fischer}{Steiner
  et~al.}{2013}]{Steiner:2012rk}
Steiner A.~W.,  Hempel M.,   Fischer T.,  2013, \mn@doi [Astrophys.J.]
  {10.1088/0004-637X/774/1/17}, 774, 17

\bibitem[\protect\citeauthoryear{{Stone}, {Tomida}, {White}  \&
  {Felker}}{{Stone} et~al.}{2020}]{2020ApJS..249....4S}
{Stone} J.~M.,  {Tomida} K.,  {White} C.~J.,   {Felker} K.~G.,  2020, \mn@doi
  [\apjs] {10.3847/1538-4365/ab929b}, \href
  {https://ui.adsabs.harvard.edu/abs/2020ApJS..249....4S} {249, 4}

\bibitem[\protect\citeauthoryear{Tanaka \& Hotokezaka}{Tanaka \&
  Hotokezaka}{2013}]{Tanaka:2013ana}
Tanaka M.,  Hotokezaka K.,  2013, \mn@doi [Astrophys.J.]
  {10.1088/0004-637X/775/2/113}, 775, 113

\bibitem[\protect\citeauthoryear{Tanaka et~al.,}{Tanaka
  et~al.}{2017}]{Tanaka:2017qxj}
Tanaka M.,  et~al., 2017, \mn@doi [Publ.Astron.Soc.Jap.] {10.1093/pasj/psx121},
  69

\bibitem[\protect\citeauthoryear{Tanaka, Kato, Gaigalas, Rynkun, Radziute
  et~al.}{Tanaka et~al.}{2018}]{Tanaka:2017lxb}
Tanaka M.,  Kato D.,  Gaigalas G.,  Rynkun P.,  Radziute L.,   et~al., 2018,
  \mn@doi [Astrophys.J.] {10.3847/1538-4357/aaa0cb}, 852, 109

\bibitem[\protect\citeauthoryear{Tanaka, Kato, Gaigalas  \& Kawaguchi}{Tanaka
  et~al.}{2020}]{Tanaka:2019iqp}
Tanaka M.,  Kato D.,  Gaigalas G.,   Kawaguchi K.,  2020, \mn@doi
  [Mon.Not.Roy.Astron.Soc.] {10.1093/mnras/staa1576}, 496, 1369

\bibitem[\protect\citeauthoryear{Tanvir, Levan, Fruchter, Hjorth, Wiersema,
  Tunnicliffe  \& de Ugarte~Postigo}{Tanvir et~al.}{2013}]{Tanvir:2013pia}
Tanvir N.,  Levan A.,  Fruchter A.,  Hjorth J.,  Wiersema K.,  Tunnicliffe R.,
   de Ugarte~Postigo A.,  2013, \mn@doi [Nature] {10.1038/nature12505}, 500,
  547

\bibitem[\protect\citeauthoryear{Tanvir et~al.,}{Tanvir
  et~al.}{2017}]{Tanvir:2017pws}
Tanvir N.,  et~al., 2017, \mn@doi [Astrophys.J.Lett.]
  {10.3847/2041-8213/aa90b6}, 848, L27

\bibitem[\protect\citeauthoryear{Troja et~al.,}{Troja
  et~al.}{2017}]{Troja:2017nqp}
Troja E.,  et~al., 2017, \mn@doi [Nature] {10.1038/nature24290}, 551, 71

\bibitem[\protect\citeauthoryear{Troja et~al.,}{Troja
  et~al.}{2019}]{Troja:2019ccb}
Troja E.,  et~al., 2019, \mn@doi [Mon.Not.Roy.Astron.Soc.]
  {10.1093/mnras/stz2255}, 489, 2104

\bibitem[\protect\citeauthoryear{Typel, Ropke, Klahn, Blaschke  \&
  Wolter}{Typel et~al.}{2010}]{Typel:2009sy}
Typel S.,  Ropke G.,  Klahn T.,  Blaschke D.,   Wolter H.,  2010, \mn@doi
  [Phys.Rev.C] {10.1103/PhysRevC.81.015803}, 81

\bibitem[\protect\citeauthoryear{Villar et~al.,}{Villar
  et~al.}{2017}]{Villar:2017wcc}
Villar V.~A.,  et~al., 2017, \mn@doi [Astrophys.J.Lett.]
  {10.3847/2041-8213/aa9c84}, 851, L21

\bibitem[\protect\citeauthoryear{Vincent, Foucart, Duez, Haas, Kidder, Pfeiffer
   \& Scheel}{Vincent et~al.}{2020}]{Vincent:2019kor}
Vincent T.,  Foucart F.,  Duez M.~D.,  Haas R.,  Kidder L.~E.,  Pfeiffer H.~P.,
    Scheel M.~A.,  2020, \mn@doi [Phys.Rev.D] {10.1103/PhysRevD.101.044053},
  101

\bibitem[\protect\citeauthoryear{Waxman, Ofek, Kushnir  \& Gal-Yam}{Waxman
  et~al.}{2018}]{Waxman:2017sqv}
Waxman E.,  Ofek E.~O.,  Kushnir D.,   Gal-Yam A.,  2018, \mn@doi
  [Mon.Not.Roy.Astron.Soc.] {10.1093/mnras/sty2441}, 481, 3423

\bibitem[\protect\citeauthoryear{Waxman, Ofek  \& Kushnir}{Waxman
  et~al.}{2019}]{Waxman:2019png}
Waxman E.,  Ofek E.~O.,   Kushnir D.,  2019, \mn@doi [Astrophys.J.]
  {10.3847/1538-4357/ab1f71}, 878, 93

\bibitem[\protect\citeauthoryear{Weiss, Hillebrandt, Thomas  \& Ritter}{Weiss
  et~al.}{2004}]{weiss2004cox}
Weiss A.,  Hillebrandt W.,  Thomas H.-C.,   Ritter H.,  2004, Cox and Giuli's
  Principles of Stellar Structure

\bibitem[\protect\citeauthoryear{Wollaeger, van Rossum, Graziani, Couch,
  Jordan~IV, Lamb  \& Moses}{Wollaeger et~al.}{2013}]{wollaeger2013radiation}
Wollaeger R.~T.,  van Rossum D.~R.,  Graziani C.,  Couch S.~M.,  Jordan~IV
  G.~C.,  Lamb D.~Q.,   Moses G.~A.,  2013, The Astrophysical Journal
  Supplement Series, 209, 36

\bibitem[\protect\citeauthoryear{Wollaeger et~al.,}{Wollaeger
  et~al.}{2018}]{Wollaeger:2017ahm}
Wollaeger R.~T.,  et~al., 2018, \mn@doi [Mon.Not.Roy.Astron.Soc.]
  {10.1093/mnras/sty1018}, 478, 3298

\bibitem[\protect\citeauthoryear{Yang et~al.,}{Yang
  et~al.}{2015}]{Yang:2015pha}
Yang B.,  et~al., 2015, \mn@doi [Nature Commun.] {10.1038/ncomms8323}, 6, 7323

\bibitem[\protect\citeauthoryear{Zhu, Lund, Barnes, Sprouse, Vassh, McLaughlin,
  Mumpower  \& Surman}{Zhu et~al.}{2021a}]{Zhu_2021}
Zhu Y.~L.,  Lund K.~A.,  Barnes J.,  Sprouse T.~M.,  Vassh N.,  McLaughlin
  G.~C.,  Mumpower M.~R.,   Surman R.,  2021a, \mn@doi [The Astrophysical
  Journal] {10.3847/1538-4357/abc69e}, 906, 94

\bibitem[\protect\citeauthoryear{Zhu, Lund, Barnes, Sprouse, Vassh, McLaughlin,
  Mumpower  \& Surman}{Zhu et~al.}{2021b}]{Zhu:2020eyk}
Zhu Y.,  Lund K.,  Barnes J.,  Sprouse T.,  Vassh N.,  McLaughlin G.,  Mumpower
  M.,   Surman R.,  2021b, \mn@doi [Astrophys.J.] {10.3847/1538-4357/abc69e},
  906, 94

\makeatother
\end{thebibliography}

\appendix

\section{Impact of opacity formula}
\label{appendix_impact_of_opacity_formula}

\begin{figure}
\centering
\includegraphics[scale=0.5]{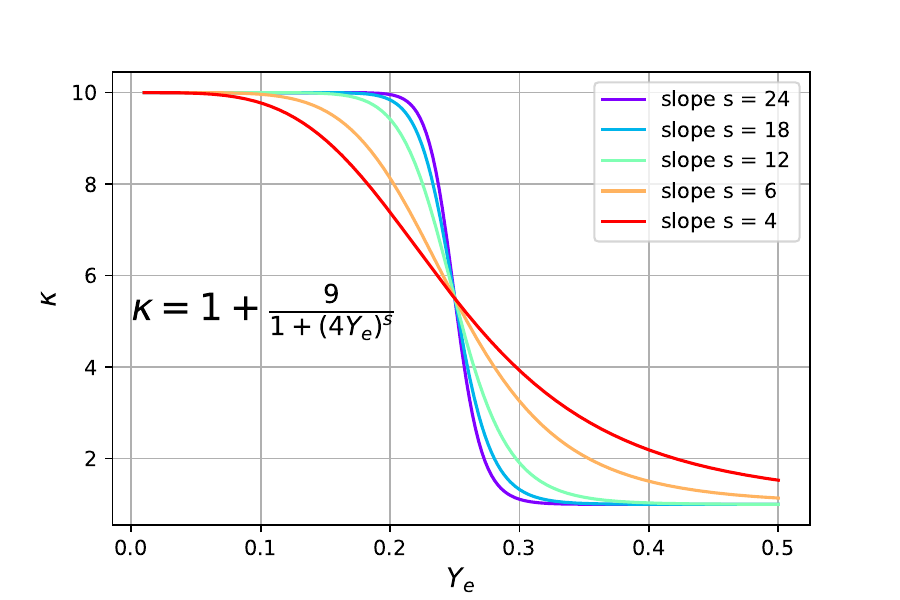}
\caption{Opacity as a function of initial $Y_e$. Different from Eq. \ref{opacity_ye}, the slope of the opacity transition near $Y_e = 0.25$ is a free parameter here, which is indicated by $s$. $s$ ranges from 4 to 24, and $s = 12$ is the baseline adopted in the main body of the paper.}
\label{opacity_slope}
\end{figure}

We study the sensitivity of kilonova light curves to the opacity formula mentioned in \S \ref{subsection_opacity}. We fix the maximum and minimum opacity to 10 cm$^2$ g$^{-1}$ and 1 cm$^2$ g$^{-1}$ respectively, and also fix the intermediate point ($Y_e = 0.25, \kappa = 5.5$ cm$^2$ g$^{-1}$). We explore the impact of the slope of the transition near $Y_e = 0.25$, which is indicated by parameter $s$ in the following formula:
\begin{equation}
	\label{opacity_ye_slope}
	\kappa = 1 + \frac{9}{1+(4Y_e)^{s}}~~ \mathrm{ [cm^2 g^{-1}]}.
\end{equation}
Fig.~\ref{opacity_slope} shows the range of the slope we test, with $s = 12$ being the baseline used in the body of the paper. $s = 24$ results in the sharpest transition, while $s = 4$ produces the mildest transition so that opacity cannot reach its minimum at $Y_e = 0.5$. 

\begin{figure}
\centering
\includegraphics[scale=0.5]{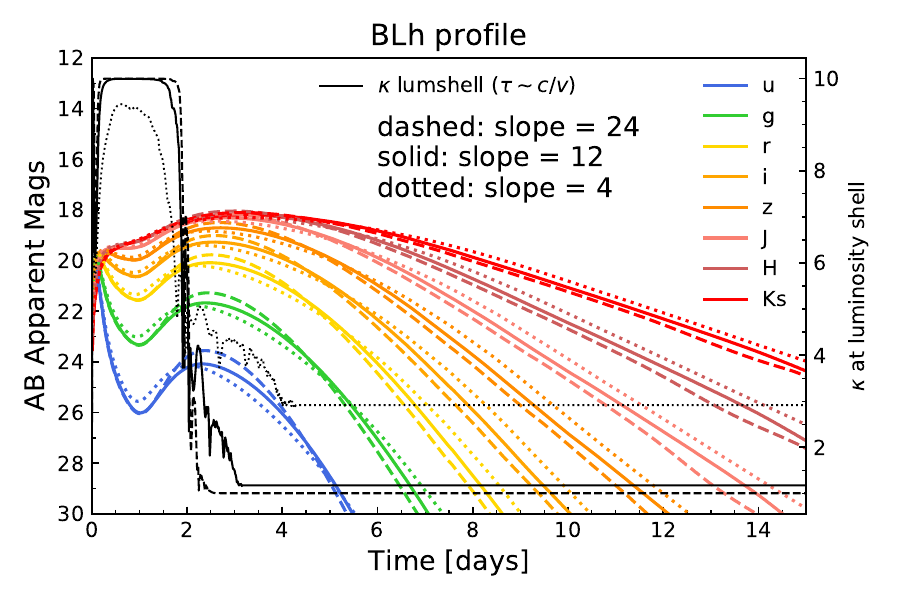}
\caption{AB magnitudes and the opacity at the luminosity shell for BLh model.
%% The dashed, solid, and dotted lines show the results of $s$ = 24, 12, and 4 respectively.
Although the magnitudes change a little, the morphology of the light curves remains unchanged in general.}
\label{ABmags_opacity_blh_opacity_slope}
\end{figure}

Fig.~\ref{ABmags_opacity_blh_opacity_slope} shows the AB magnitudes and the opacity at the luminosity shell for the BLh binary, using $s$ = 24, 12, and 4. The definition of luminosity shell is given in \S \ref{subsection_general_features}. The outermost fast high-$Y_e$ component is not affected by the modification of the opacity formula, producing the first peak of the light curve. For $s$ = 4, the effect of lanthanide curtain is alleviated, but still present due to the very low $Y_e$ ($\sim$0.15) of the component shown in Fig.~\ref{blh_profile}. As  shown by the opacity at luminosity shell, the opacity plateau is only a little smaller than 10 cm$^2$ g$^{-1}$, so most of the radiation is trapped inside. At late times, for $s$ = 4, the opacity of the inner high-$Y_e$ component increases compared to the baseline. Therefore, it's natural that the radiation is inhibited, and that the kilonova is redder and becomes transparent later. However, these are only minor changes to the light curves, and in general, the results are not sensitive to the opacity formula. 

\begin{figure}
\centering
\includegraphics[scale=0.5]{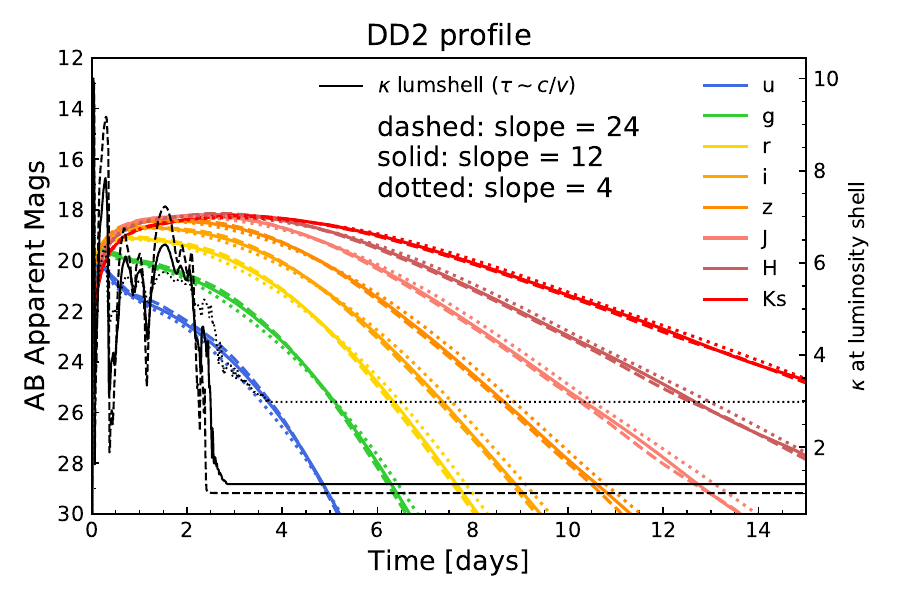}
\caption{AB magnitudes and the opacity at the luminosity shell for DD2 model.
%% The dashed, solid, and dotted lines show the results of $s$ = 24, 12, and 4 respectively.
The light curves are insensitive to the slope of the transition in the opacity formula.}
\label{ABmags_opacity_DD2_opacity_slope}
\end{figure}

We do not report the results for the SFHo binary, but the opacity formula also has little impact on SFHo results. The DD2 binary is shown in Fig.~\ref{ABmags_opacity_DD2_opacity_slope}. For this model, we find that the slope of the opacity profile has essentially no impact. In fact, the outer part of the DD2 profile has a $Y_e$ near 0.25 (Fig.~\ref{DD2_profile}), so formulae with different slopes result in similar opacities.

%% \newtxt{
%% We conclude that for realistic models including BLh, DD2, and SFHo, variations in the slope in the opacity formula do not change the light curves significantly.}

\section{Boundary Velocity}
\label{appendix_boundary_velocity}
    At the outer boundary, the \texttt{SNEC} code sets pressure, temperature, and density to zero. Among them, only the pressure $p_{\text{imax}}$ is important since the other quantities are not actually used in the evolution. However, the $p_{\text{imax}}=0$ boundary condition can lead to a large pressure gradient at the boundary when the simulation begins. At that time, the ejecta is very hot ($\sim 10^9$ K) and $p_{\text{imax-1}}$ is dominated by the radiation pressure, which is proportional to $T_{\text{imax-1}}^4$. This discontinuity causes the velocity near outer boundary to increase to very large values, sometimes even exceeding the speed of light (e.g. BLh). 
    
\begin{figure}
\centering
\includegraphics[scale=0.5]{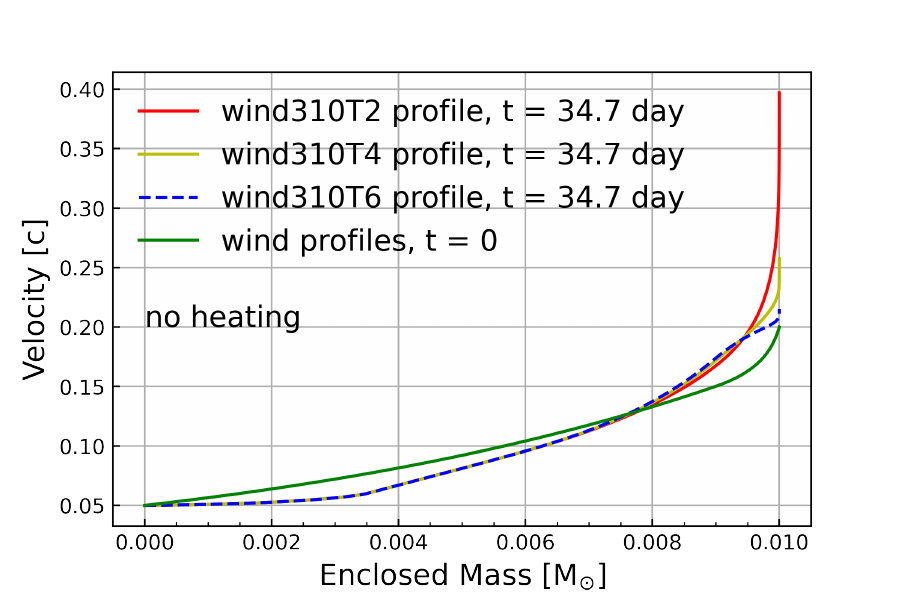}
\caption{Initial and final velocity as a function of mass for the wind310TX profiles. In this test, r-process heating is turned off to preclude its effects on the velocity. The green line shows the initial velocity distribution, while the other lines show the final velocity distribution using the modified temperature profiles with power law index = 2, 4, 6 respectively. A power law factor large enough for temperature effectively reduces the pressure gradients at the outer boundary, and thus mitigate the boundary velocity divergence problem.}
\label{wind310Tx_noheating}
\end{figure}

    For the wind profiles, one of the solutions is to modify the initial temperature distribution. Instead of using the uniform $10^9$ K, we use a powerlaw decay near the outer boundary, which is already introduced in Eq.~(\ref{wind_initial_teperature}). We tested various powerlaw indexes, and find that $\alpha \gtrsim$ 6 is enough to solve the problem (Fig.~\ref{wind310Tx_noheating}).   
    
\begin{figure}
\centering
\includegraphics[scale=0.5]{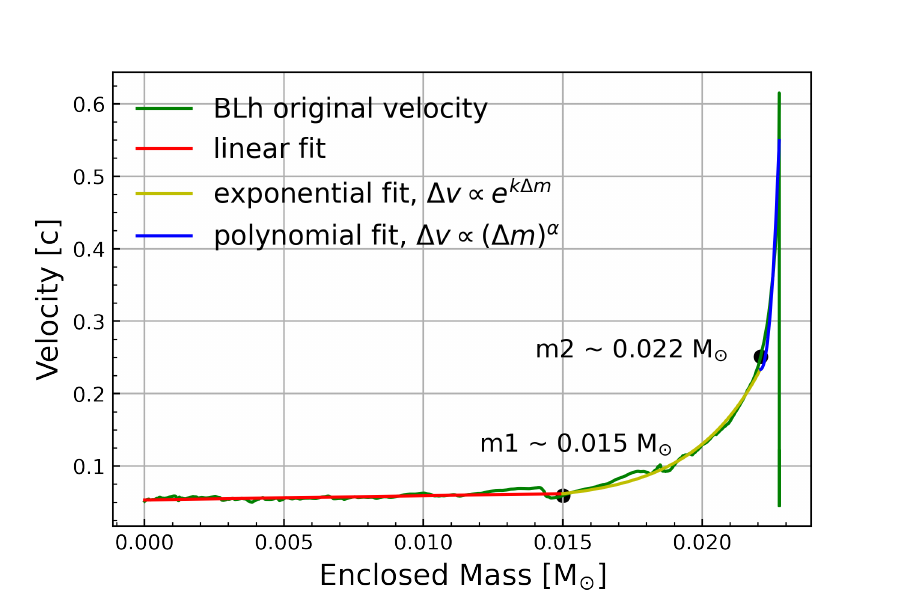}
\caption{Velocity profile for the BLh and the BLh with modified velocity (BLh-mvel) profiles. We use a piecewise function to fit the initial velocity in BLh profile. We set $m_1$ and $m_2$ to 0.015 and 0.022 $M_{\odot}$ respectively. When $m<m_1$, velocity in BLh-mvel profile grows linearly with $m$. When $m$ is between $m_1$ and $m_2$, ($v - v(m_1)$) is proportional to $e^{m - m_1}$. When $m>m_2$, we use the function $v - v(m_2) = C (m - m_2)^{\alpha}$ to fit.}
\label{blh-mvel_profile}
\end{figure}

\begin{figure}
\centering
\includegraphics[scale=0.5]{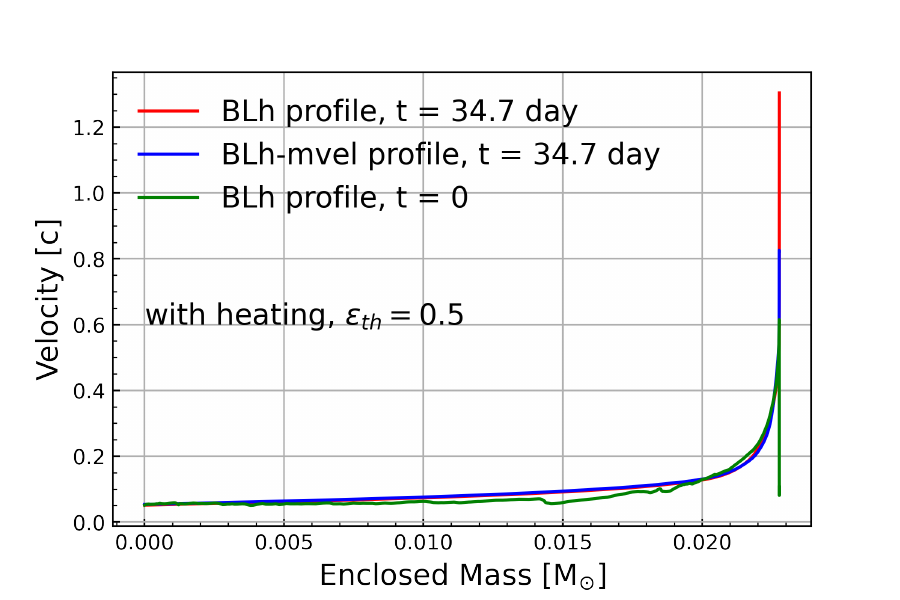}
\caption{Velocity as a function of mass for the BLh and BLh-mvel profiles. Due to unphysical pressure gradient at the outer boundary and r-process heating, the velocity at the outer boundary can even exceed the speed of light. The problem is less severe for BLh-mvel than BLh profile. Since the mass and energy near outer boundary only accounts for a very small part of the whole ejecta, we find that it does not affect light curves.}
\label{blh_blh-mvel_boundary_vel}
\end{figure}

    For realistic profiles, the problem can be alleviated by smoothing the initial velocity distribution. Figure \ref{blh-mvel_profile} shows the piecewise fit for BLh profile ($m < m_1$: linear; $m_1 < m < m_2$: exponential; $m > m_2$: polynomial). We call the new profile BLh-with-modified-velocity profile, or BLh-mvel profile. We show light curves produced with this modified profile in \S \ref{subsection_hydrodynamics}. With the BLh-mvel profile, the maximum velocity at the outer boundary is reduced to around 0.8~c (Figure \ref{blh_blh-mvel_boundary_vel}). 
    
    The above changes to the initial profiles indicate that the boundary velocity problem is profile-dependent. However, the final light curves are largely unaffected by these dynamics close to the outer boundary. This is because the region affected by the outer boundary encloses a small amount of material, as shown in Fig.~\ref{blh_blh-mvel_boundary_vel}. The increase of the kinetic energy due to the boundary velocity problem is not large enough to visibly affect the light curves, as can be observed by comparing the light curves obtained with the BLh and the BLh-mvel profiles shown in Fig.~\ref{CodeValidation-Hydro_blh-mvel2}.

\section{Method of BLh extrapolation}
\label{appendix_method_of_blh_extrapolation}
\begin{figure}
\centering
\includegraphics[scale=0.5]{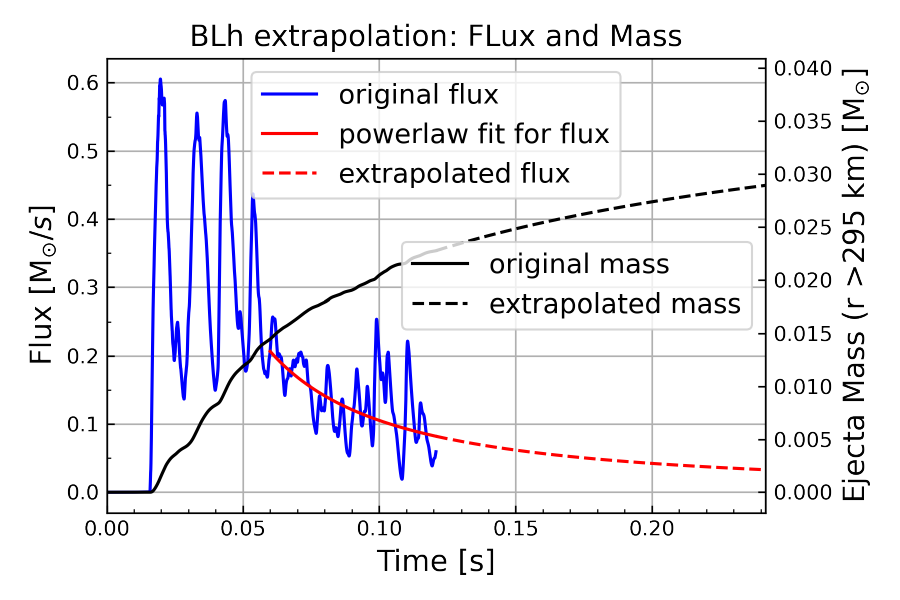}
\caption{Mass extrapolation for the BLh profile. We use a power law to fit ejecta flux after 0.06~s, and then extrapolate it to $t_{\text{end}}$. From the integration of the flux at 295~km, we obtain the mass of ejecta outside 295~km as a function of time (black lines, solid: original data; dashed: extrapolated).}
\label{FluxMass_blh_extrapolation}
\end{figure}

We extrapolate the BLh profile by fitting all thermodynamic quantities in time and then extrapolating them. Specifically, we integrate the outflow rate from the \texttt{WhiskyTHC} simulations to obtain the mass of the material that has crossed an extraction sphere with $r$ = 295~km as a function of time. We denote the mass of the ejecta still enclosed by $r$ = 295 km at time $t$ as $m(t)$. The mass of the material that has crossed the extraction sphere at any given time is as $M_{\text{tot}} - m(t)$. We use a power law to fit the mass flux after 0.06~s, and then extrapolate it to $t_{\text{end}}$. Figure \ref{FluxMass_blh_extrapolation} shows the case in which $t_{\text{end}} = 0.24$~s, that is twice the original simulation time for the BLh binary. Note that here the time is given from the beginning of \texttt{WhiskyTHC} simulations and includes the period before the merger. With the extrapolation, the total ejecta mass increases from 0.022~$M_{\odot}$ to 0.029~$M_{\odot}$. 

\begin{figure}
\centering
\includegraphics[scale=0.5]{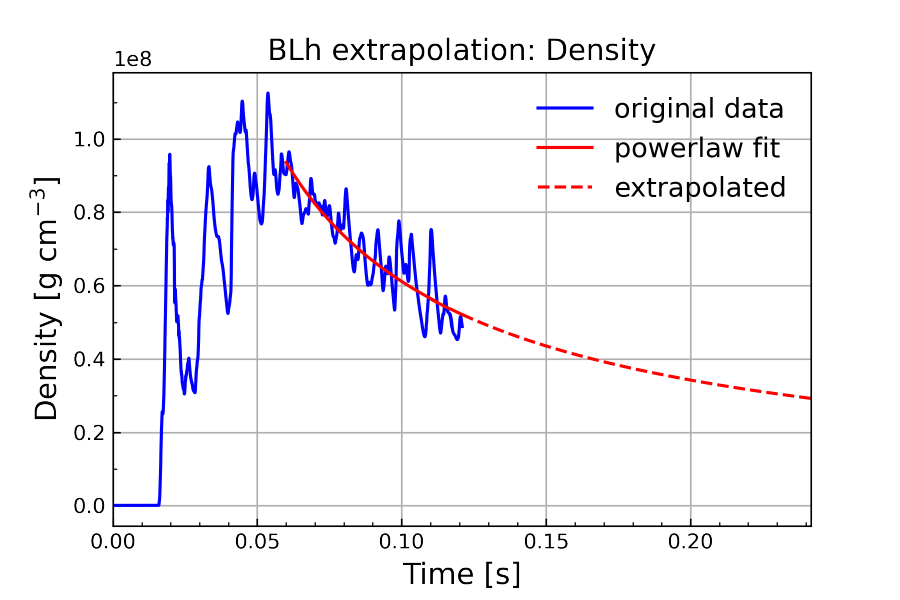}
\caption{Density extrapolation for the BLh profile. We first reconstruct the density at 295~km as a function of time according to the BLh profile and its mass flux at 295~km (blue line). Then, we use a power law to fit the density after 0.06~s (red solid line). Finally, we extrapolate the power law to $t_{\text{end}}$ (red dashed line), e.g. 0.24~s in the figure.}
\label{Density_blh_extrapolation}
\end{figure}

For each profile we have density, velocity, temperature, etc, as a function of enclosed mass. For instance, the density profile is $\rho(m)$. Since we know the function $m(t)$, we can use it to calculate the time at which each Lagrangian fluid element crosses the extraction sphere. From this we can obtain $\rho(t)$ on the extraction sphere. We fit $\rho(t)$ after 0.06 s with a power law and extrapolate it to $t_{\text{end}}$ (see Fig.~\ref{Density_blh_extrapolation}). As a last step, we convert the extrapolated $\rho(t)$ back to $\rho(m)$ and get the new profile.

This extrapolation methodology is not necessarily limited to power law extrapolation. Indeed, we use power law fits for the mass flux and the density, a linear function for the entropy, and a constant for temperature, velocity, initial $Y_e$, and expansion timescale.

\section{Energy Conservation for BLh profile}
\label{appendix_energy_conservation_blh}
\begin{figure}
\centering
\includegraphics[scale=0.5]{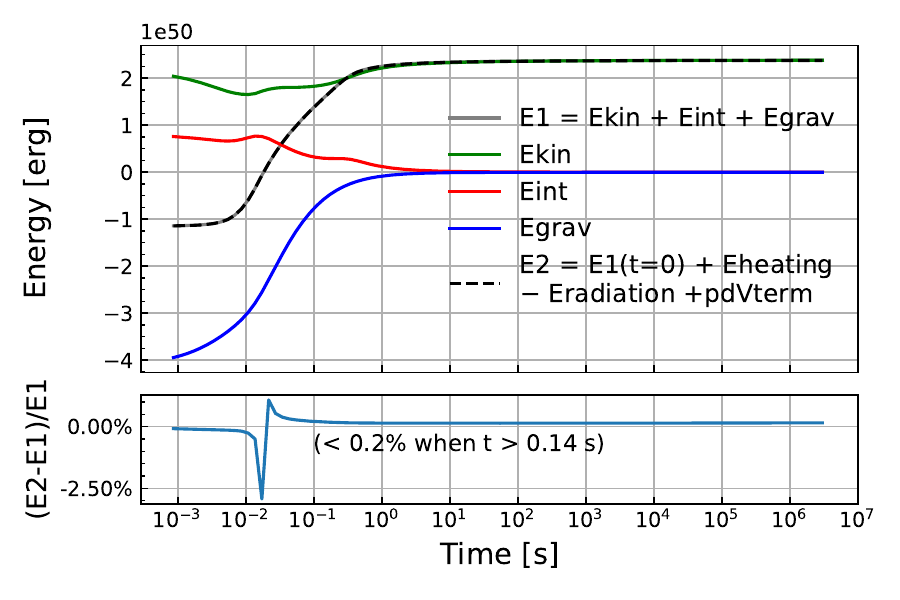}
\caption{Same as Fig.~\ref{E1E2_wind310T6_ye0.1} but for the BLh profile. The total energy is negative initially due to large gravitational energy, but soon becomes positive as a result of the mechanical work done on the inner boundary. The large spike in the relative difference between E1 and E2 is caused by total energy changing sign. The difference drops to below 0.2\% after 0.14 s, so energy is well conserved.}
\label{E1E2_blh}
\end{figure}

\begin{figure}
\centering
\includegraphics[scale=0.5]{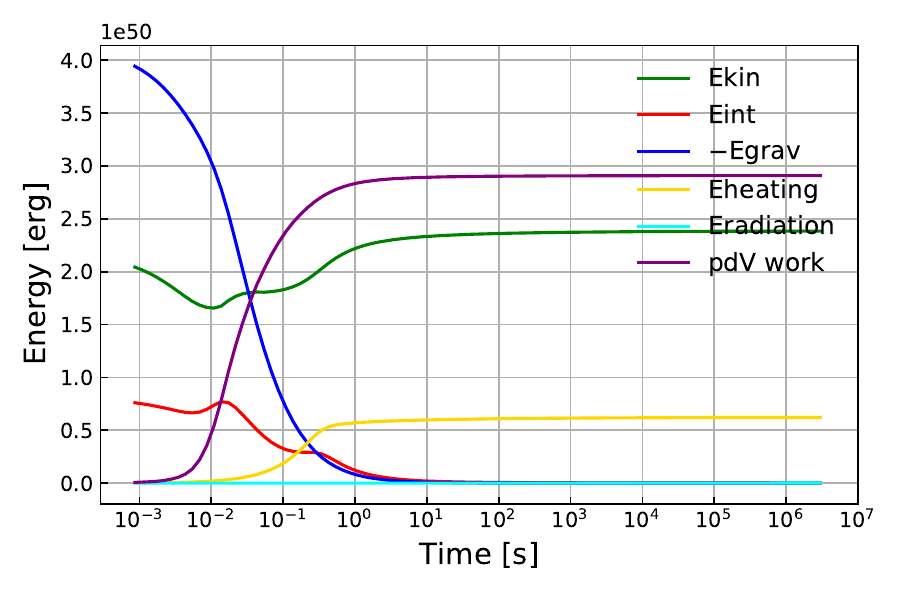}
\caption{Different energy terms as a function of time for the BLh model. The total energy is at first dominated by gravitational energy. $p{\rm d}V $ work at inner boundary and the r-process heating increase the total energy of the ejecta from negative to positive. Afterwards, the total energy is dominated by kinetic energy, like for the wind profiles. Only a small fraction of the energy is radiated as most of the specific internal energy is lost to expansion.}
\label{Energy_blh}
\end{figure}

We check energy conservation for the optimal wind profiles in \S \ref{subsection_energy_conservation}. Here, we repeat this analysis for the BLh profile. Other simulation profiles behave in a similar way. As shown in Fig.~\ref{E1E2_blh}, the total energy is initially negative, because the profile is initially still gravitationally bound. However, the mechanical work done on the inner boundary by pressure forces and r-process heating unbinds the ejecta. This is expected, since we use the Bernoulli criterion to identify the ejecta in the merger simulations \citep[e.g.,][]{Kastaun:2014fna}. After this initial phase that lasts about one second, the total energy of the ejecta is dominated by the kinetic energy (see Figure \ref{Energy_blh}), as was the case for the wind profiles. When the total energy crosses zero, there is a jump in the relative difference between E1 (the total energy of the ejecta) and E2 (initial ejecta energy + r-process heating + $p{\rm d}V$ work - radiated energy). After 0.14~s, the relative difference between E1 and E2 drops to below 0.2\%. We conclude that \texttt{SNEC} conserves energy very well with the adopted setup.
 
% Don't change these lines
\bsp	% typesetting comment
\label{lastpage}

\end{document}